\documentclass[a4paper,12pt]{article} 
\usepackage{graphicx}
\usepackage{times}
\usepackage{graphicx}
\usepackage{latexsym}
\usepackage{amsmath,amsfonts}
\usepackage{amssymb,mathrsfs}
\usepackage{authblk}
\usepackage{setspace,textcomp}

\usepackage[numbers,sort&compress]{natbib}
\usepackage{epsfig}
\usepackage{cite}
\newcommand{\dd}{\mathrm d}

\usepackage{color}
 \usepackage{hyperref}
\begin{document}
\title{Transonic behaviour and stability analysis of quasi-viscous black hole accretion}

\author[a]{Deepika B. Ananda\thanks{Present Address: Nicolaus Copernicus Astronomical Centre
of the Polish Academy of Sciences, Warsaw, Poland}}
\author[b]{Soumini Chaudhury}
\author[b]{Tapas K. Das\thanks{email: tapas@hri.res.in}}
\author[c]{Ishita Maity\thanks{Deceased}}
\author[d]{Sankhasubhra Nag}
\affil[a]{\small Indian Institute of Science Education and Research, Pune 411 008, India}
\affil[b]{\small Harish-Chandra Research Institute, Allahabad 211 019, India}
\affil[c]{\small Louisiana State University, Louisiana 70803, USA}
\affil[d]{\small Department of Physics, Sarojini Naidu College for Women, Kolkata 700 028, India}
\date{}
\maketitle
\begin{abstract}

\noindent
Analytical studies of black hole accretion usually presumes the
stability of the stationary transonic configuration.
Various authors in the past several decades demonstrated the validity
of such an assumption for inviscid
hydrodynamic flow. Inviscid approximation is a reasonable approach for low angular
momentum advection dominated flow in connection to certain
supermassive black holes at the centres of the galaxies (including our own) fed from a number of stellar donors. Introduction of a weak viscosity, as a first order linear correction involving the
viscosity parameter, however, may sometimes provide a more detail
understanding of the observed black hole spectra. The transonic behaviour of the stationary solutions have been studied for the aforementioned quasi-viscous accretion for all possible geometric configurations 
of axisymmetric flow. For a sufficiently low range of the viscosity
parameter, transonic solutions containing one or three critical points have been found for allowed ranges in the
astrophysical parameters under the post-Newtonian pseudo-Schwarzschild scheme.
With the introduction of such viscosity parameter, the only feasible
critical points are of saddle and spiral types in
contrary to the inviscid case where centre type points were formed instead of spiral ones. Introduction of linear perturbations on stationary flow solutions and their time evolution in both standing and radially propagating wave forms have been examined (completely analytically) in detail. 
Our analysis shows that similar kind of secular instability exists in all the considered disk models at large distance in the asymptotic limit, however, the model itself is valid only within a certain length scale and the disks sustain within that length scale only for at least a considerable time scale. 
\end{abstract}
%

\section{Introduction}

Viscous transport of angular momentum is a necessary phenomena to allow the matter to fall in through the event horizon for a rotation dominated axially symmetric accretion flow onto astrophysical black holes \citep{FKR02,Kato_08}. A proper viscosity prescription is imperative for the study of such flow having the Keplerian distribution of angular momentum.

Forty two years after the discovery of the standard accretion disk model 
\citep{shakura_sunayev_73,NT_73}, exact modeling of viscous transonic black hole accretion (including appropriate heating and cooling mechanism), however, is still quite an arduous task. This is true even for flow dynamics analyzed from a purely Newtonian framework - let alone for the general relativistic accretion disks. Over the past several years, various authors have pointed out that for low angular momentum advection dominated accretion (which can model the flow structure onto Sgr A*, for instance), large radial velocity of matter close to the black hole may imply that the infall time scale is much smaller compared to the corresponding viscous time scale \citep{Yuan2014}. This further indicates that the low angular momentum advective accretion may practically be considered to possess almost constant angular momentum. For such sub-Keplerian hot accretion disk, the assumption of inviscid flow is surely not unjustified from an astrophysical point of 
view. Such practically inviscid accretion flow may manifest multi-transonic behavior where steady standing shock waves may be developed \citep{Sonali_1,Sonali_4,Sonali_5,Sonali_6,Sonali_7,Sonali_8,Sonali_9,Sonali_10,Sonali_11,Das_2002,Sonali_13,Sonali_14,BT_2012,multi_auth_2015}.

Far away from the accretor, especially where the flow is subsonic, inflow time scale may become comparable to the viscous time scale. Introduction of viscous transport of angular momentum may be of considerable relevance to probe the accretion dynamics in such region even for low angular momentum accretion flow to assist the inwards drift of the accreting matter. However, the exact prescription for viscosity in an accretion disk is still a matter of much
debate \citep{FKR02,Papaloizou_Lin_1995}. It has really been demonstrated that for low angular momentum, sub-Keplerian accretion disks, viscosity can be implemented as a perturbative effect, up to a certain length scale, which involves a first order correction in the standard $\alpha$-viscosity parameter \citep{shakura_sunayev_73} about the zeroth order inviscid solution, introduced as a `quasi-viscous' disk model in \citep{JKB_2007,JKB_2009}. This method reduces the problem of solving a second-order nonlinear differential equation, Navier-Stokes equation \citep{Landau_1987}, to that of an effective first-order differential equation. Such quasi-viscous prescription helps to capture the important properties of the axisymmetric accretion on  a comparatively large length scales, without compromising on the fundamentally simple and elegant features of the inviscid flow model \citep{Chadrasekhar_1981}. One obtains the stationary integral flow solutions without 
solving the energy conservation equations -- to avoid the complex formulation of the corresponding heating and cooling mechanism. 

Existing works on the 
quasi-viscous flow, however, concentrates on one particular geometrical configuration of flow -- axially symmetric accretion in hydrostatic equilibrium along the vertical direction. Such a flow configuration is usually associated with thin disk models. The quasi-viscous formalism, however, requires that the flow possess with ab-initio non-zero value of the radial flow velocity along the equatorial plane. Such a flow configuration with reasonably low angular momentum is better described by a quasi-spherical flow rather than a thin disk structure with its radius dependent flow thickness. A conical wedge shaped geometrical configuration of the rotating flow serves as the appropriate description of the accretion profile under consideration. Such a geometry is be a better alternative of the flow described by a standard thin disk structure. 

Results obtained so far in the literature for quasi-viscous disk accretion for flow in vertical equilibrium are to be compared with the calculations performed using a more 
general theoretical set up capable of studying various features of hot sub-Keplerian accretion disk in all three different flow geometries considered in the literature, i.e., flow with constant thickness, conical wedge shaped flow, and flow in hydrostatic equilibrium along the vertical direction (see, e.g., section 4 of ref.~\citep{Arpita_2014}). It is to be noted that the corresponding expressions for the flow thickness in any of the flow configurations in the existing literature
are derived from a set of idealised assumptions. In reality, the rigorous
derivation of the flow thickness may be accomplished by
using the framework of non-LTE radiative transfer 
(e.g., \citep{Hubeny_Hubeny_1998,Davis_Hubeny_2006}) or, by using the
Grad-Shafranov equations for the MHD-related aspects of the flow
(e.g., \citep{Beskin_Tchekhovsokoy_2005,Beskin_1997,Beskin_2009}).

In the present work, we provide a generalized formalism capable of studying the quasi-viscous flow in all possible geometric configuration of the accreting flow structure under the influence of all available post-Newtonian pseudo-Schwarzschild black hole potentials proposed in the literature. We find the multi-critical properties of such flow and demonstrate how to categorise                 the nature of the corresponding critical points -- i.e., how to identify which critical point is saddle type and which critical point is spiral type by following a dynamical systems approach \citep{JS_1999}. However, a global understanding of the flow topologies is obtained upon performing a complete numerical investigation of the nonlinear
stationary equations describing the dependence of the velocity on the radial distance. While this is a general practice, it is still possible to semi-quantitatively capture the essential behaviour of the transonic solution without resorting to numerical
techniques through an alternative means reported in the recent literature on accretion
\citep{Sonali_14,Chaudhury_2006,Mondal_2007,Goswami_2007}. Equipped with the mathematical formalism of a general dynamical systems approach, these studies have shown that one can derive a clear analytical conception of some of the global
features of the flow by analysing the local properties of the critical points.

Most of the aforementioned works dealing with multi-critical flow essentially address the stationary transonic flow solutions corresponding to the steady state accretion. Transient phenomena are not quite unusual in the large scale astrophysical set up. Also, recently
some of the authors \citep{JKB_2007,JKB_2009} demonstrated that the introduction of a very small
amount of viscosity may give rise to secular instabilities as well. It is thus essential to ensure that the information collected from the stationary integral flow solutions are reliable by establishing that the steady states are stable --- at least for the astrophysically relevant time scales. Such a task can be accomplished by perturbing the set of equations describing the steady state quasi-viscous accretion disk and thereby demonstrating that such perturbation does not diverge with time. We perform a linear stability analysis of the stationary quasi-viscous flow solutions to analyse such perturbation.

We observe that for all three flow geometries considered here, a certain length scale can be defined in terms of the $\alpha$- viscosity parameter as well as the density and the temperature of the ambient medium - beyond which the quasi-viscous prescription does not hold good in the sense that the first order correction becomes comparable with its corresponding unperturbed value of the constant specific angular momentum, $\lambda_0$. Numerical estimate for such length scale varies with different geometric configurations of flow. We generalised these studies to all the disk models available in the literature. 


\section{Basic Formalism}
The accreting material is assumed to possess low angular momentum and is considered to be an advective adiabatic fluid with a weak viscous effect driven by a gravitational field described in terms of a generalised pseudo-Schwarzschild potential, $\Phi(r)$. 

For an inviscid flow, the system is governed by the equation of continuity and the Euler equation. For a viscous flow one needs to incorporate the angular momentum conservation equation as well to include the effect of viscosity. Due to axisymmetry of the problem and by averaging all the relevant physical quantities over local height, all the governing equations are formulated and solved on the equatorial plane itself.

For a fluid accreting in the form of a thin disk like structure onto some accretor, the continuity equation is obtained as,
\begin{equation}
\dfrac{\partial}{\partial t}(\Sigma)+ \dfrac{1}{r}\dfrac{\partial}{\partial r}(\Sigma vr)=0.
\label{eqc}
\end{equation}
Where, $\Sigma= \rho H$ is the local surface density of the disk, $\rho$ being the local height averaged volume density, $H$ being half of the local disk height. $v$ is the radial drift velocity of the fluid along the equatorial plane, where $r$ is the radial position on the equatorial plane.

The earlier works~\citep{JKB_2007,JKB_2009} under this quasiviscous approach in accretion disk problem dealt with the disk model where the disk height was determined at each value of $r$, from the hydrostatic equilibrium condition along the lateral or vertical direction. Apart from that, in the literature, the conical disk model is used where the quasi-spherical flow through a certain solid angle, maintaining azimuthal symmetry and the symmetry with respect to the equatorial plane, is considered. Lastly the simplest and primitive model of a disk with a constant thickness or height has been considered here. 

The dependence of $H$ on radial distance $r$ varies in different disk models. These are \citep{Chakrabarti2001},
\begin{subequations}\begin{align}
H &=c_s(r)\sqrt{\frac{r}{\gamma\Phi^{\prime}(r)}} &\textrm{for vertical equilibrium model (V)},\\
H &= \Theta r,\;\;\;\; \Theta \textrm{ being a constant} &\textrm{for conical flow model (C)},\\
H &= H_0 &\textrm{for constant height disk (H)}.
\end{align}\end{subequations}
Here, $c_s(r)=\sqrt{\partial P/\partial\rho}=\sqrt{{\gamma P}/{\rho}}$ is the local polytropic sound speed in the flow medium with equation of state as $P=K\rho^{\gamma}$ \citep{Chandra_1939}. Now these can be generalised into a single parametric form,
\begin{equation}
H=h\rho^{\epsilon}\bar{g}(r), \label{H}                                                                                                                                                                                                                                                                                                                                                                                                                                                                                                    \end{equation} 
such that, \[\begin{array}{l l l l}
\textrm{for V} & h=\sqrt{K}, & \epsilon=(\gamma -1)/2, & \bar{g}(r)=\sqrt{r/\Phi'(r)}; \\                                                                                                                                                                                                                                                                                                                                                                                                                                                                                                                                                            
\textrm{for C} & h=\Theta,   & \epsilon=0, & \bar{g}(r)=r; \\
\textrm{for H} & h=H_0,  & \epsilon=0, & \bar{g}(r)=1.
\end{array}\]

The conservation equation of specific angular momentum in the flow is given by \citep{FKR02},
\begin{equation}
\rho H \dfrac{\partial}{\partial t}(r^{2}\bar\omega)+ \rho v H \dfrac{\partial}{\partial r}(r^{2}\bar\omega) = \dfrac{1}{2 \pi r}\left( \dfrac{\partial G}{\partial r}\right)
\label{eqcam}
\end{equation}
where, $\bar\omega$ is the local angular velocity of the flow, $ G = 2 \pi \nu \Sigma r^{3} \dfrac{\partial \bar\omega}{\partial r} $ is the torque and $ \nu = \alpha c_{s} H $ in accordance to Shakura \& Sunyaev prescription for kinematic viscosity \citep{shakura_sunayev_73}. From the above equation it is clear that even in stationary condition i.e., under the condition $\dfrac{\partial}{\partial t}(r^{2}\bar\omega)=0$, the effective specific angular momentum $r^{2}\bar\omega$ (=$\lambda_{\rm eff}$,  say) is not a constant, rather is dependent on $r$. It would however be independent of $r$ (a constant $\lambda_0$), for inviscid flow, i.e., when $\alpha$ = 0. Hence, in the weak viscosity limit $\lambda_{\rm eff}$ may be expressed as a position dependent perturbative correction to $\lambda_{0}$ proportional to $\alpha$, the Shakura \& Sunyaev (1973) dimensionless viscosity parameter as, 
\begin{equation}
\lambda_{\rm eff}=r^{2} \bar\omega = \lambda_{0}[1 + \alpha F(c_{s},v,r)],
\label{angmom_breakup}
\end{equation}
such that $\alpha F<<1$. 

The radial Euler equation for such systems appears to be (with the prime on $\Phi(r)$ denoting a spatial derivative), 
\begin{equation}
\dfrac{\partial v}{\partial t} + v \dfrac{\partial v}{\partial r} + \dfrac{1}{\rho} \dfrac{\partial P}{\partial r} + \Phi^{'}(r) - \dfrac{\lambda_{\rm eff}^{2}}{r^{3}}=0
\label{euler_eq}
\end{equation}

For the generalised case, equation \eqref{eqc} may be recast  as, 
\begin{equation}
\dfrac{\partial}{\partial t}(g_{1}(\rho))+ g_{2}(r) \dfrac{\partial}{\partial r}f(\rho ,v,r)=0\label{modcon}\end{equation} where
$
g_{1}(\rho)= \rho^{1+\varepsilon},~ 
g_{2}(r)= \dfrac{1}{hr \bar{g}(r)}~ \textrm{and }$ 
$f(\rho ,v,r) = \dfrac{g_{1}(\rho)v}{g_{2}(r)}=\rho vrH $. 

Therefore, using $\rho=(\gamma K)^{-n}c_s^{2n}$ and $n = \dfrac{1}{\gamma -1}$ \citep{Chandra_1939}, $H$ and $f$ can be expressed as,
\begin{eqnarray}
H &=& \dfrac{(\gamma K)^{-n \epsilon}c_s^{2n \epsilon}}{r g_2(r)},
\label{gen_H}
\end{eqnarray}
\begin{eqnarray}
f &=& \rho v r H =  \dfrac{(\gamma K)^{-n (1+\epsilon)}c_s^{2n(1+ \epsilon)}v}{g_2(r)}.
\end{eqnarray}
Here, $f$ is inherently negative as $v$ is negative for accretion flows. 

Now, equation \eqref{eqcam} may similarly be written as,
\begin{equation}
\dfrac{1}{v}\dfrac{\partial}{\partial t}(r^{2}\bar\omega)+\dfrac{\partial}{\partial r}(r^{2}\bar\omega) = \dfrac{\alpha}{f} \dfrac{\partial}{\partial r}(f_{1}\dfrac{\partial \bar\omega}{\partial r}).
\label{modang}
\end{equation}
where $ f_{1} = \rho c_{s} H^{2} r^{3} $ and $ \dfrac{f_{1}}{f} = \dfrac{r^2 c_{s} H}{v} $. Now, putting the  $\lambda_{\rm eff}$ from  equation \eqref{angmom_breakup} in equation \eqref{modang} we get,
\begin{equation}
F(c_{s},v,r) = -2\left[\dfrac{f_{1}}{fr^{3}}+ \int \dfrac{f_{1}}{fr^{3}} \left(\dfrac{1}{f} \dfrac{\partial f}{\partial r} \right) \mathrm dr \right]
\label{Fsol}
\end{equation}
by assuming $\dfrac{\partial}{\partial t}(\lambda_{\rm eff}) \approx 0$ and neglecting all higher orders of $\alpha$ and retaining only the first order, one may introduce the `quasi-viscous' approximation.

Under the same approximation the radial Euler equation becomes, 
\begin{equation}
\dfrac{\partial v}{\partial t} + v \dfrac{\partial v}{\partial r} + \dfrac{1}{\rho} \dfrac{\partial P}{\partial r} + \Phi^{'}(r) - \dfrac{\lambda_{0}^{2}}{r^{3}}-\dfrac{2\alpha \lambda_{0}^2F}{r^{3}}=0.
\label{euler_eq1}
\end{equation}
The equation~\eqref{euler_eq1} thus becomes the effective dynamical equation in our quasi-viscous model.


\section{Stationary Solutions}
For $ \dfrac{\partial }{\partial t} = 0 $, 
the equations \eqref{modcon} and \eqref{euler_eq1} take the form 
\begin{eqnarray}
\dfrac{\dd f}{\dd r} = 0, \label{conf}
\label{stcon} \\
v \dfrac{\dd v}{\dd r} + \dfrac{1}{\rho} \dfrac{\dd P}{\dd r} + \Phi^{'}(r) - \dfrac{\lambda_{0}^{2}}{r^{3}}-\dfrac{2\alpha \lambda_{0}^2F}{r^{3}}=0.
\label{strad1}
\end{eqnarray}

It is clear that from equation \eqref{stcon}, the first integral, $f$ (=$\rho v r H$), is a constant; while $-4\pi f$ is equal to the mass accretion rate ($\dot{M}$). However, due to presence of the $\alpha$ dependent term, the second equation \eqref{strad1} cannot provide a closed algebraic form for the first integral which otherwise would represent total specific energy for an inviscid flow. 

Also under stationary conditions, the expression of $F$, using the condition~\eqref{conf} in equation~\eqref{Fsol}, is found to be
\begin{equation}
F_{s} = -2 \dfrac{f_{1}}{fr^{3}} = -2 \dfrac{c_{s}H}{v r}\label{Fs}.
\end{equation}
Hence replacing $F$ by $F_s$ in equation \eqref{strad1} and using $\dfrac{1}{\rho} \dfrac{\mathrm d P}{\mathrm d r}=\dfrac{2 c_s}{\gamma-1} \dfrac{\mathrm d c_s}{\mathrm d r}$, we obtain,
\begin{eqnarray}
v \dfrac{\mathrm d v}{\mathrm d r} + \dfrac{2 c_s}{\gamma-1} \dfrac{\mathrm d c_s}{\mathrm d r} + \Phi^{'}(r) - \dfrac{\lambda_{0}^{2}}{r^{3}}-\dfrac{2\alpha \lambda_{0}^2F}{r^{3}}=0.
\label{strad2}
\end{eqnarray}

Again, by using $\dfrac{1}{\rho} \dfrac{\mathrm d \rho}{\mathrm d r}=\dfrac{2}{c_s(\gamma-1)} \dfrac{\mathrm d c_s}{\mathrm d r}$, equation \eqref{stcon} reduces to, 
\begin{eqnarray}
\dfrac{\mathrm dc_{\rm s}}{\mathrm dr} = -\dfrac{c_{\rm s}}{2n(1+\epsilon)}\left[\dfrac{1}{v}\dfrac{\mathrm dv}{\mathrm dr} -\dfrac{g_2'}{g_2}\right].
\label{stcon2}
\end{eqnarray}

Stationary flow equations \eqref{strad2} and \eqref{stcon2} can be combined to obtain the velocity gradient as,
\begin{eqnarray}
\dfrac{\mathrm dv}{\mathrm dr} &=& \dfrac{\dfrac{\lambda_0^2}{r^3}+\dfrac{2 \alpha \lambda_0^2}{r^3}F_s-\Phi' -\dfrac{c_s^2}{1+ \epsilon}\dfrac{g_2'}{g_2}}{v-\dfrac{c_s^2}{v(1+\epsilon)}}=\dfrac{N(c_s,v,r)}{D(c_s,v,r)} \;\;\;\;\textrm{  (let's ~say)}.
\label{dvdr}
\end{eqnarray}


\subsection{Critical point analysis}
\label{CP_find}
Critical point condition (all corresponding quantities have been labelled with subscript `$c$') is given by the condition as $N=D=0$. Therefore, we have,
\begin{subequations}
\begin{eqnarray}
\dfrac{\lambda_0^2}{r_c^3}+\dfrac{2 \alpha \lambda_0^2}{r_c^3}F_{s_c}-\Phi_c' &=&\dfrac{c_{s_c}^2}{1+ \epsilon}\left.\dfrac{g_{2}'}{g_{2}}\right|_c
\label{neq0}\\
v_c^2 &=& \dfrac{c_{s_c}^2}{1+\epsilon}.
\label{vccsc}
\end{eqnarray}
\end{subequations}
Equation \eqref{neq0} is solved for the critical point radius, $r_c$, for any constant values of $\gamma$ and $\alpha$ with $F_{s_c}(c_{s_c})$ (using equations \eqref{Fs} and \eqref{gen_H}) and with a given choice for the two flow parameters, i.e., $\lambda_0$ and the entropy accretion rate ($\cal \dot{M}$); while $\cal \dot{M}$ is related to the conserved mass accretion rate ($\dot{M}=-4\pi f$) through $c_{s_c}$ as expressed in \citep{Chakra_1990},
\begin{eqnarray}
{\cal \dot{M}} &=& \dot{M}(\gamma K)^{n}  = \dfrac{4 \pi (\gamma K)^{-n \epsilon}c_{s_c}^{2n(1+ \epsilon)+1}}{\sqrt{1+ \epsilon}g_2(r_c)}=\dfrac{4 \pi h(\gamma K)^{-n \epsilon}c_{s_c}^{2n(1+ \epsilon)+1}r_c\bar{g}(r_c)}{\sqrt{1+ \epsilon}}, 
\label{smdot}
\end{eqnarray} using equation~\eqref{vccsc}. Hence one gets, \[c_{s_c}=\left[\frac{(\gamma K)^{n\epsilon}\sqrt{1+\epsilon}}{4\pi h r_c\bar{g}(r_c)}\dot{\cal M}\right]^\frac{1}{2n(\epsilon+1)+1}.\] Again using equations~\eqref{Fs} and \eqref{H}, it is obtained that, \[F_{s_c}=\frac{2\sqrt{1+\epsilon}H_c}{r_c}\] and \[H_c=h(\gamma K)^{-n\epsilon}c_{s_c}^{2n\epsilon}\bar{g}(r_c).\]

It may be noted that for the models `C' and `H', $\epsilon=0$ and for the model `V, $n\epsilon=1/2$ and $h=\sqrt{K}$, so that for all three disk models the $c_{s_c}$, $H_c$ and $F_{s_c}$ and hence location of the critical point ($r_c$), do not involve $K$.

Once we determine the $r_c$ values, the critical values (or values at fixed points) for flow velocity, sound speed, and local height can subsequently be determined at each evaluated critical radius.

Now to find the radial velocity gradient at the critical point one has to take limit of $dv/dr$ as $r$ tending to the critical point and making use of L'Hospital rule one may find a quadratic equation for $\mathrm dv/\mathrm dr|_c$ as \citep{BT_2012},
\[\mathscr{A}\left.\frac{dv}{dr}\right|_c^2+\mathscr{B}\left.\frac{dv}{dr}\right|_c+\mathscr{C}=0;\]
where,
\begin{subequations}
\begin{eqnarray}
\mathscr{A} &=& \Xi +2,\\
\mathscr{B} &=& -2\Xi v_c \left. \dfrac{g_2'}{g_2}\right|_c + \dfrac{\bar\Delta}{v_c} \left(2+(2n\epsilon +1)\Xi \right),\\
\mathscr{C} &=& \dfrac{3\lambda_0^2}{r_c^4}+\Phi_c''+ v_c^2 \left.\dfrac{g_2''}{g_2}\right|_c + v_c^2 (\Xi -1)\left(\left.\dfrac{g_2'}{g_2}\right|_c \right)^2 + \bar\Delta \left(\dfrac{10}{r_c}+(2-\Xi(2n\epsilon +1))\left.\dfrac{g_2'}{g_2}\right|_c \right),
\end{eqnarray}\label{abc}
\end{subequations}
with the definitions, 
\begin{eqnarray}
\Xi &=& \dfrac{\gamma -1}{1+ \epsilon} \\ \textrm{and }
\bar\Delta &=& \dfrac{\alpha \lambda_0^2 F_{sc}}{r_c^3}.
\end{eqnarray}

Hence the solution we have,
\begin{eqnarray}
\left.\dfrac{\mathrm dv}{\mathrm dr}\right|_c = \dfrac{-\mathscr B \pm \sqrt{\mathscr B^2-4\mathscr A\mathscr C}}{2\mathscr A}.
\label{vc}
\end{eqnarray}

Using equation \eqref{vc} in equation \eqref{stcon2} one can obtain the value of $\dfrac{\mathrm dc_{\rm s}}{\mathrm dr}$ at the critical point.
One may check that the condition, $\alpha=0$, implies $\bar\Delta = 0$, which gives back the inviscid results. After knowing $r_c, v_c, c_{s_c},\mathrm dv/\mathrm dr|_c$ and $\mathrm dc_s/\mathrm dr|_c $, one may use standard numerical algorithm such as Runge-Kutta method on the equations~\eqref{dvdr} and \eqref{stcon2} to draw the phase portrait explicitly, but the instead of doing that we follow a more elegant method (see refs.~\citep{Chaudhury_2006,Sonali_14}) to extract the qualitative features of it using the techniques of dynamical systems described in the following sections.


\subsection{Categorization of critical points}
\label{CP_nature}
In what follows we provide a detailed description of the procedure to be followed to understand the nature of the critical points and the topological behaviour of the phase orbits around them. To accomplish such task equation~\eqref{dvdr} is parametrised by some arbitrary parameter $\tau$ (similar approach in invisid cases as followed 
in \citep{JKB_2009,Chaudhury_2006,Mondal_2007,Goswami_2007,Sonali_14,Afshordi_Paczynski_2003}) and presented as a set of coupled autonomous first-order differential equations as,  
 
\begin{eqnarray}
\dfrac{\mathrm dv^2}{\mathrm d \tau} &=& 2v^2\left[\dfrac{\lambda_0^2}{r^2}+\dfrac{2 \alpha \lambda_0^2}{r^2}F_s-r\Phi' -\dfrac{c_s^2}{1+ \epsilon}\dfrac{g_2'}{g_2}r \right]\\
\dfrac{\mathrm dr}{\mathrm d \tau} &=& r\left(v^2-\dfrac{c_s^2}{(1+\epsilon)}\right).
\end{eqnarray}

The system of above equations are then linearised about the fixed points (critical points) as a next step. By applying linear perturbations about the critical point quantities as $v^2=v_c^2+\delta v^2$, $r=r_c+\delta r$, $c_s^2=c_{s_c}^2+\delta c_s^2$ and $F_s=F_{s_c}+\delta F_s$, it is possible to derive a set of two autonomous first-order linear differential equations in the $\delta v^2~-~\delta r$ plane as,
\begin{eqnarray}
\dfrac{\mathrm d}{\mathrm d \tau}(\delta v^2) &=& {\cal A} \delta v^2 +{\cal B} \delta r\label{delvsq}\\
\dfrac{\mathrm d}{\mathrm d \tau}(\delta r) &=& {\cal C} \delta v^2 +{\cal D} \delta r\label{delr}
\end{eqnarray}
using the interrelations,
\begin{eqnarray}
\dfrac{\delta c_s^2}{c_{s_c}^2} &=& \left[ -\dfrac{\delta v^2}{2 v_c^2} +\left.\dfrac{g_2'}{g_2}\right|_c \delta r\right] \dfrac{1}{n(1+ \epsilon)} ~~~{\rm and}\\
\dfrac{\delta F_s}{F_{s_c}} &=& \left( \dfrac{2n \epsilon +1}{2}\right)\dfrac{\delta c_s^2}{c_{s_c}^2} -\dfrac{\delta v^2}{2v_c^2}-\dfrac{2\delta r}{r_c}-\left.\dfrac{g_2'}{g_2}\right|_c \delta r.
\end{eqnarray}

The coefficients have the following forms,
\begin{subequations}
\begin{eqnarray}
{\cal A}&=& 2v_c^2 \left[-\dfrac{\alpha \lambda_0^2 F_{s_c}}{r_c^2v_c^2}\left( 1+ \dfrac{2n \epsilon +1}{2n(1+\epsilon)} \right) +\dfrac{r_c}{2n(1+ \epsilon)} \left.\dfrac{g_2'}{g_2}\right|_c \right]
\label{coeff_A}\\
{\cal B} &=&  2v_c^2 \left[ -\dfrac{2 \lambda_0^2}{r_c^3}-\Phi'-\Phi'' r_c-\dfrac{8 \alpha \lambda_0^2 F_{s_c}}{r_c^3} -\dfrac{c_{s_c}^2}{(1+\epsilon)}\left.\dfrac{g_2'}{g_2}\right|_c-\dfrac{\alpha \lambda_0^2 F_{s_c}}{r_c^2}\left.\dfrac{g_2'}{g_2}\right|_c \left(\dfrac{2n-1}{n(1+\epsilon)}\right) \right]\nonumber\\
&+&2v_c^2 \left[\dfrac{c_{s_c}^2}{(1+\epsilon)}r_c\left(\left.\dfrac{g_2'}{g_2}\right|_c\right)^2\left(1-\dfrac{1}{n(1+\epsilon)} \right)-\dfrac{c_{s_c}^2r_c}{(1+\epsilon)}\left.\dfrac{g_2''}{g_2}\right|_c\right]\\
{\cal C} &=& r_c \left(1+\dfrac{1}{2n(1+\epsilon)}\right) \\
{\cal D} &=&-\dfrac{r_c c_{s_c}^2}{n(1+\epsilon)^2}\left.\dfrac{g_2'}{g_2}\right|_c.
\end{eqnarray}
\end{subequations}

After decoupling the equations \eqref{delvsq} \& \eqref{delr} with suitable linear transformation and assuming solutions of type $\sim e^{\Omega \tau}$ for the transformed dependent variables, the eigenvalues $\Omega$ signifying the growth rates may turn out to be, \[\Omega^2-({\cal A}+{\cal D})\Omega+({\cal A}{\cal D}-{\cal B}{\cal C})=0;\] solution of which occur in the form, 

\begin{eqnarray}
\Omega=\dfrac{P \pm \Delta}{2}.
\label{Omega_defi}
\end{eqnarray} 

Here, $P$, $Q$ and $\Delta$ are defined as,
\begin{subequations}
\begin{eqnarray}
{P}={\cal A}+{\cal D}\\
\label{P_defi}
Q={\cal A}{\cal D}-{\cal B}{\cal C}\\
\Delta^2=P^2-4Q.
\label{delta_defi}
\end{eqnarray}
\end{subequations}

The nature of the possible critical points can be predicted by investigating the form of $\Omega$~\citep{JS_1999}. If $Q<0$, the critical point has to be a saddle. For $Q>0$, if $P=0$ the point is a centre. The previous two varieties are the only ones found in an inviscid flow. But if $P\neq 0$ the critical point must become a node and in addition if $\Delta^2<0$ the point will turn out to be a spiral (node). This analysis therefore delivers the qualitative features of the phase portrait without explicitly obtaining the stationary integral solutions. 

For our present work, under the condition $\alpha\neq 0$, only saddle and spiral type critical points are obtained. In this paper, we denote the two real eigenvalue solutions by $\rm \Omega_{1,2}$ (relevant for saddle type critical points) and denote $\rm \Omega_{\rm re}$ by ${P}/{2}$ and $\rm \Omega^2_{\rm im}$ by ${\Delta^2}/{4}$ (relevant for centre or spiral type critical points).   
\section{Overall Methodology} \label{method}
The overall plan of this work is as follows.
We have already constructed the set of linear energy momentum conservation 
equations including the presence of the quasi viscous effects, as well 
as the mass conservation equation. From the modified Euler-like equation
and the continuity equation, we found out the equation \eqref{neq0} for locating the critical 
points, velocity gradient in eq.~\eqref{vc} and sound speed gradient, conditions for categorising 
the types of the critical points in a generalised parametrised form in the previous section. In the following 
sections, based on the generalised conditions, we find the critical points 
for various geometric configurations of axially symmetric 
accretion flow. Using the corresponding critical point conditions, the 
possibility of having multi-transonic flow is discussed, and it is 
pointed out that stationary shocks may form as a consequence of having 
multiple critical points. The topological nature of the phase orbits 
for the stationary transonic integral solutions is discussed. Once we acquire 
the details of the stationary properties of the flow profile, we get into 
the stability analysis of such accretion configurations. To accomplish 
such task, we linear-perturb the stationary flow solutions to check 
whether such perturbation diverges, and conclude on the nature of the 
stability of the flow based on such observation.

For inviscid polytropic flow, one integrates the Euler and the continuity 
equation to obtain two global first integrals of motion -- the conserved specific flow 
energy (the Bernoulli constant) and the mass accretion rate, respectively, 
see, e.g., \citep{FKR02,Das_2002} for further 
details. The term `global' signifies that such first integrals remain 
conserved along the entire streamline in the steady state, i.e., along the 
entire integral flow solution (phase orbit) characterised by a set of 
specified initial boundary conditions. For quasi-viscous accretion, however, 
the modified Euler-like equation~\eqref{euler_eq1} becomes an integro-differential 
equation, and hence can not be integrated analytically. No such complications
arise while obtaining the integral solution of the continuity equation. One 
thus obtains a globally conserved mass accretion rate but the impossibility 
of obtaining the integral solution of the Euler like equation does not allow 
to have any specific energy term for the quasi viscous flow which is 
conserved along the streamline. Such energy like variable can only be 
defined, as is explained in detail in the subsequent sections.

Presence of non-analytically integrable term in the Euler like equation alters 
the solution scheme for obtaining the flow profile. For inviscid flow, total 
specific energy was globally conserved and hence the overall flow behaviour 
could be parameterized by $\left[{\cal E},\lambda,\gamma\right]$ where 
${\cal E}$ is the globally conserved specific energy of the flow, 
$\lambda$ is constant specific angular momentum and $\gamma$ is the fixed 
value of the adiabatic index since the atomicity of the gas has been 
considered to be invariant. For quasiviscous flow, 
neither ${\cal E}$ nor $\lambda$ will be considered as an appropriate boundary 
condition. Since we do not incorporate any heating or cooling mechanism in 
our work (viscosity has been introduced as a perturbative effect), $\gamma$ 
can still be used as the initial boundary condition for characterising the 
flow. Instead of $\left[{\cal E},\lambda\right]$, we will be using 
$\left[{\dot {\cal M}},\lambda_0\right]$ to characterise the integral 
solutions. 

\subsection*{Multi-transonic behaviour}
Equation~\eqref{neq0} may be solved for a fixed set of $\left[\dot{\cal M}, \lambda_0, \gamma\right]$ to obtain the location of the critical point. It is not necessary that the solution will be unique. One may have at most at most three real roots for the equation~\eqref{neq0} --- i.e., at most three critical points --- may form, all of them being situated outside the black hole event horizon. In subsequent sections, we explicitly demonstrate that the critical points formed for our our quasi-viscous model can be of saddle type or of spiral type. A real physical transonic solution can pass through a saddle type critical point only. For  a multi-critical flow, usually two saddle type critical points embrace a spiral type critical point.

If the stationary integral solutions passing through two different saddle type points can be connected by a stationary shock, true multi-transonic flow profiles are formed. All these issues have been demonstrated  analytically in subsequent sections.

\subsection*{`Accretion' type and `Wind' type}
For astro-physically relevant set of $\left[\dot{\cal M}, \lambda_0, \gamma\right]$, we obtain the integral solutions passing either through one or through three critical points. Solutions containing three critical points can be categorised into two distinct groups. If we define \[{\cal E}_{\rm eff}=\frac{v^2}{2}+\frac{c_s^2}{\gamma-1}+\Phi(r)+\frac{\lambda_0^2}{2r^2}\] to be the total specific energy of the flow which is not globally conserved but can be evaluated on a saddle type critical point, one finds that ${\cal E}_{\rm eff}$ is different at two different (inner and outer) saddle type critical points. For \[{\cal E}_{\rm eff}^{\rm inner}<{\cal E}_{\rm eff}^{\rm outer}\] the ingoing solutions passing through the outer sonic point is vulnerable to make transition onto the integral solution passing through the inner sonic point if a stationary shock forms. We then have a muti-transonic shocked accretion.

On the other hand, if \[{\cal E}_{\rm eff}^{\rm inner}>{\cal E}_{\rm eff}^{\rm outer},\] outgoing flow solutions (wind type solution according to the conventional classification of phase topology, see ref.\citep{Das_2002} )passing through the inner saddle type critical point may launch albeit discontinuously through a standing shock, onto the wind type solution passing through the outer sonic point. Hence one obtains multi-transonic wind solutions with shocks in wind instead of accretion branch. In this case, the conventional ingoing accretion solution will be mono-transonic, in spite of obtaining three critical pints by solving the equation~\eqref{neq0} for  a fixed set of initial boundary condition characterised by $\left[ \dot{\cal M}, \lambda_0, \gamma\right]$.

                                                                                                                                                                                                                     \section{Expressions for different flow models:}
 
In Table~\ref{Table:expressions_VCH}, we display the 
the expressions for different variables (already introduced in the previous sections) corresponding to three disk geometries.

\begin{table}[h]
\centering
\begin{tabular*}{1.0\linewidth}{@{\extracolsep{\fill}}lccc}
\hline
Quantities  & V & C & H\\
\hline
H & $c_s \sqrt{r/{\gamma \Phi'}}$ & $\Theta r$ & $H_0$\\
h & $\sqrt{K}$ & $\Theta$ & $H_0$\\
$\epsilon$ & $(\gamma-1)/2$ & 0 & 0\\
$\bar{g}(r)$ & $\sqrt{r/\Phi'}$ & $r$ & 1\\
$g_1$ & $\rho^{\frac{\gamma +1 }{2}}$ &$\rho$ & $\rho$ \\
$g_2$ & $\sqrt{\Phi'/{Kr^3}}$ & $1/{r^2 \Theta}$ & $1/{r H_0}$\\
f & $\rho^{\frac{\gamma +1 }{2}} v \sqrt{Kr^3/\Phi'}$ & $\rho v r^2 \Theta$ & $\rho v r H_0$\\
\hline
\end{tabular*}
\caption{Expressions for various important variables and constants corresponding to the vertical equilibrium geometry (V), conical
geometry (C), and constant-height disk geometry (H).}
\label{Table:expressions_VCH}
\end{table}
The space gradient of flow velocity for various models is evaluated from equation~\eqref{dvdr} as, 
\begin{equation}
\dfrac{\mathrm dv}{\mathrm dr} = \left\lbrace \begin{array}{lr}
\dfrac{v \left[\dfrac{\lambda_0^2}{r^3}-\Phi'+\dfrac{\beta^2c_s^2}{2}\left( \dfrac{3}{r_c}-\dfrac{\Phi''}{\Phi'}\right)-\dfrac{4\alpha \lambda_0^2 c_s^2}{v r^3}\sqrt{\dfrac{1}{r \Phi' \gamma}} \right]}{v^2-\beta^2 c_s^2}  & (\rm V)\\
\dfrac{v \left[\dfrac{2c_s^2}{r}+\dfrac{\lambda_0^2}{r^3}-\Phi'-\dfrac{4\alpha \lambda_0^2 c_s \Theta}{v_0 r^3} \right]}{v^2-c_s^2} & (\rm C)\\
\dfrac{v \left[\dfrac{c_{s}^2}{r}+\dfrac{\lambda_0^2}{r^3}-\Phi'-\dfrac{4\alpha \lambda_0^2 c_{s} H_0}{v r^4} \right]}{v^2-c_{s}^2} & (\rm H)
\end{array}\right.
\end{equation}
where, $\beta^2 = \dfrac{2}{\gamma +1}$. 
The space gradient of sound speed is expressed, from equation~\eqref{stcon2} by, 
\begin{equation}
\dfrac{\mathrm dc_{s}}{\mathrm dr} = \left\lbrace\begin{array}{lr}
-c_s\left(\dfrac{\gamma -1}{\gamma +1} \right)\left[ \dfrac{1}{v}\dfrac{\mathrm dv}{\mathrm dr}+\dfrac{3}{2r}-\dfrac{\Phi''}{2\Phi'}\right] & ({\rm V})\\ 
-c_s\left(\dfrac{\gamma -1}{2} \right)\left[ \dfrac{1}{v}\dfrac{\mathrm dv}{\mathrm dr}+\dfrac{2}{r}\right] & ({\rm C})\\ 
-c_{s}\left(\dfrac{\gamma -1}{2} \right)\left[ \dfrac{1}{v}\dfrac{\mathrm dv}{\mathrm dr}+\dfrac{1}{r}\right] & ({\rm H})
                                                   \end{array}\right.
\end{equation}


The real root(s) of the following equations, evaluated from equations~\eqref{neq0}, \eqref{Fs}, will be the location(s) of the critical point(s) in the corresponding flow models,
\begin{subequations}\begin{align} \begin{array}{rccr}
\frac{\lambda_0^2}{r_c^3}-\Phi'(r_c)-\frac{4\alpha\lambda_0^2}{r_c^4}\sqrt{\frac{\gamma+1}{2\gamma}}\sqrt{\frac{r_c}{\Phi'(r_c)}}\left[\sqrt{\frac{\gamma(\gamma+1)}{2}}\sqrt{\frac{\Phi'(r_c)}{r_c^3}}\frac{\dot{\cal M}}{4\pi}\right]^{\frac{\gamma-1}{2\gamma}} & & & \\
 -\frac{r_c\Phi_c''-3\Phi_c'}{(\gamma+1)r_c\Phi_c'}\left[\sqrt{\frac{\gamma(\gamma+1)}{2}}\sqrt{\frac{\Phi'(r_c)}{r_c^3}}\frac{\dot{\cal M}}{4\pi}\right]^{\frac{\gamma-1}{\gamma}}&=&0 & \textrm{for `V'}
\end{array} & \\
\begin{array}{lccr}
\frac{\lambda_0^2}{r_c^3}-\frac{4\alpha\lambda_0^2\Theta}{r_c^3}-\Phi'(r_c)+\frac{2}{r_c}\left[\frac{\dot{\cal M}}{4\pi\Theta r_c^2}\right]^\frac{2(\gamma-1)}{\gamma+1} &=& 0 & \textrm{for `C'} 
\end{array} & \\
\begin{array}{lccr}
\frac{\lambda_0^2}{r_c^3}-\frac{4\alpha\lambda_0^2H_0}{r_c^4}-\Phi'(r_c)+\frac{1}{r_c}\left[\frac{\dot{\cal M}}{4\pi H_0 r_c}\right]^\frac{2(\gamma-1)}{\gamma+1} &=& 0 & \textrm{for `H'} \end{array} &
\end{align}\end{subequations}

The flow velocity and the sound speed are related at critical points and the relations can be evaluated from equation~\eqref{vccsc}, for different flow geometries.
For the `C' and `H' case, sonic and critical points bear identical meaning, whereas, for `V' they differ by a constant factor $\sqrt{2/(\gamma+1)}$.

The slopes of the transonic flow lines at a critical point (saddle) are evaluated by L'Hospitals rule as obtained in eqs.~\eqref{abc}, where the constants $\mathscr A$, $\mathscr B$ and $\mathscr C$ for different flow models are obtained as,
\begin{subequations}
\begin{eqnarray}
\mathscr A &=&\left\lbrace\begin{array}{lr}
                      \dfrac{4 \gamma}{\gamma +1}  & (\rm V) \\
\gamma+1                    & (\rm C) \\
\gamma+1                    & (\rm H) \\
                        \end{array}\right.\\
\mathscr B &=& \left\lbrace\begin{array}{lr}
 -2 \xi \beta c_{s_c}\left[ \dfrac{\Phi_c''}{\Phi_c'}-\dfrac{3}{r_c}\right]-\dfrac{4\alpha \lambda_0^2}{\beta^2\sqrt{\gamma \Phi_c'}r^{\frac{5}{2}}}\left[1+2\xi \right] & (\rm V)\\
\dfrac{2(\gamma+1)}{r_c}\left[2\xi c_{s_c}-\dfrac{\alpha \lambda_0^2 \Theta}{v_c r_c^2} \right]  & (\rm C)\\
\dfrac{2(\gamma+1)}{r_c}\left[\xi c_{s_c}-\dfrac{\alpha \lambda_0^2 H_0}{v_c r_c^3} \right]    & (\rm H) \\
\end{array}\right.\\
\mathscr C &=& \left\lbrace\begin{array}{lr}
\dfrac{3\lambda_0^2}{r_c^4}+\Phi_c''+\dfrac{\beta^2 c_{s_c}^2}{2}\left[\dfrac{\Phi_c'''}{\Phi_c'}-\dfrac{1}{2}\left(\dfrac{\Phi_c''}{\Phi_c'}\right)^2-\dfrac{3\Phi_c''}{\Phi_c'r_c}+\dfrac{15}{2r_c^2} +\xi \left(\dfrac{\Phi_c''}{\Phi_c'}-\dfrac{3}{r_c}\right)^2\right] &  \\
\;\;\;\;\;\;\;\;\;\;\;\;\;\;\;\;\;\;\;\;\;\;\;\;\;\;\;\;\;\;\;\;\;\;\;\;\;\;\;\; -\dfrac{2\alpha\lambda_0^2c_{s_c}}{\beta\sqrt{\Phi_c'\gamma}r_c^{2.5}}\left[\dfrac{10}{r_c}+\left(\dfrac{\Phi_c''}{\Phi_c'}-\dfrac{3}{r_c}\right)(1-2\xi)\right]  & (\rm V)  \\
\dfrac{3\lambda_0^2}{r_c^4}+\Phi''_c +\dfrac{2(2\gamma-1) c_{s_c}^2}{r_c^2}-\dfrac{4\alpha \lambda_0^2 \Theta}{r_c^4}(\gamma +2)                             & (\rm C) \\
\dfrac{3\lambda_0^2}{r_c^4}+\Phi''_c +\dfrac{\gamma c_{s_c}^2}{r_c^2}-\dfrac{2\alpha \lambda_0^2 H_0}{r_c^5}(\gamma +7)                          & (\rm H)
\end{array}\right.
\end{eqnarray}\end{subequations}
where we write, $\dfrac{\gamma -1}{\gamma +1}=\xi$.


The nature of critical points for each flow geometry is evaluated by determining the coefficients ($\mathcal{A}, \mathcal{B}, \mathcal{C}, \mathcal{D}$) of the equations~\eqref{delvsq}-\eqref{delr}
and for various flow geometries those are found to be,
\begin{subequations}
\begin{eqnarray}
{\cal A} 
&=& \left\lbrace \begin{array}{lr}
                 2v_c^2\left[\xi\dfrac{a}{2}+\dfrac{2 \alpha \lambda_0^2 \gamma c_{s_c}^2}{v_c^3 r_c^2\sqrt{\gamma \Phi'_c r_c}} \right] & (\rm V)\\
\dfrac{2\alpha \lambda_0^2 \Theta(\gamma +1)}{r_c^2}+ 2(1- \gamma) c_{s_c}^2 & (\rm C)\\
\dfrac{2\alpha \lambda_0^2 H_0(\gamma +1)}{r_c^3}+ (1- \gamma) c_{s_c}^2 & (\rm H)\end{array}\right. \\
{\cal B} 
&=& \left\lbrace \begin{array}{lr} -2v_c^2 \left[\left( \dfrac{2 \lambda_0^2}{r_c^3}+\Phi'_c + \Phi''_c r_c\right)+\dfrac{v_c^2}{2}\dfrac{\Phi''_c}{\Phi'_c}b\right] & \\
\;\;\;\;\;\;\;\;\;\;\;\;\;\; -2v_c^2 \left[\xi\dfrac{v_c^2}{2r_c}a^2-\dfrac{2\alpha \lambda_0^2 c_{s_c}^2}{v_c r_c^3\sqrt{\gamma \Phi'_c r_c}}\left(8+a\left(\dfrac{3 -\gamma}{\gamma +1} \right) \right)\right] & (\rm V)\\         
2v_c^2 \left[ \dfrac{4 \alpha \lambda_0^2 \Theta (\gamma +1)}{r_c^3}+4\dfrac{c_{s_c}^2(1-\gamma)}{r_c}-\left( \dfrac{2 \alpha \lambda_0^2}{r_c^3}+\Phi'_c + \Phi''_c r_c\right)\right] & (\rm C)\\
2v_c^2 \left[ \dfrac{2 \alpha \lambda_0^2 H_0 (\gamma +5)}{r_c^4}+\dfrac{c_{s_c}^2(1-\gamma)}{r_c}-\left( \dfrac{2 \alpha \lambda_0^2}{r_c^3}+\Phi'_c + \Phi''_c r_c\right)\right] & (\rm H) \end{array}\right. \\
{\cal C} 
&=& \left\lbrace \begin{array}{lr} \gamma \beta^2 r_c & (\rm V)\\
r_c \left( \dfrac{\gamma +1}{2}\right) & (\rm C)\\
r_c \left( \dfrac{\gamma +1}{2}\right) & (\rm H) \end{array}\right. \\
{\cal D} 
&=& \left\lbrace \begin{array}{lr} -a\left(\dfrac{\gamma -1}{\gamma +1}\right)v_c^2 & (\rm V)\\
2(\gamma -1)c_{s_c}^2 & (\rm C)\\
(\gamma -1)c_{s_c}^2 & (\rm H) \end{array}\right.
\end{eqnarray}\end{subequations}

where, $a = \left[r_c\dfrac{\Phi''_c}{\Phi'_c}-3\right]$, $b = \left[1+r_c\dfrac{\Phi'''_c}{\Phi''_c}-r_c\dfrac{\Phi''_c}{\Phi'_c}\right]$ and $\dfrac{\gamma -1}{\gamma +1}=\xi$.

\section{Qualitative understanding of the stationary flow configuration}

Results demonstrated in this section are obtained for the Paczyn´ski and
Wiita (1980) potential \citep{PW_1980},

\begin{eqnarray}
\Phi_{PW}(r) = \dfrac{-1}{2(r-1)},
\end{eqnarray}
where $r$ is the radial coordinate scaled by the Schwarzschild radius. Similar results may be obtained for any other specific pseudo-Schwarzschild black hole potential.

With the help of the mathematical procedure outlined in section \ref{CP_find}, the critical points have been obtained by solving equation \eqref{neq0} for $r_c$~(scaled by Schwarzschild radius) in terms of the system parameters, i.e., $\lambda_0$ and $\cal \dot{M}$ for a given choice of $\gamma$ and $\alpha$. We look for physically allowed single critical points and three critical points (identified as outer-middle-inner critical points) sets with variation in $\alpha$ values for all the three flow geometries.

A subset of the $\lambda_0$ - $\dot{\mathcal{M}}$ parameter space which corresponds to three critical point solutions satisfying the condition, that the local flow energy ${\cal E}_{0,\rm outer}>{\cal E}_{0,\rm inner}$ as mentioned in section~\ref{method}, where \[{\cal E}_0=\dfrac{v^2}{2} + n c_s^2 + \Phi(r) + \dfrac{\lambda_{0}^{2}}{2r^{2}},\] may support multicritical incoming solution \citep{Chakra_Das_2004}. The other option i.e. ${\cal E}_{0,outer}<{\cal E}_{0,inner}$ cannot support the multicritical incoming solution in spite of having 3 critical points on the phase plot, rather it would correspond to the single critical point incoming solution through the inner one. Accordingly, in the parameter space the subregions are labelled by $\rm A$ and $\rm A_1$ respectively.

\begin{figure*}[h!]
\centering
\begin{tabular*}{1.0\linewidth}{@{\extracolsep{\fill}}ccc}%
\hline
{\bf V} & {\bf C} & {\bf H} \\
\hline
 &$\alpha$=0.00 &\\
\hline
\epsfig{file=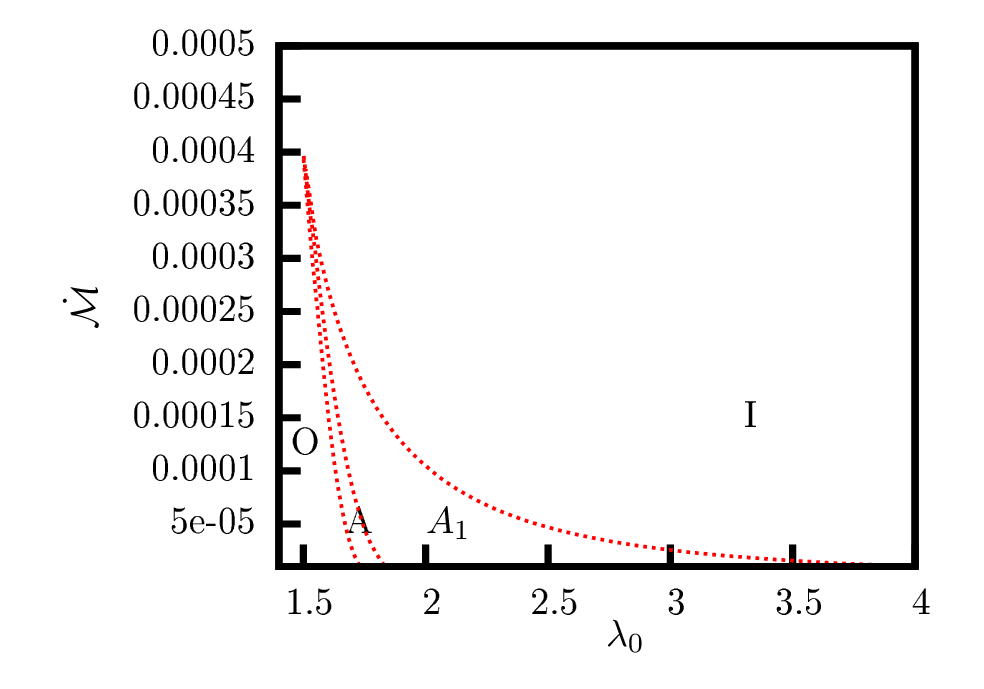,angle=0,width=2.15in}&
\epsfig{file=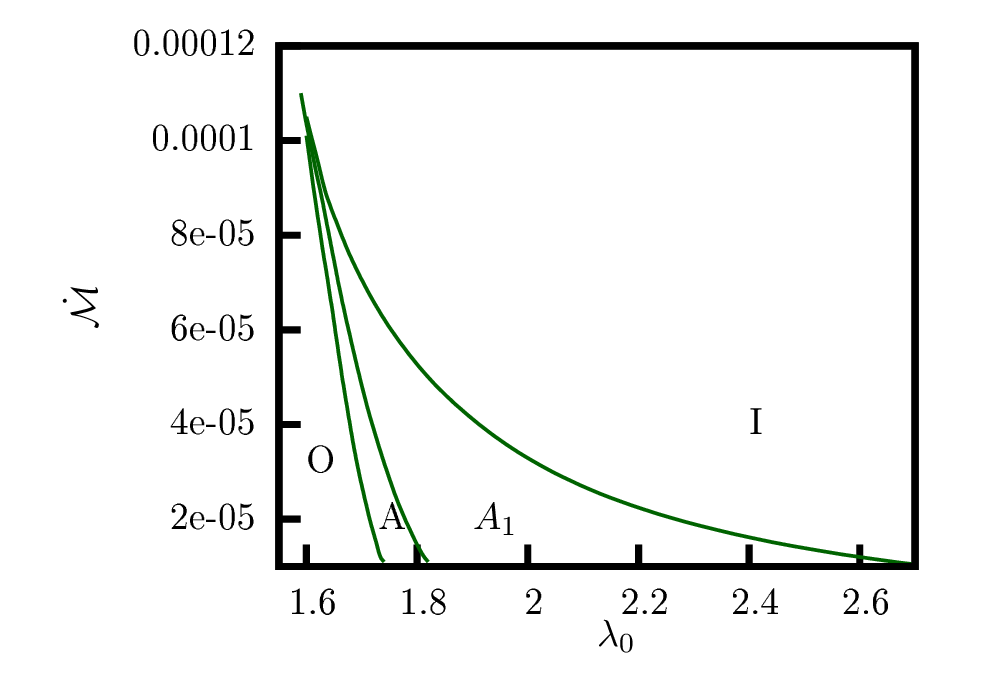,angle=0,width=2.15in}&
\epsfig{file=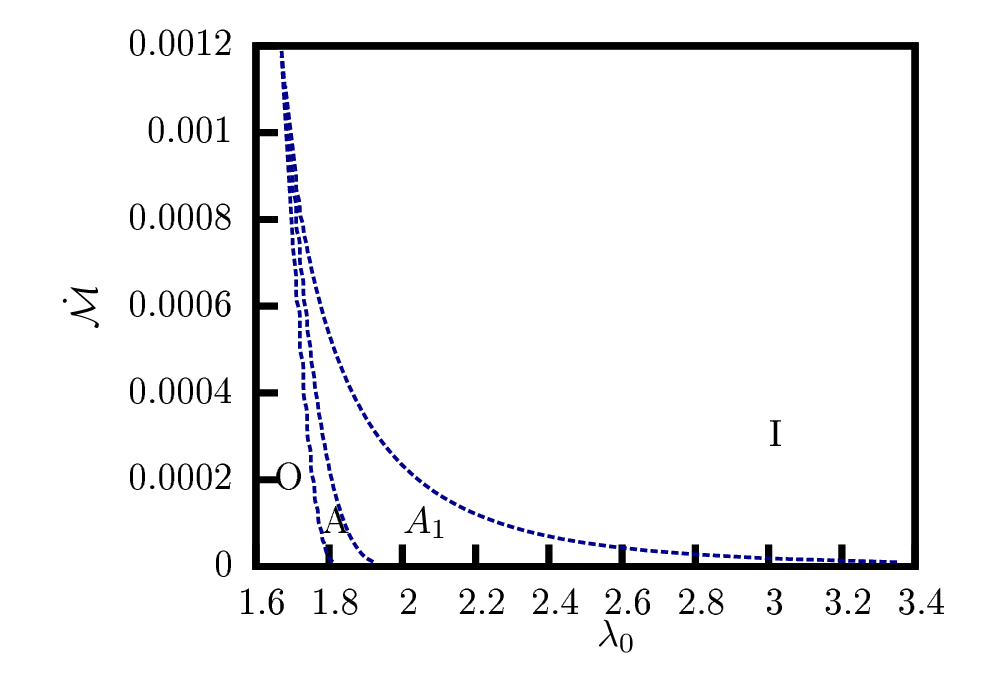,angle=0,width=2.15in}\\
%
%
& &\\
\hline
 &$\alpha$=0.01 &\\
\hline
\epsfig{file=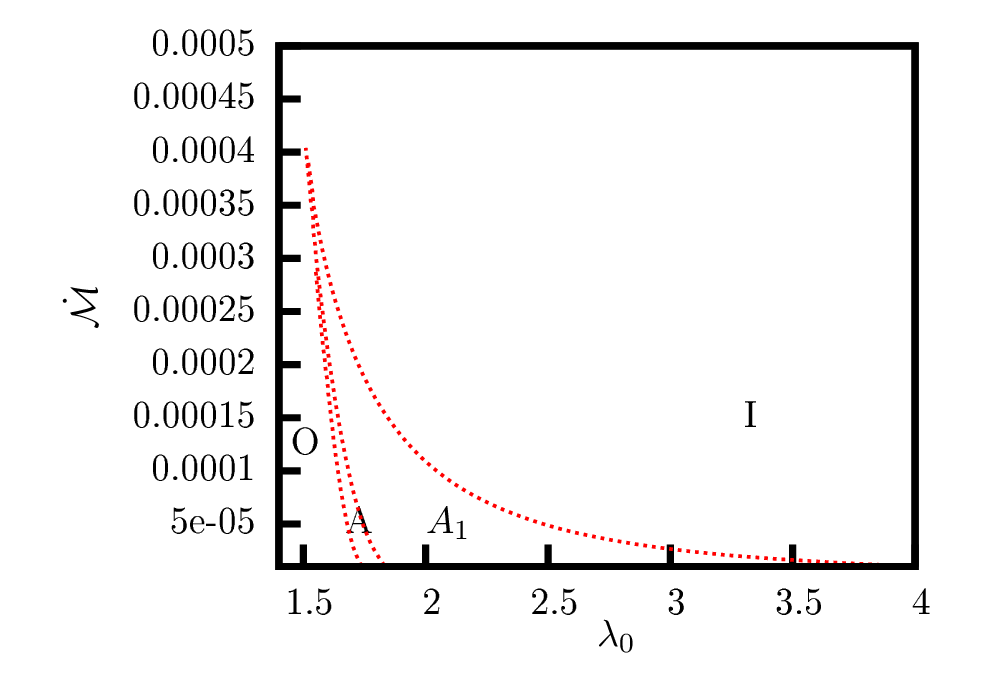,angle=0,width=2.15in}&
\epsfig{file=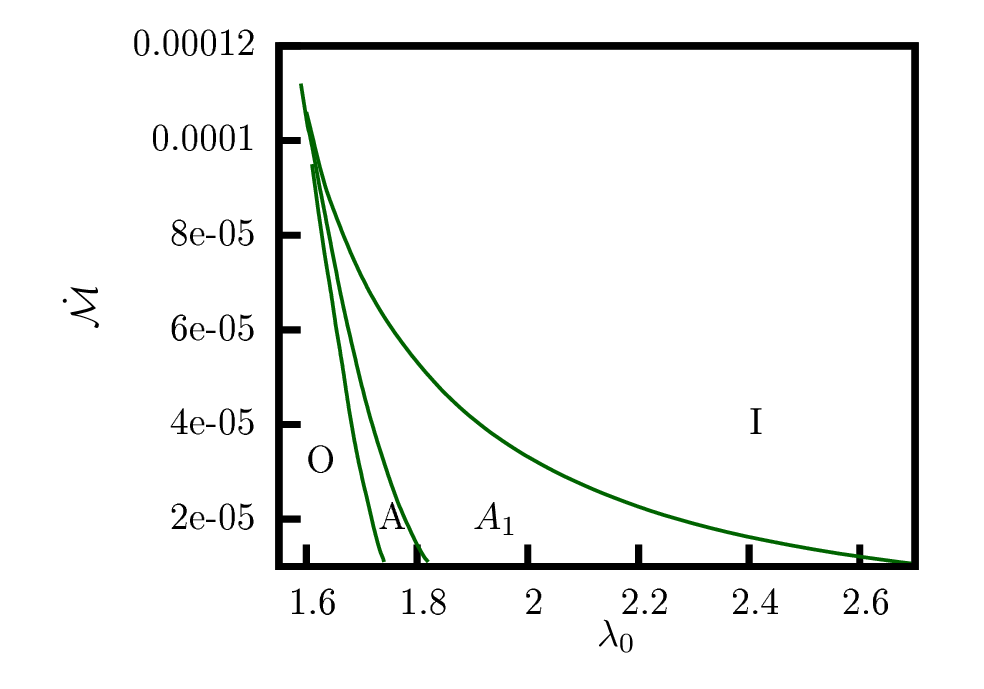,angle=0,width=2.15in}&
\epsfig{file=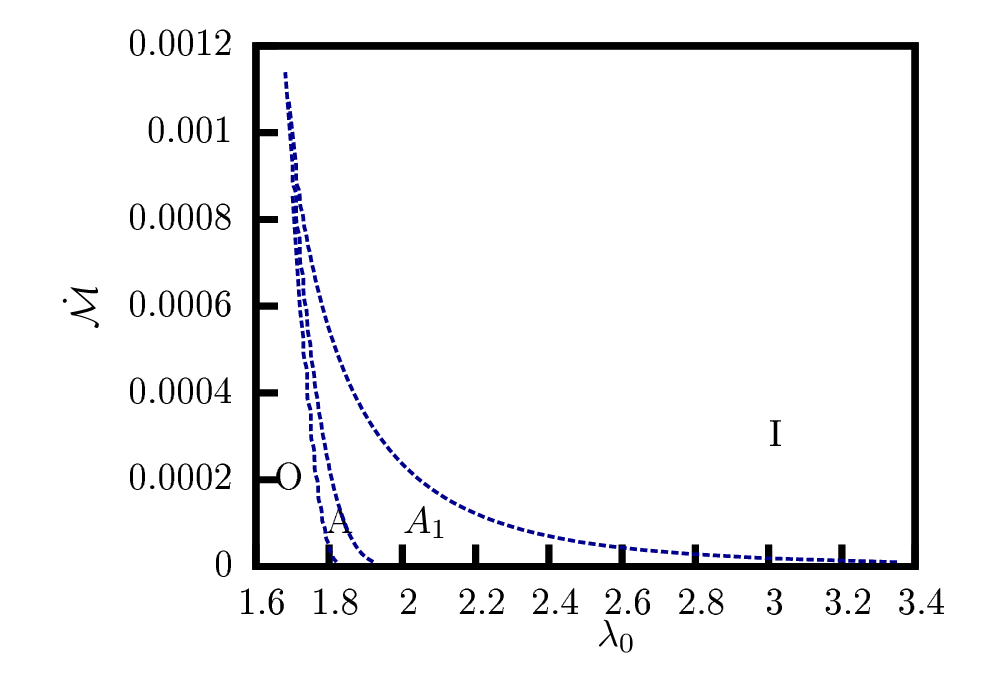,angle=0,width=2.15in}\\
%
&  &\\
\hline
&$\alpha$=0.1 &\\
\hline
\epsfig{file=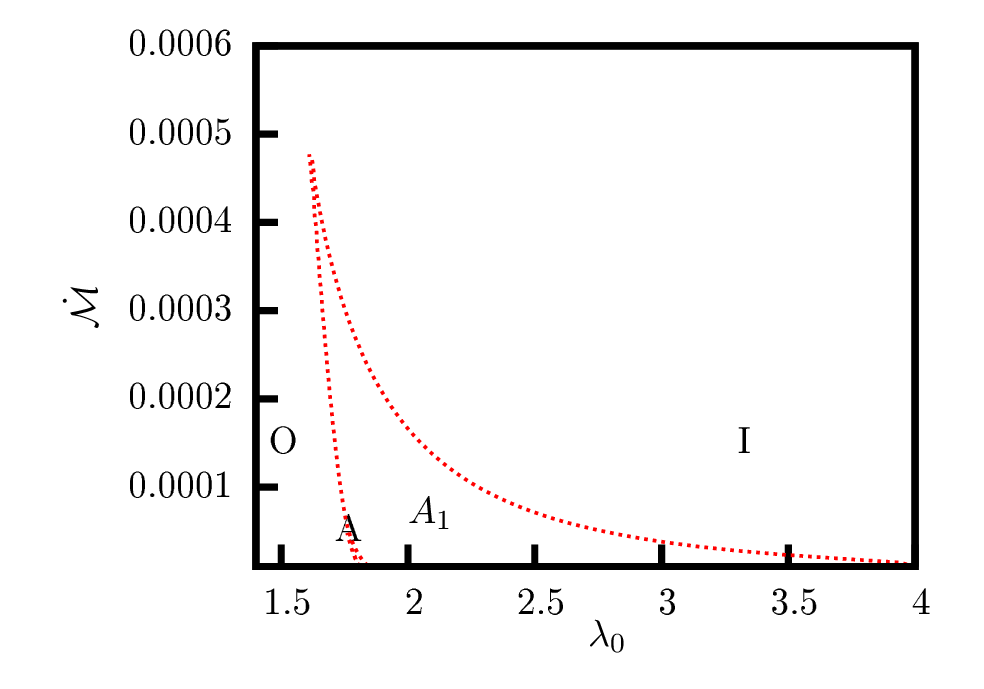,angle=0,width=2.15in}&
\epsfig{file=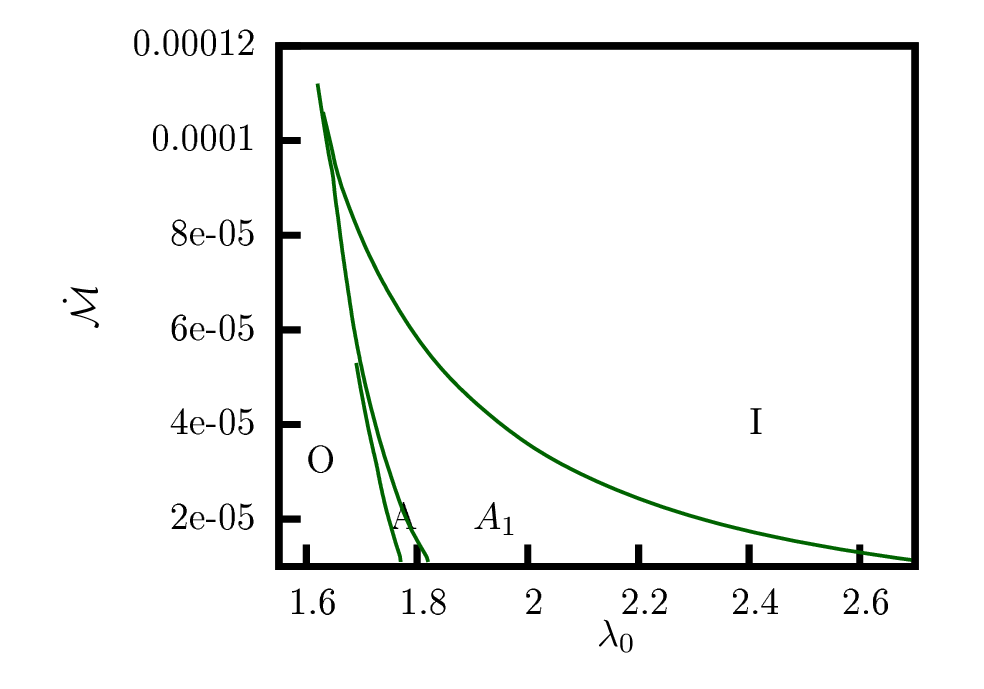,angle=0,width=2.15in}&
\epsfig{file=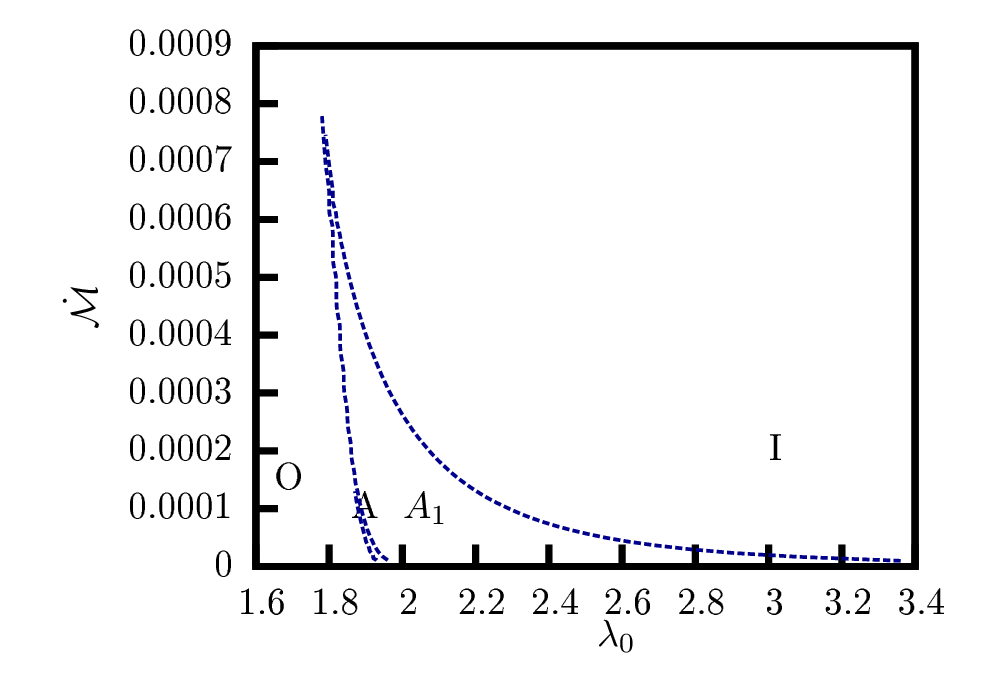,angle=0,width=2.15in}\\
 & &\\
\hline
\end{tabular*}
\caption{\small Regions of single (O for outer and I for inner) and three critical points for both multicritical ($\rm A$) and single critical ($\rm A_1$) accretion for polytropic flow in the parameter space of entropy accretion rate ($\cal \dot{M}$) and constant angular momentum $(\lambda_0$), for three different disk models (V, C \& H) and $\alpha$ - viscosity parameter values varying between 0 to 0.1 under $\gamma=\frac{4}{3}$. The dotted-red lines are for the vertical equilibrium geometry (V), the solid-green lines are for the conical
geometry (C), and the dashed-blue lines are for the constant-height disk geometry (H). For the model `H', $H_0=1$ and for `C', $\Theta=0.1$ --- the values chosen in this work for numerical demonstration of the relevant features.}
\label{Fig:2D_para_alphavary_VCH}
\end{figure*}

In figure (\ref{Fig:2D_para_alphavary_VCH}), we have shown the single (`O' for outer or `I' for inner one) and wedge shaped multitransonic i.e. three critical point solutions in the parameter space of $\cal \dot{M}$ and $\lambda_0$, a feature common to all the three different disk models (V, C \& H) considering three values of the $\alpha$ - varying between 0 and 0.1 for 
$\gamma=\frac{4}{3}$. However, the span of the 
[$\cal \dot{M}$, $\lambda_0$] for multi critical solutions varies between disk models. It covers maximum span for the $\cal \dot{M}$ parameter for the `H' case and minimum for the `V' case.  Further, for any of the disk geometries, the parameter space corresponding to three critical point solution is classified into $\rm A$ and $\rm A_1$ region based 
on the prescription mentioned above. It should be noted that a point on the boundary line between the subregions $\rm A$ and $\rm A_1$ indicates equal value for the local flow energy for the outer and inner 
critical points. The multicritical accretion region seems to contract with increasing $\alpha$ values. In particular, with higher value of $\alpha$, these regions turn to monotransonic `$O$' regions in the lower $\lambda_0$ side. 


\begin{figure*}[h!]
\centering
\begin{tabular*}{1.0\linewidth}{@{\extracolsep{\fill}}|cccc|}
\hline
{\bf CP} for ${\alpha}=0.0$ & {\bf V} & {\bf C} & {\bf H} \\
%
\hline
&
\epsfig{file=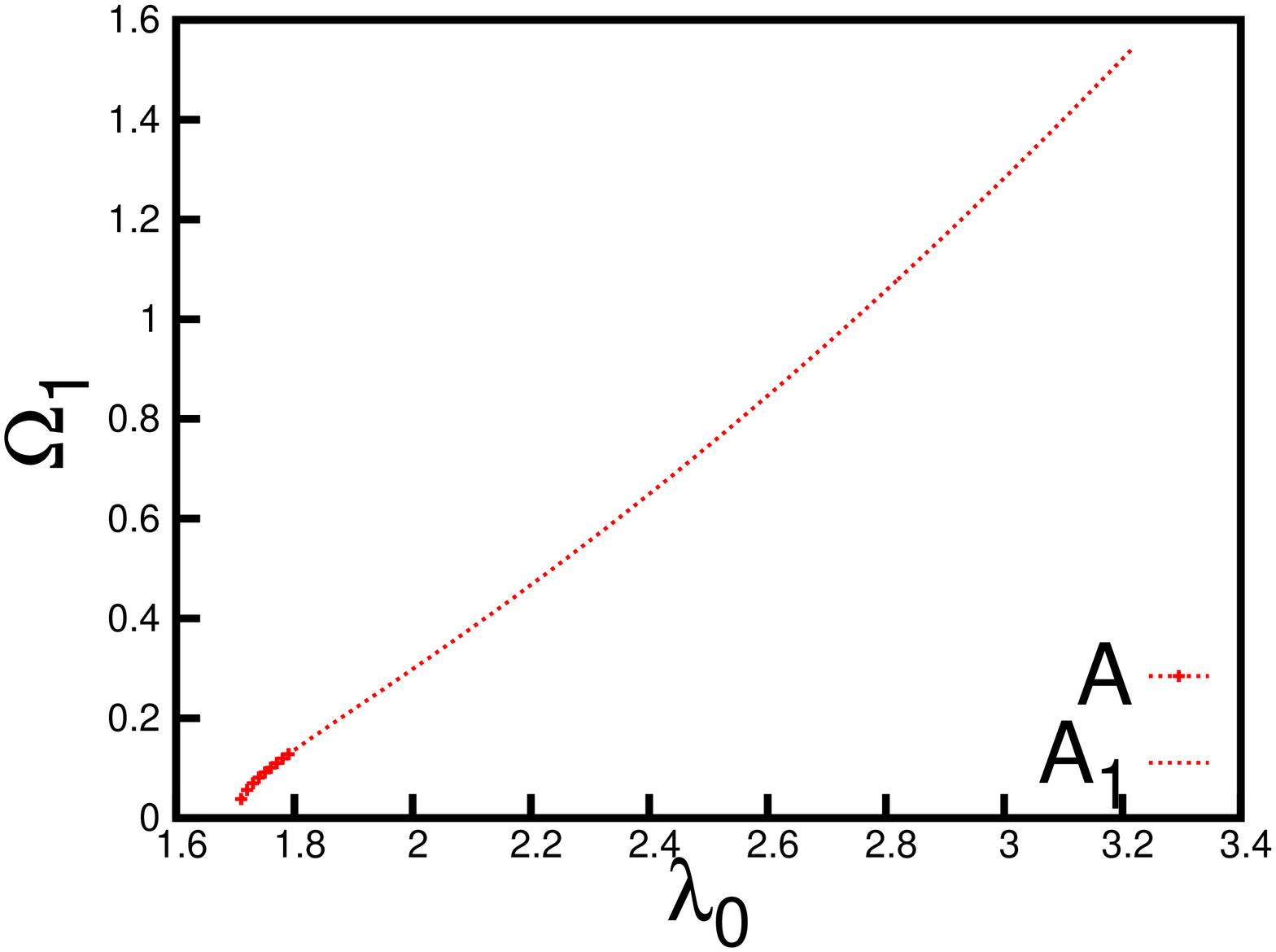,angle=270,width=1.4in}&
\epsfig{file=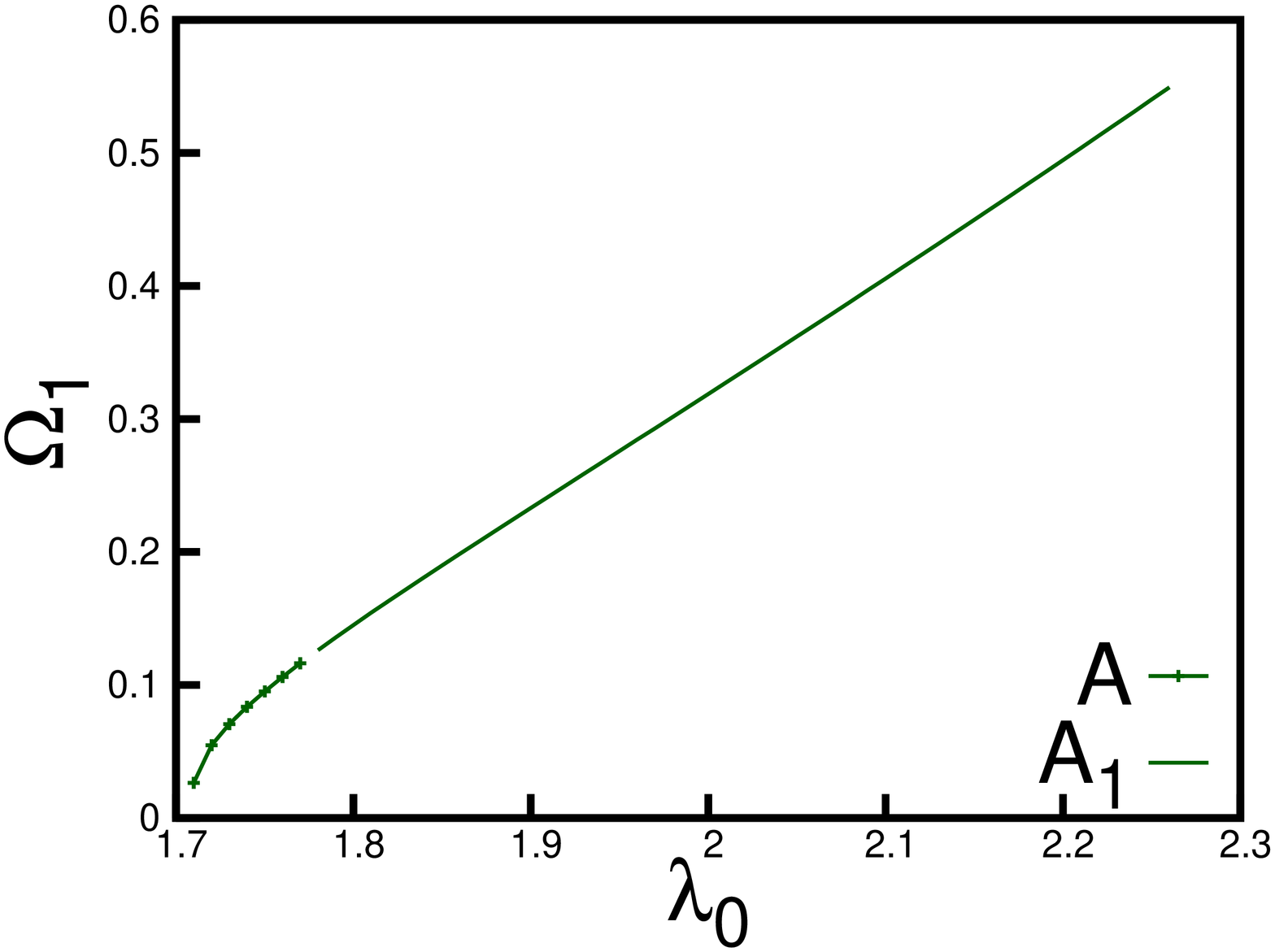,angle=270,width=1.4in}&
\epsfig{file=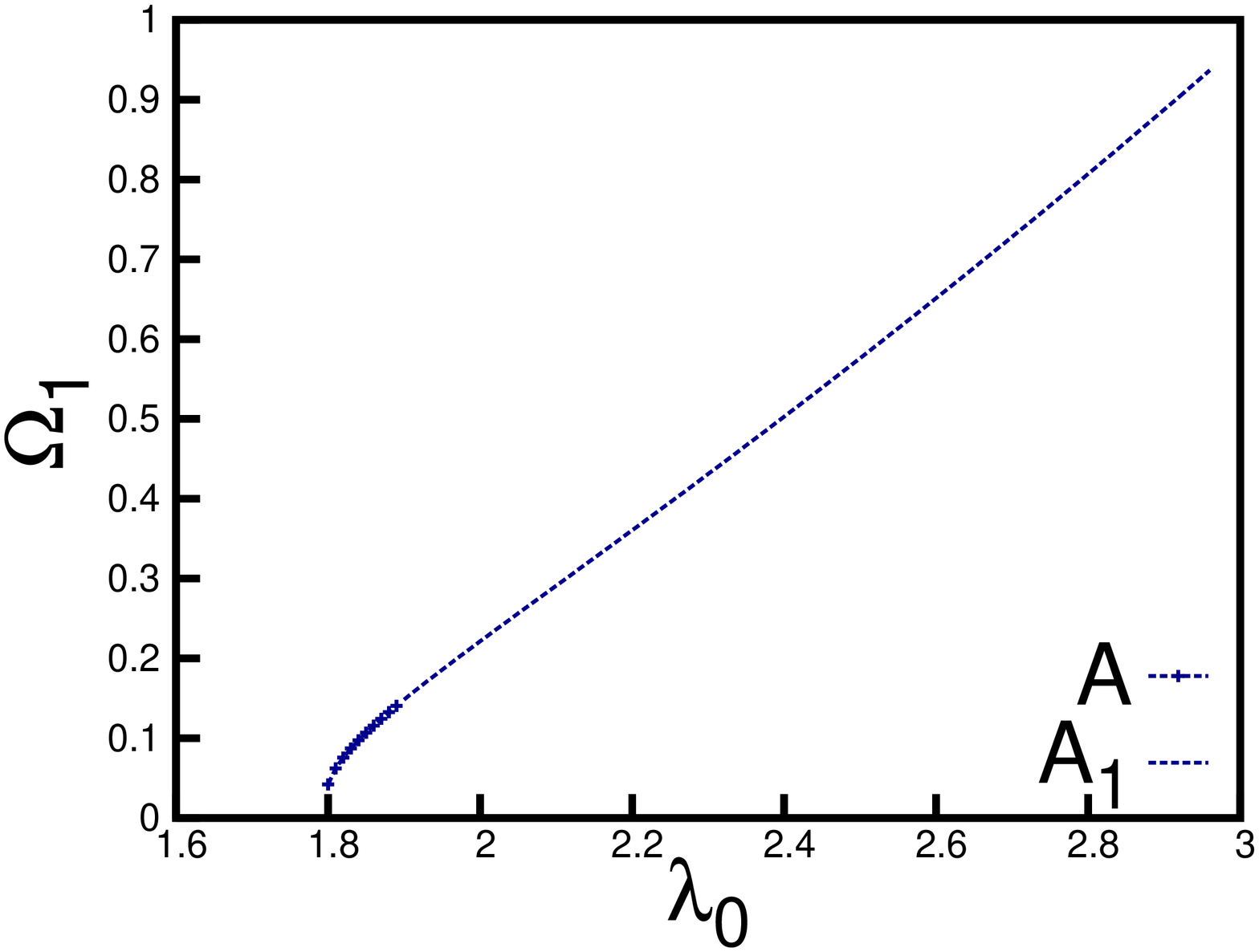,angle=270,width=1.4in}\\
%
 Inner&
\epsfig{file=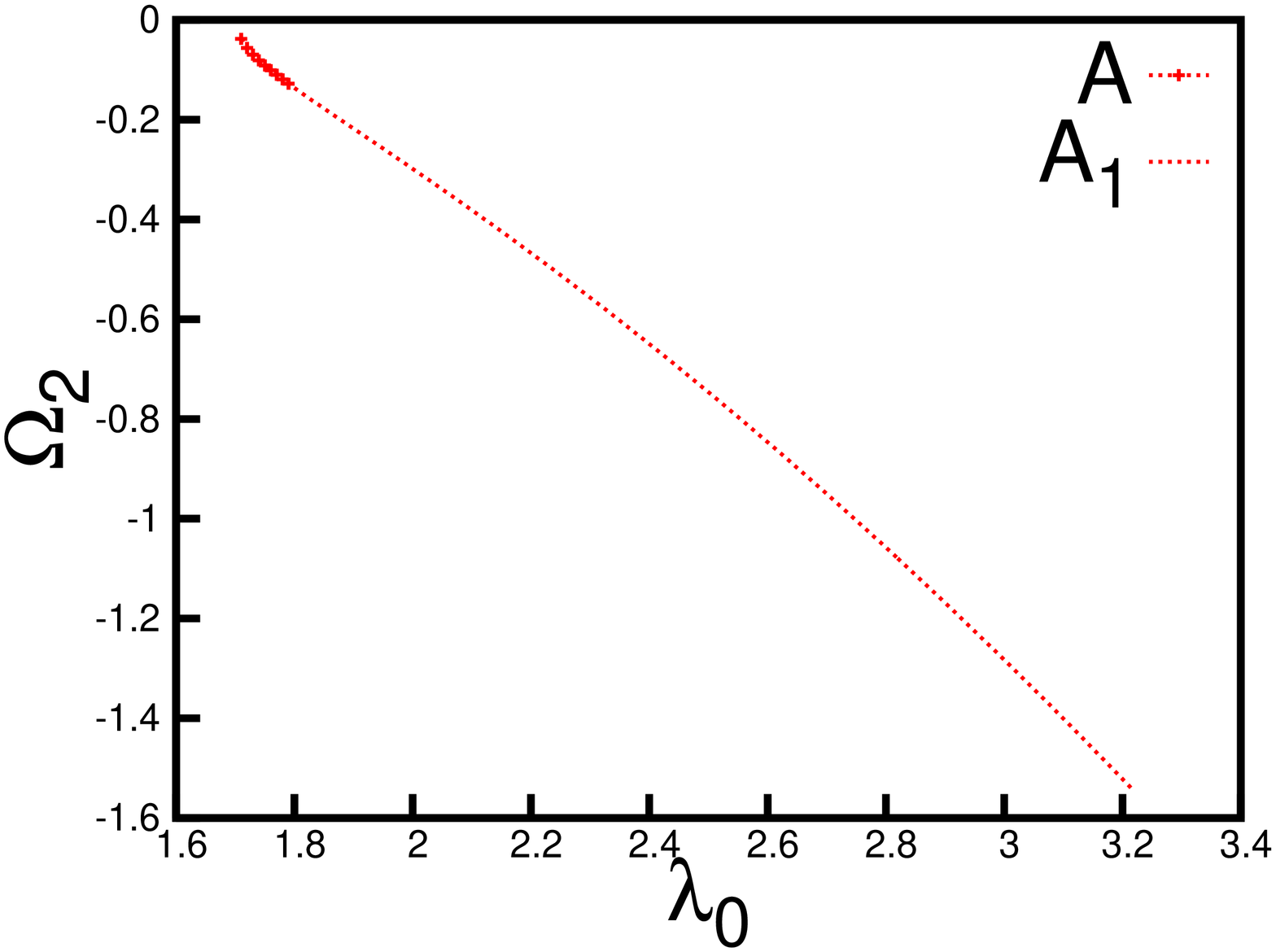,angle=270,width=1.4in}&
\epsfig{file=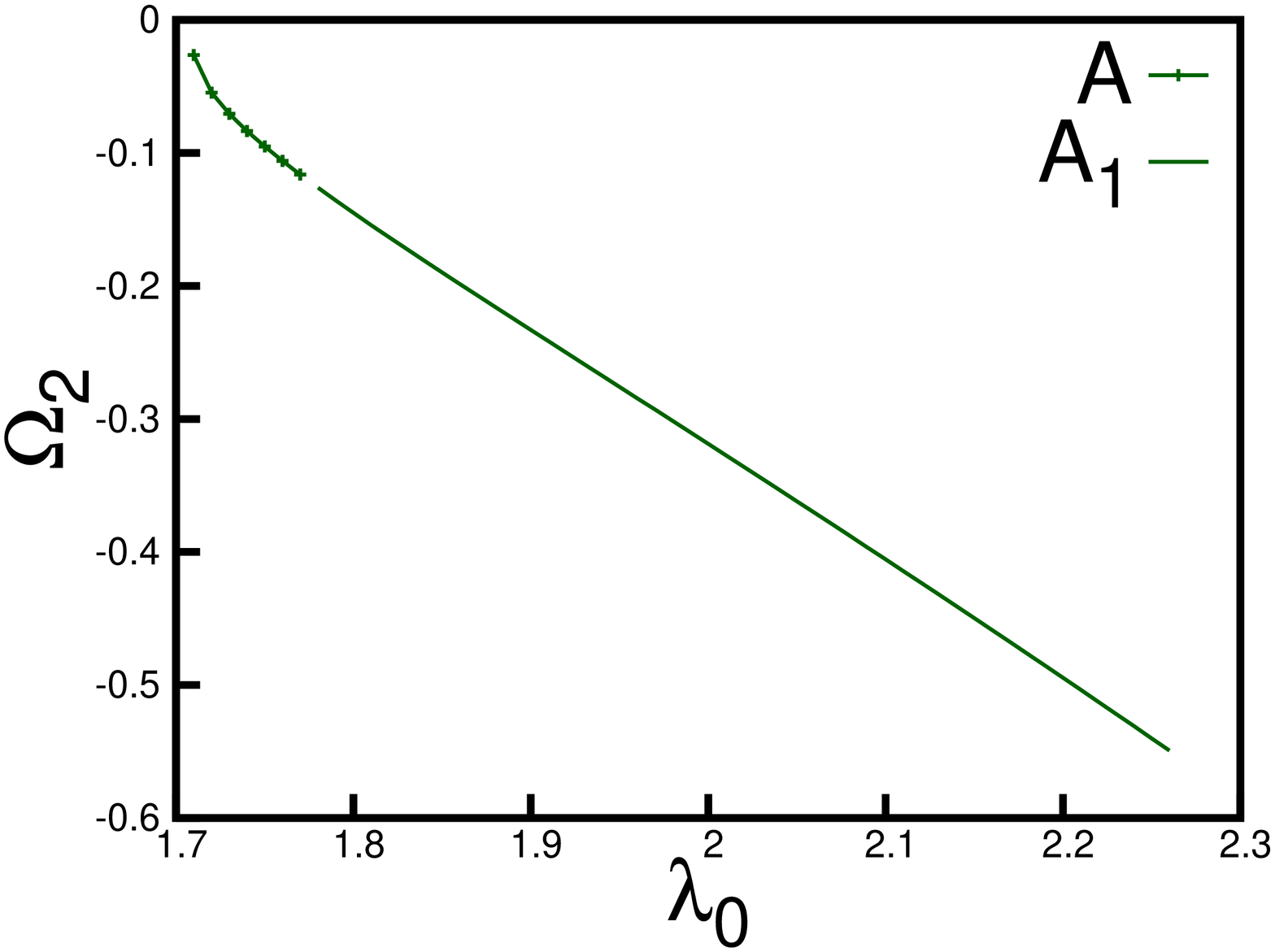,angle=270,width=1.4in}&
\epsfig{file=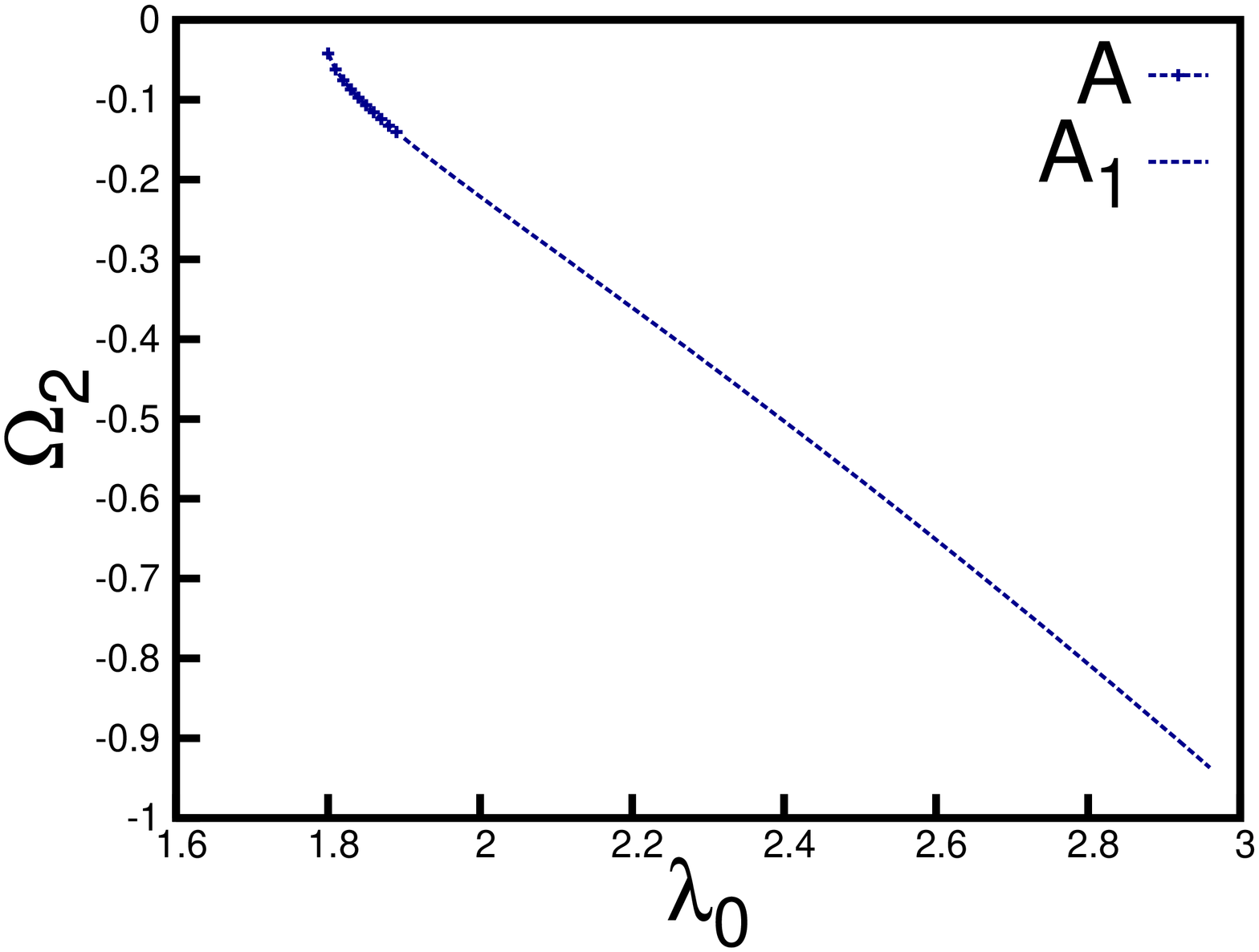,angle=270,width=1.4in}\\
\hline
&
\epsfig{file=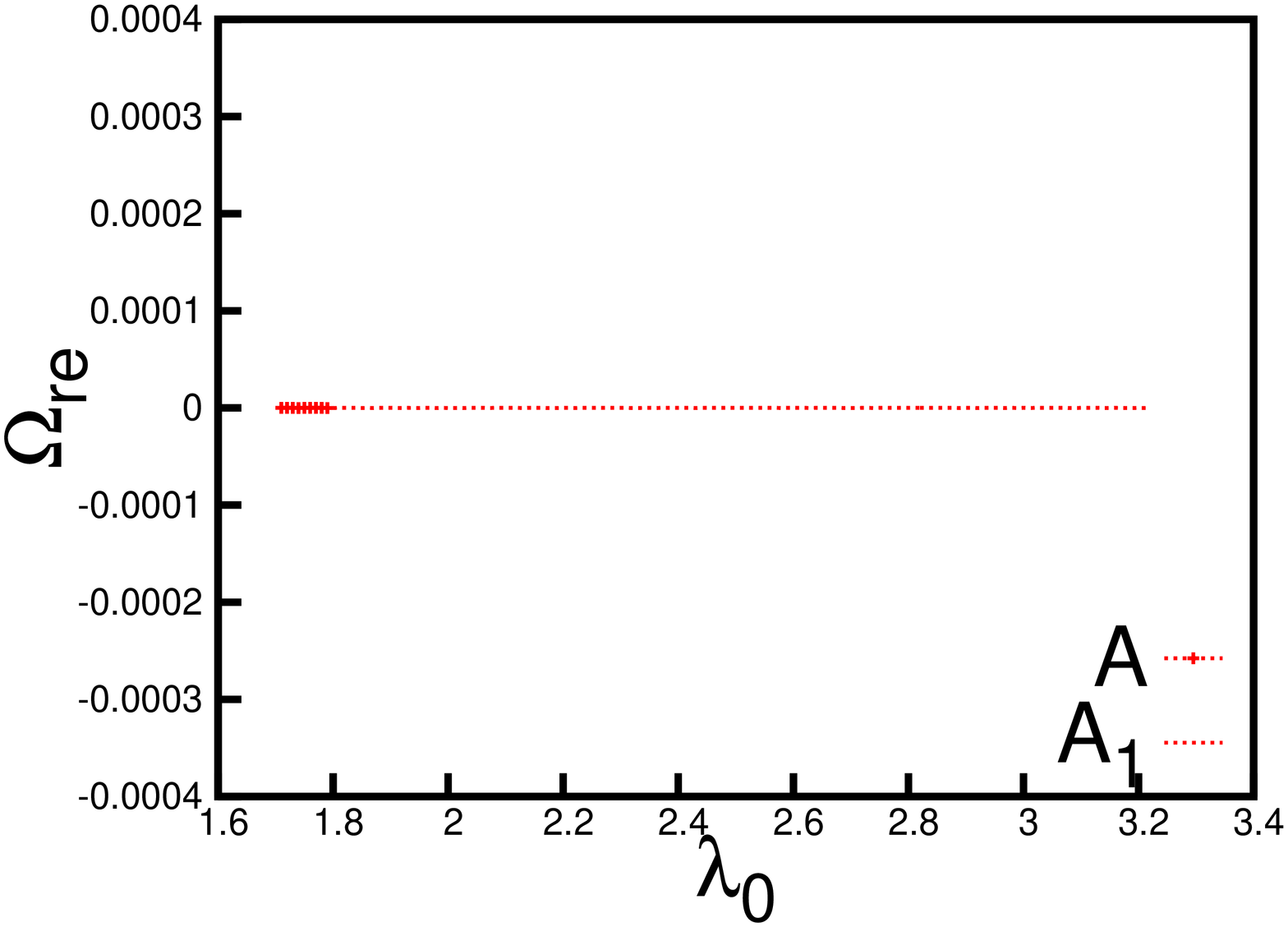,angle=270,width=1.5in}&
\epsfig{file=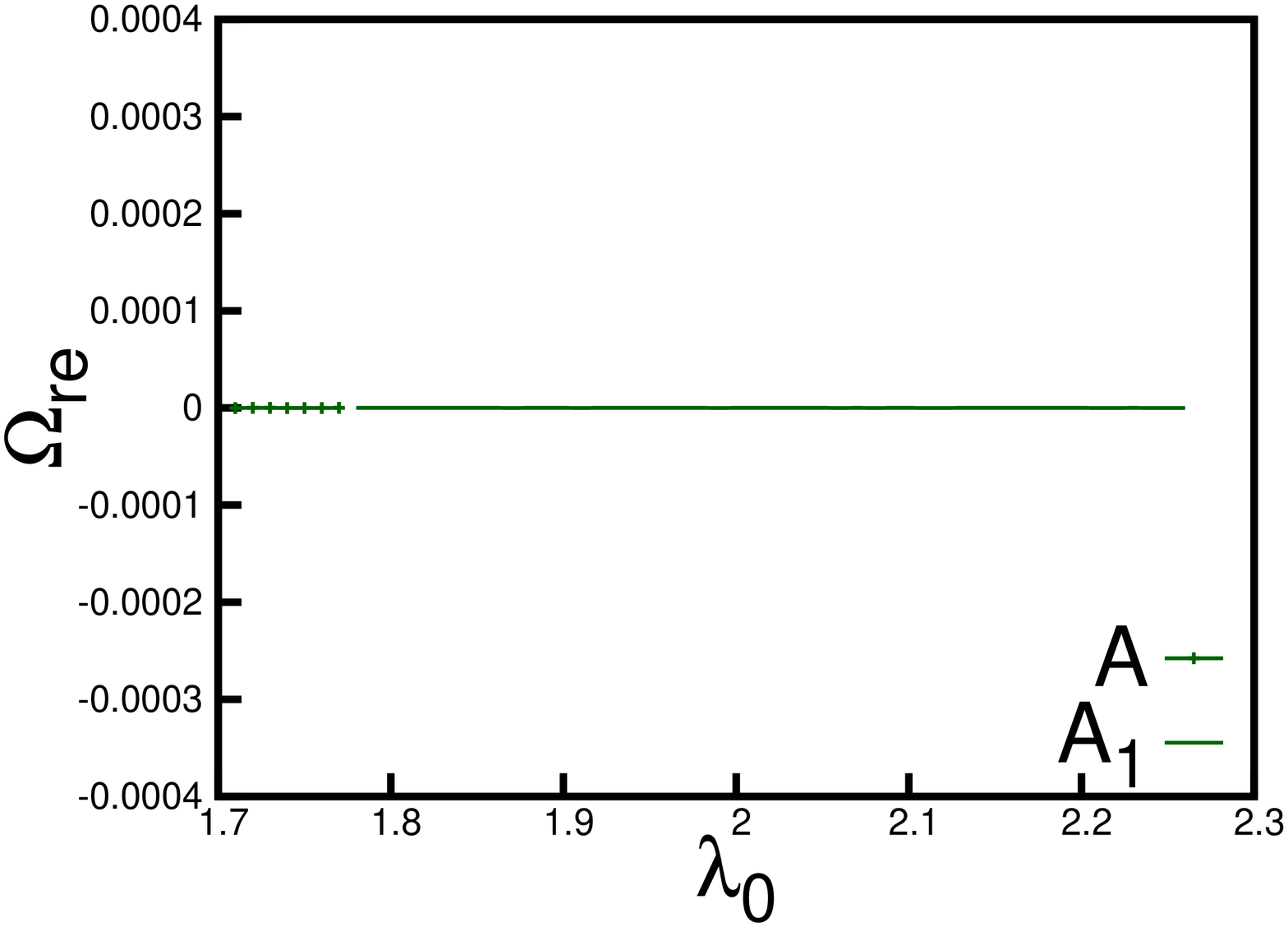,angle=270,width=1.5in}&
\epsfig{file=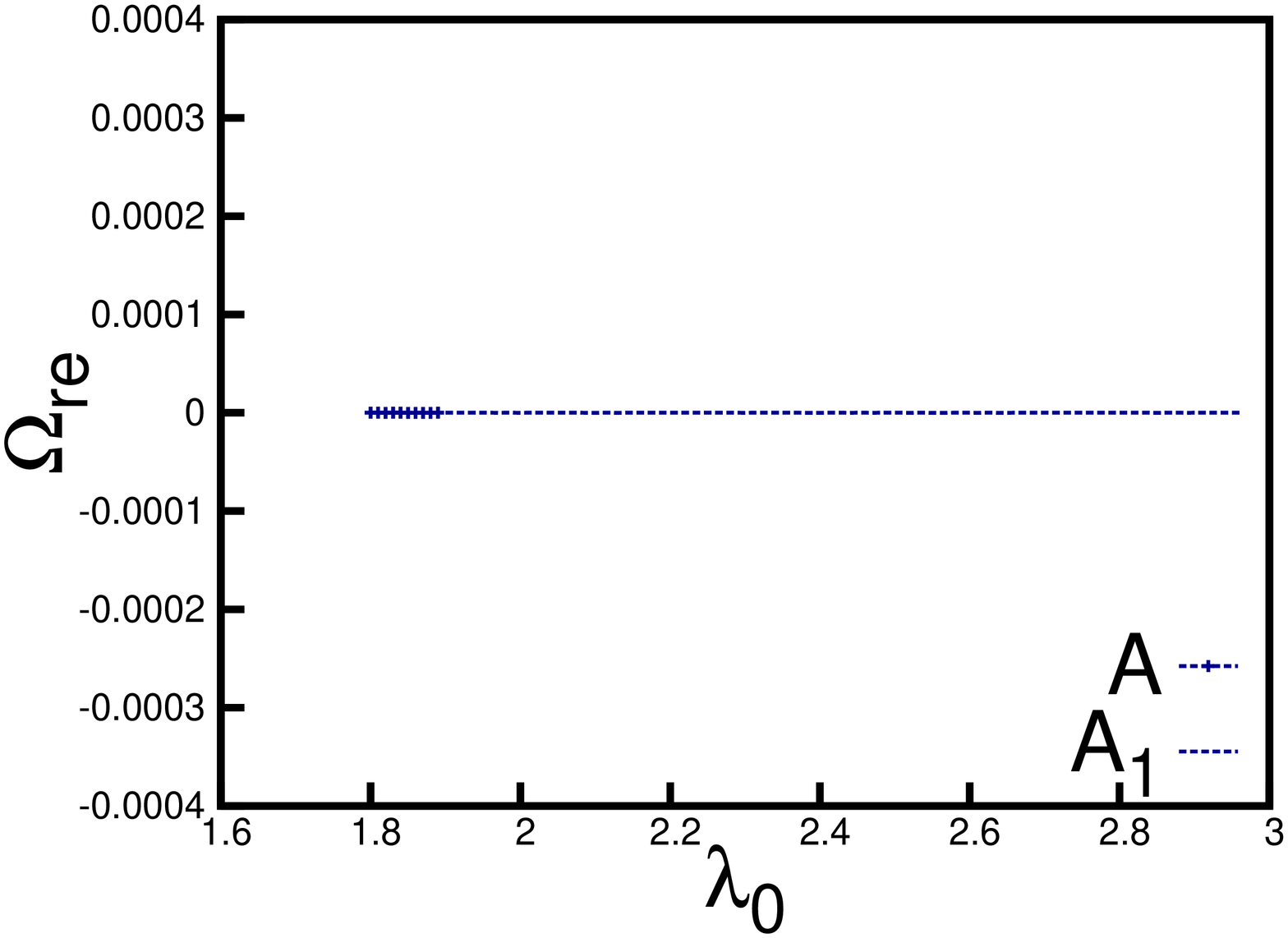,angle=270,width=1.5in}\\
Middle &
\epsfig{file=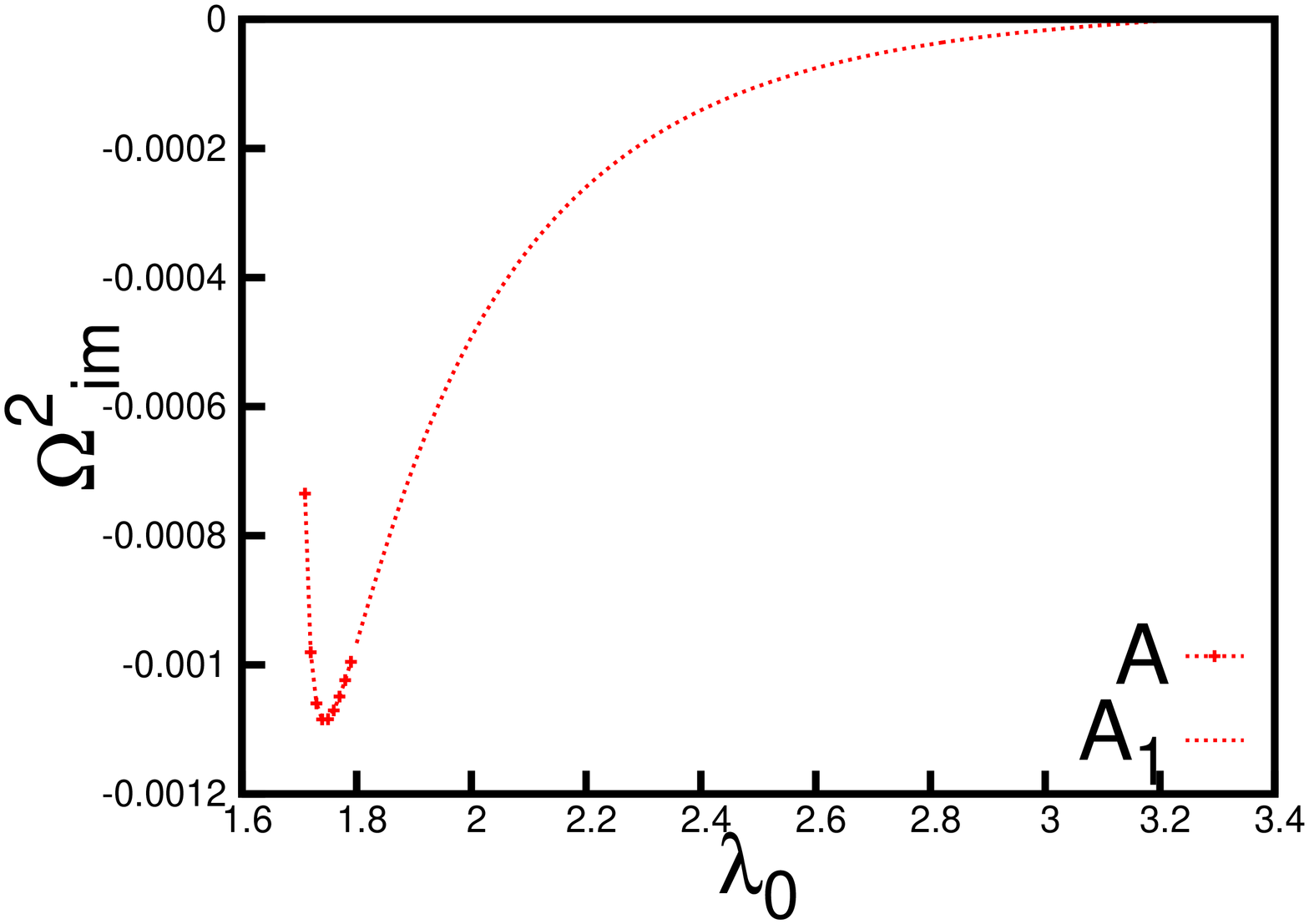,angle=270,width=1.5in}&
\epsfig{file=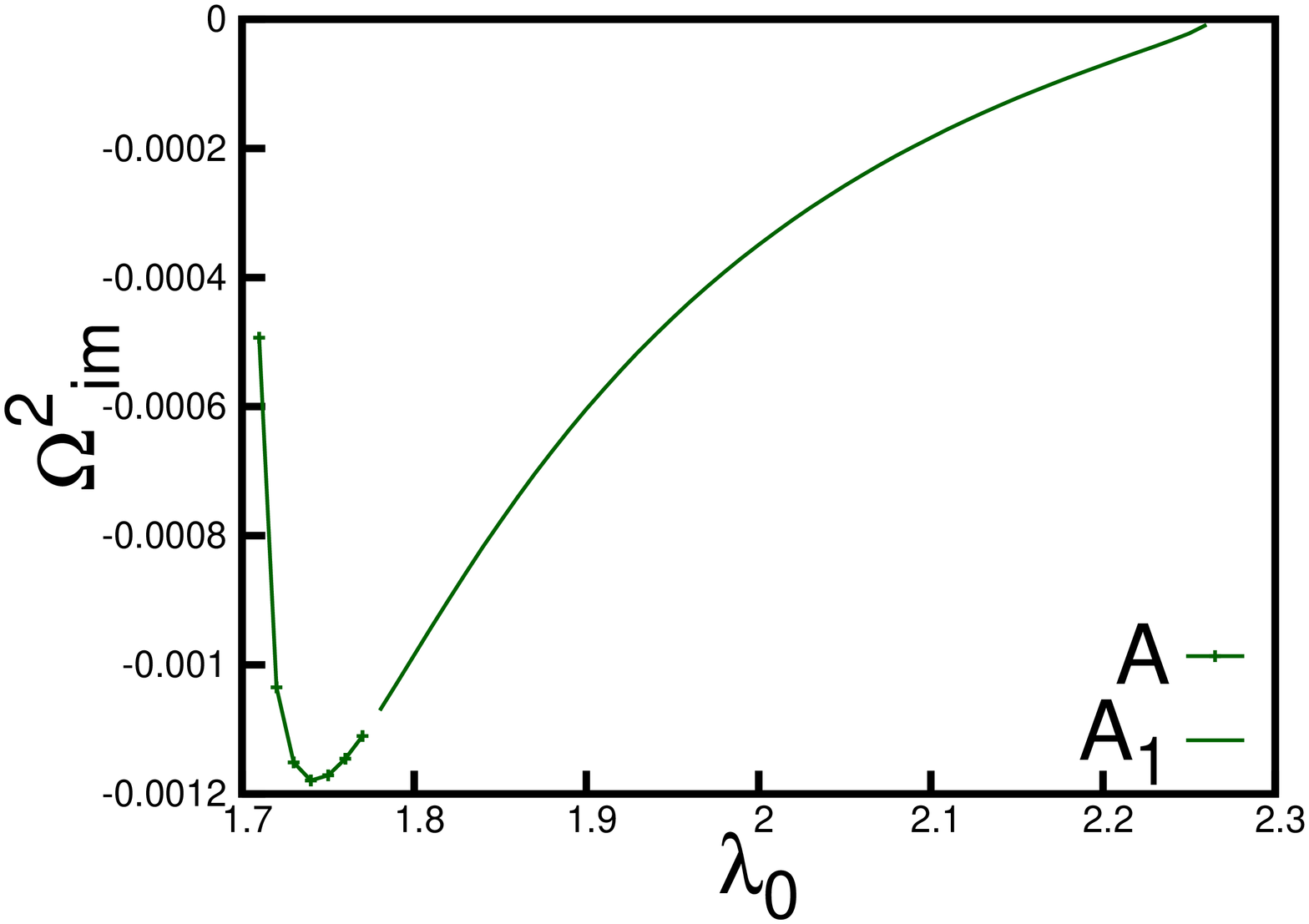,angle=270,width=1.5in}&
\epsfig{file=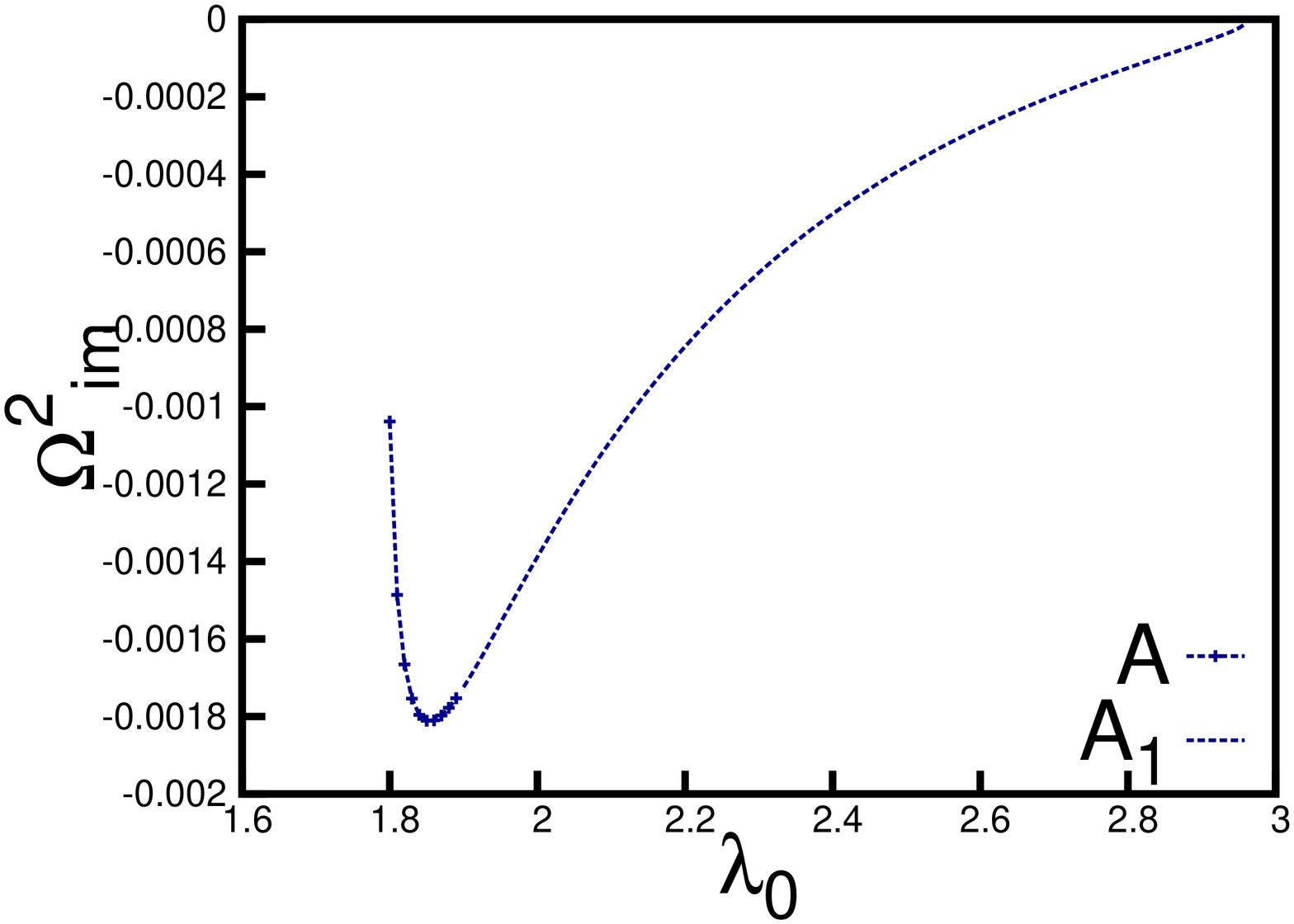,angle=270,width=1.5in}\\
%
\hline
&
\epsfig{file=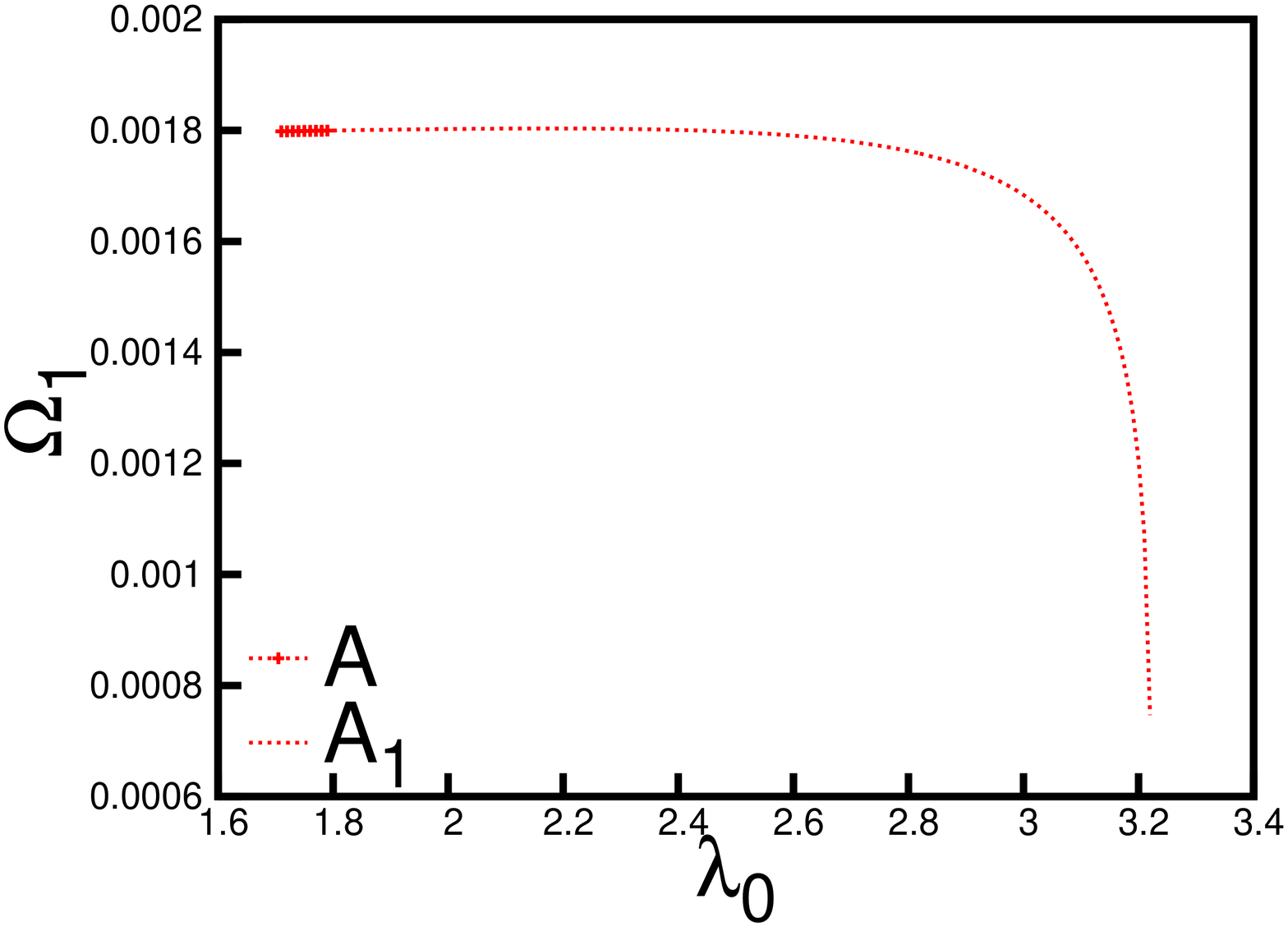,angle=270,width=1.5in}&
\epsfig{file=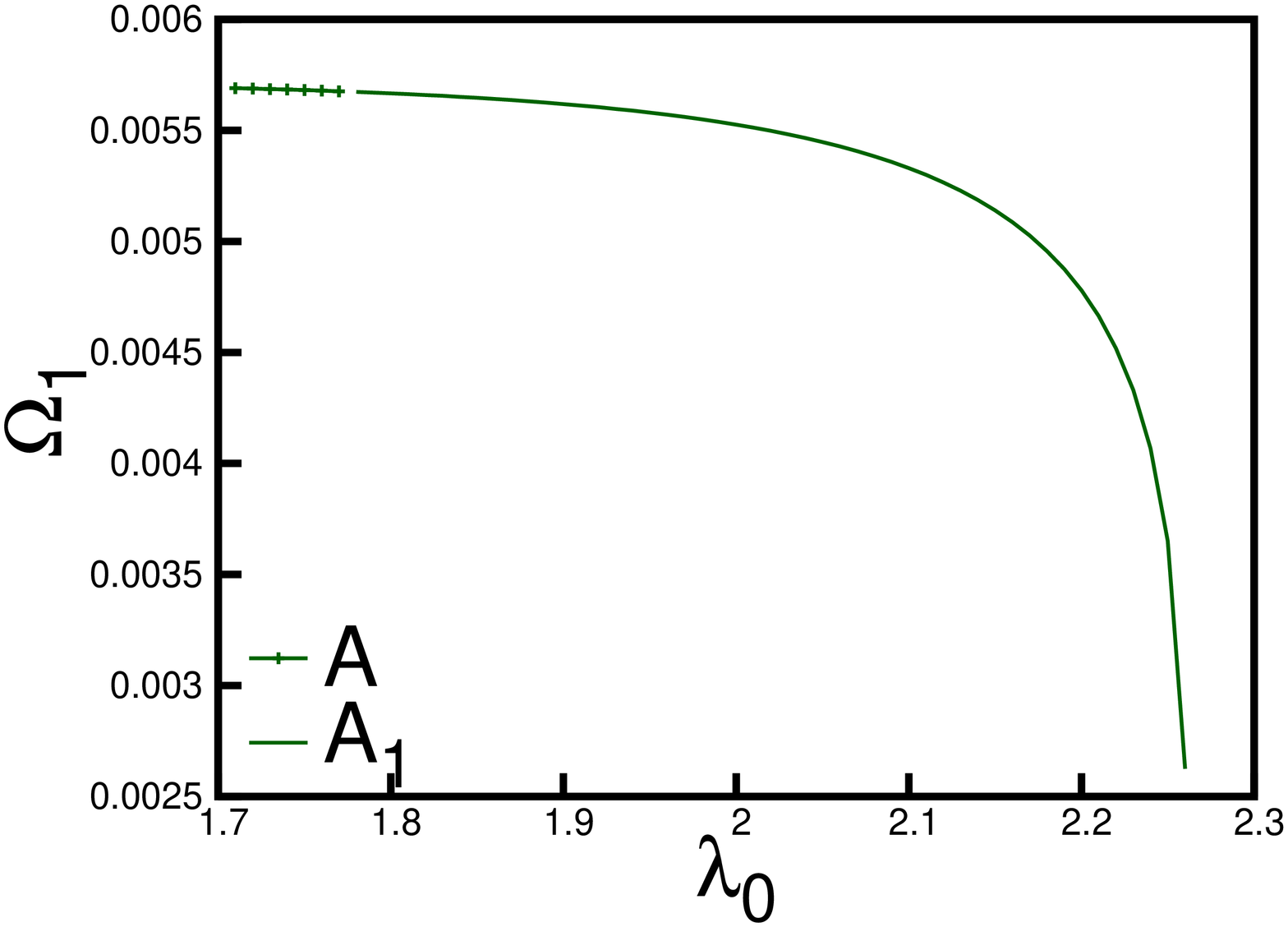,angle=270,width=1.5in}&
\epsfig{file=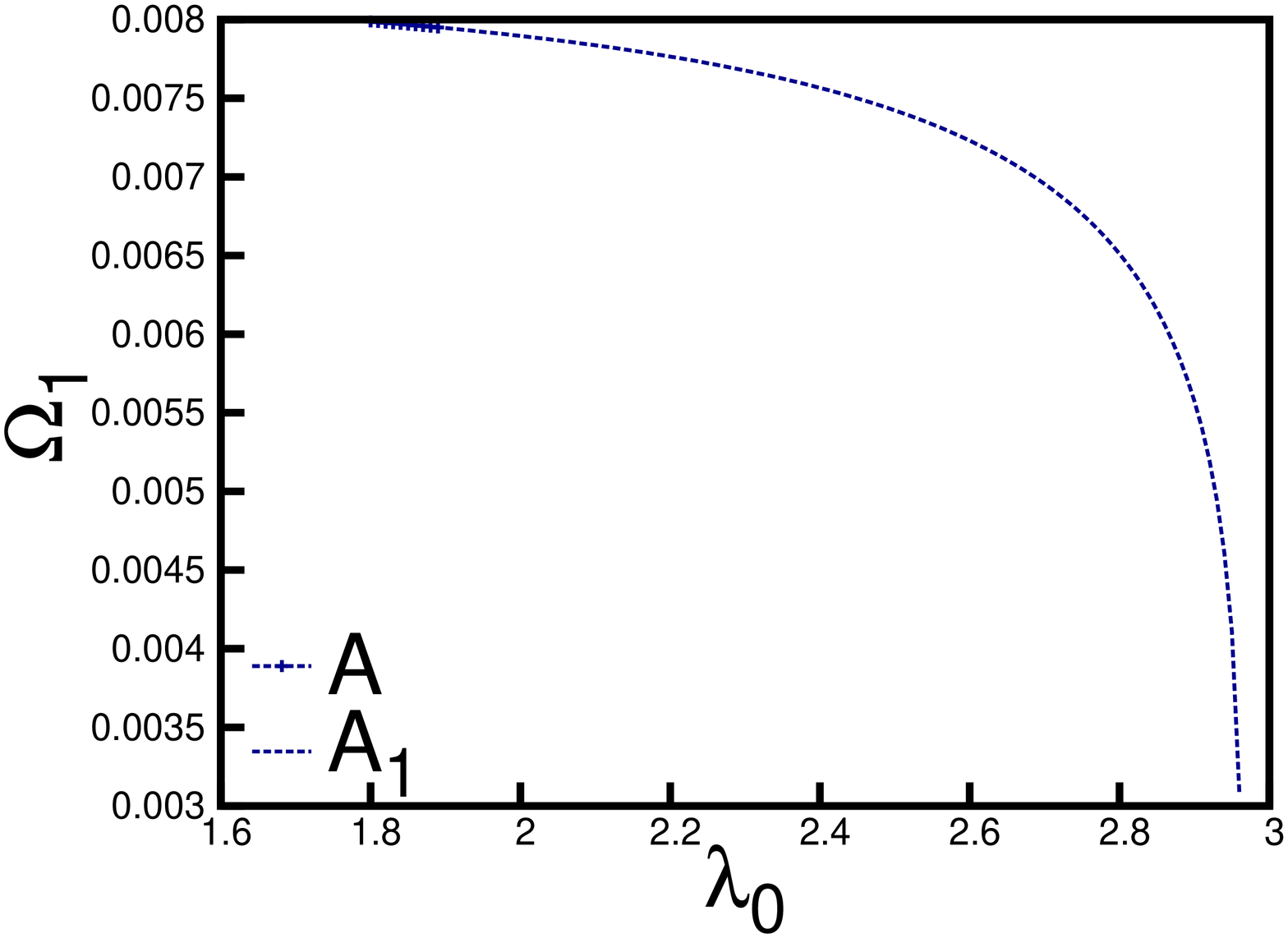,angle=270,width=1.5in}\\
Outer &
\epsfig{file=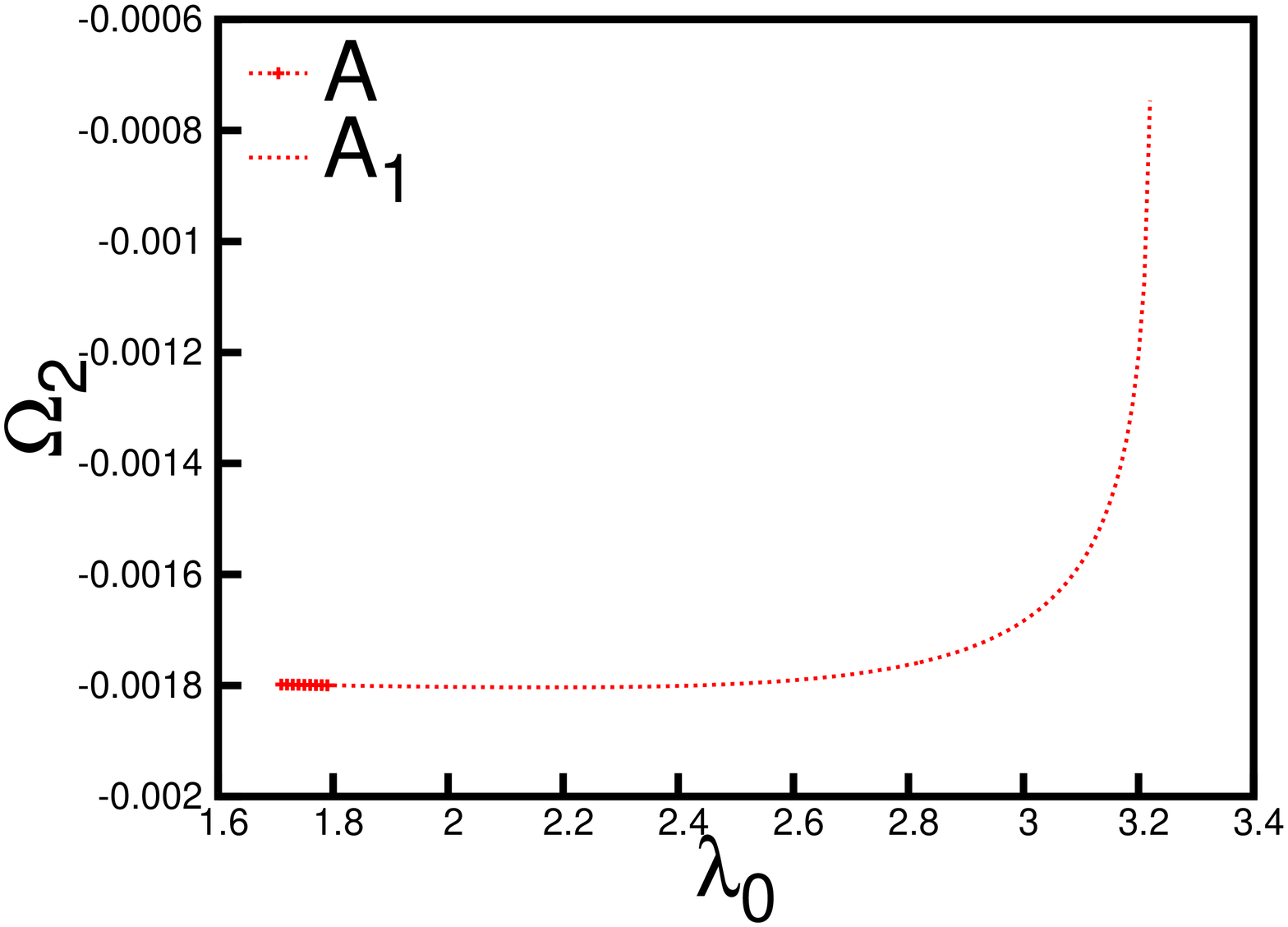,angle=270,width=1.5in}&
\epsfig{file=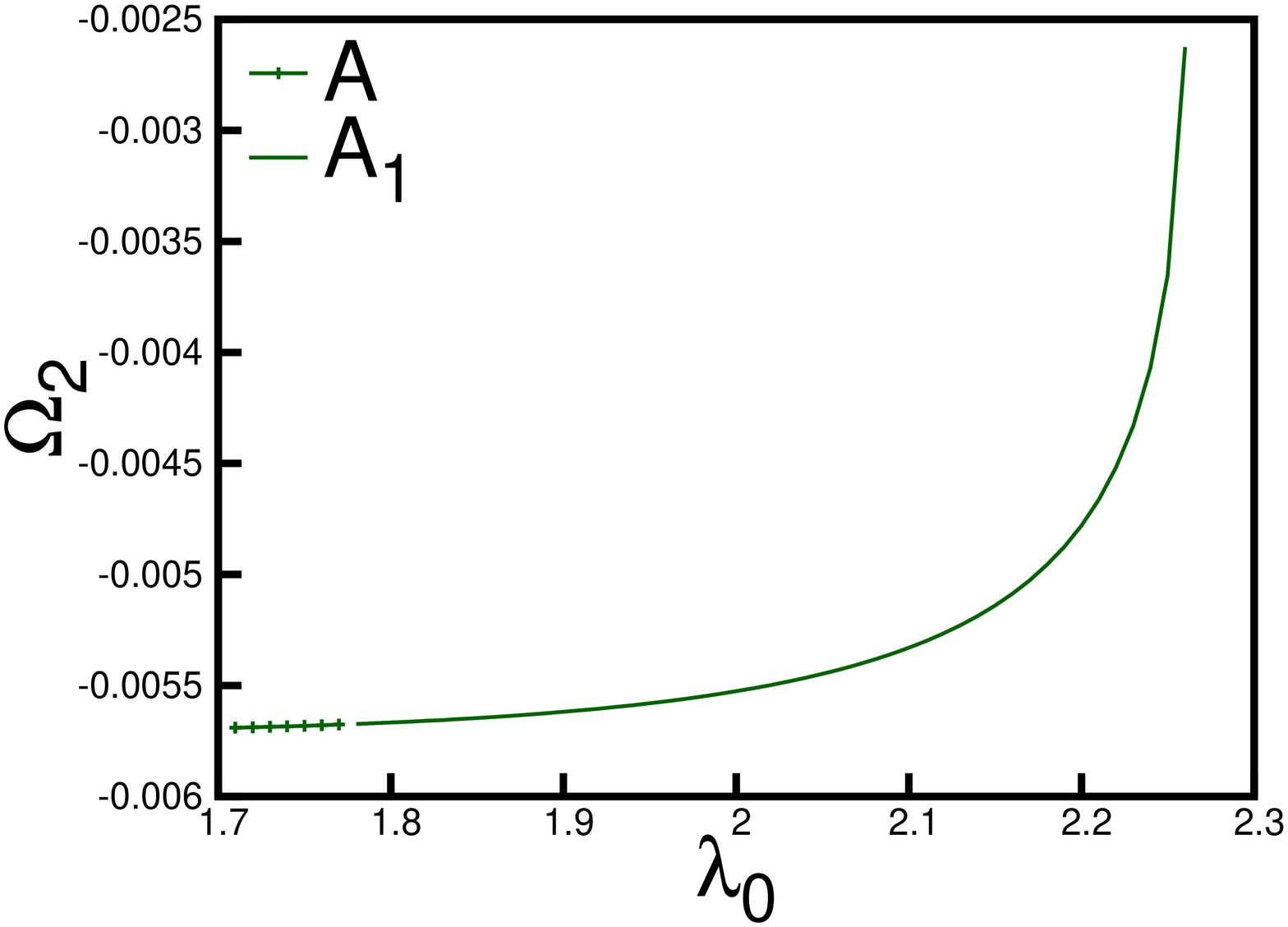,angle=270,width=1.5in}&
\epsfig{file=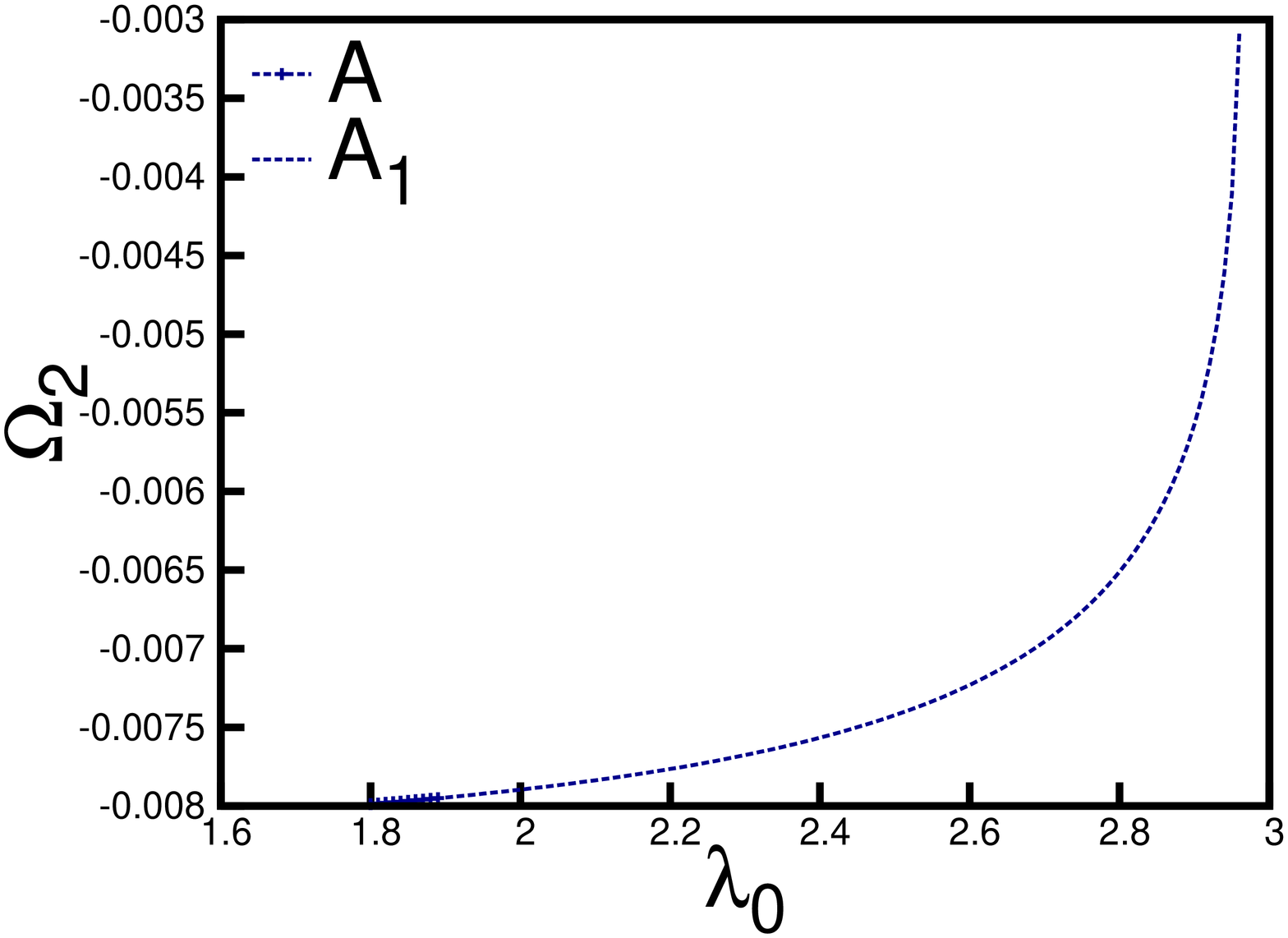,angle=270,width=1.5in}\\
%
\hline
\end{tabular*}
\caption{\small Variation of eigenvalues for the inner, middle and outer critical points (CP) for three different disk models (V, C \& H) under $\alpha$ = 0.0 (inviscid case) for $\gamma=\frac{4}{3}$ and $\cal \dot{M}$ = $2 \times 10^{-5}$. $\rm A$ and $\rm A_1$ indicate multicritical and single critical accretion solutions respectively. The dotted-red lines (dotted-red lines with points) represent multicritical (single critical) accretion solutions for vertical equilibrium geometry (V), the solid-green lines (solid-green lines with points)  represent multicritical (single critical) accretion solutions for conical geometry (C), and the
dashed-blue lines (dashed-blue lines with points) represent multicritical (single critical) accretion solutions for the constant-height disk geometry (H).}
\label{Fig:o2_a_0.0}
\end{figure*}

\begin{figure*}[h!]
\centering
\begin{tabular*}{1.0\linewidth}{@{\extracolsep{\fill}}|cccc|}
\hline
{\bf CP} for $\alpha=0.01$ & {\bf V} & {\bf C} & {\bf H} \\
%
\hline
&
\epsfig{file=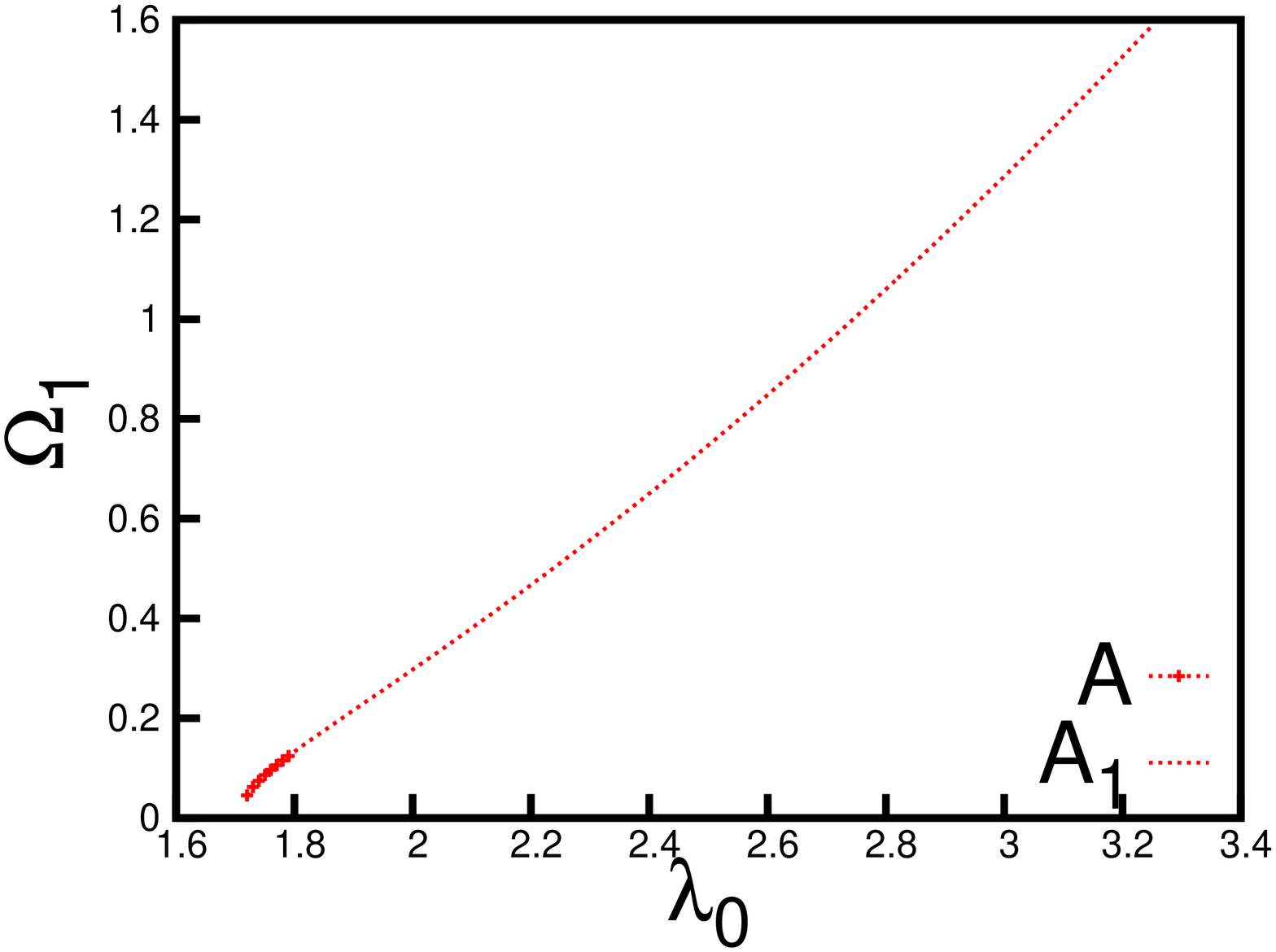,angle=270,width=1.5in}&
\epsfig{file=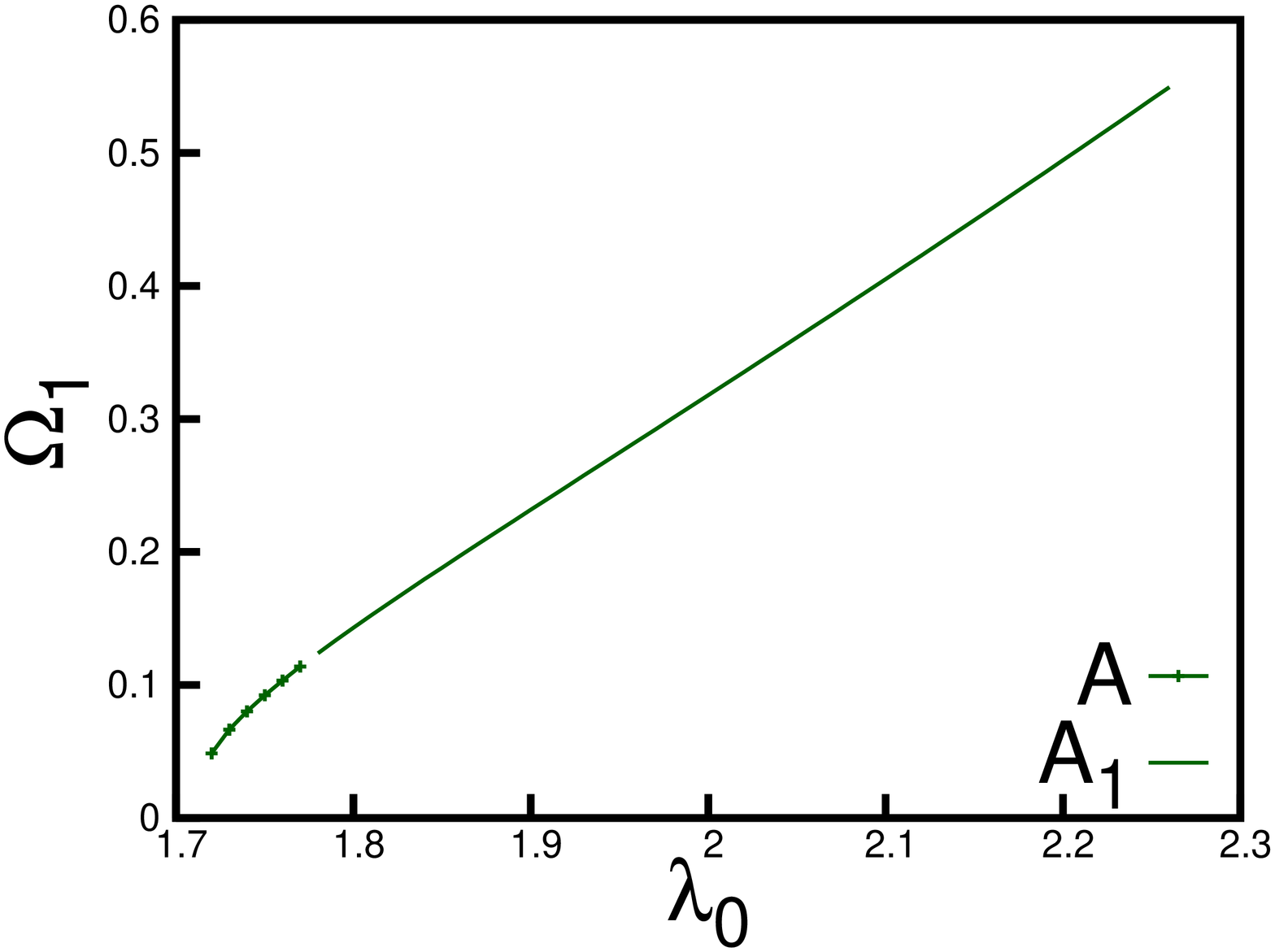,angle=270,width=1.5in}&
\epsfig{file=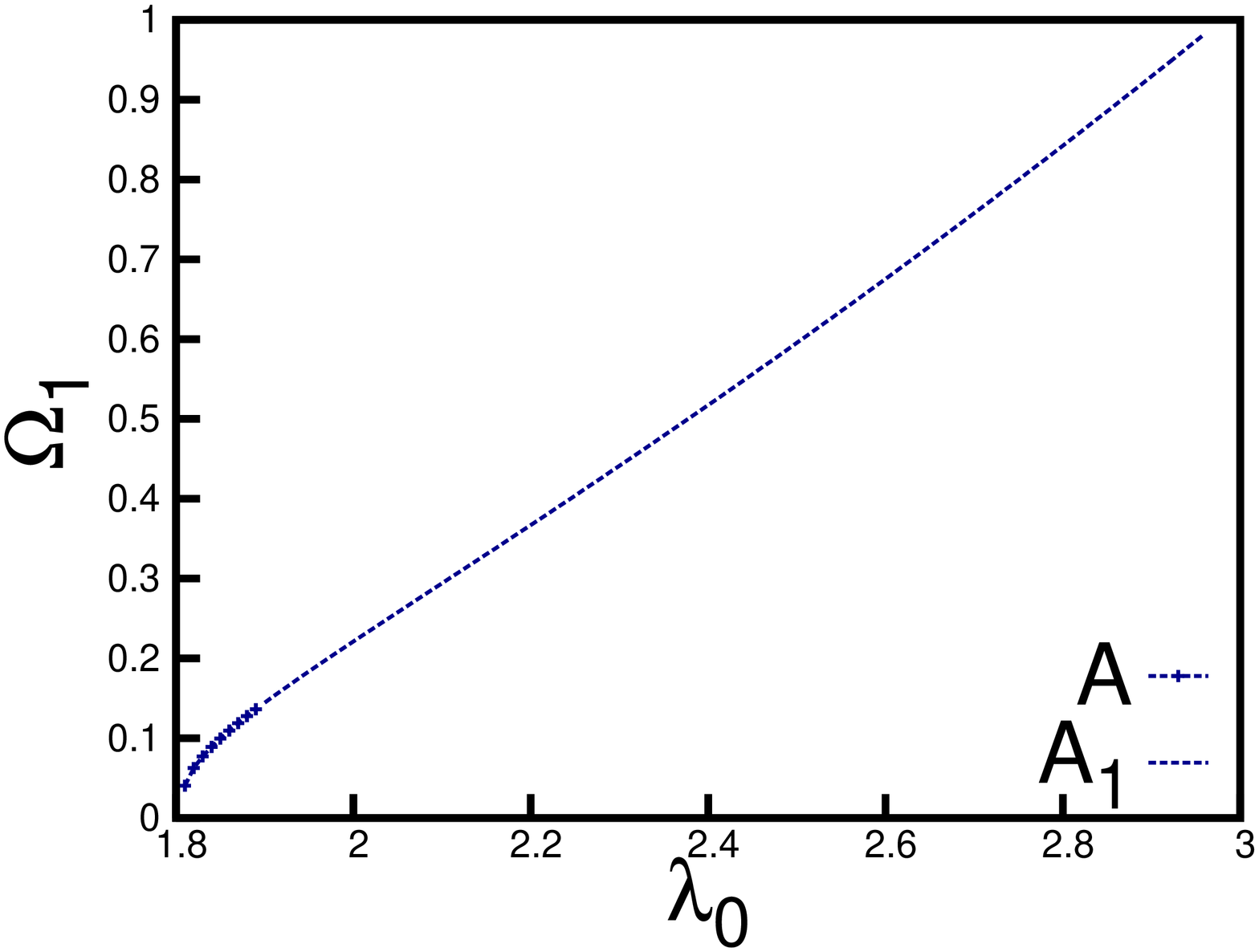,angle=270,width=1.5in}\\
%
 Inner&
\epsfig{file=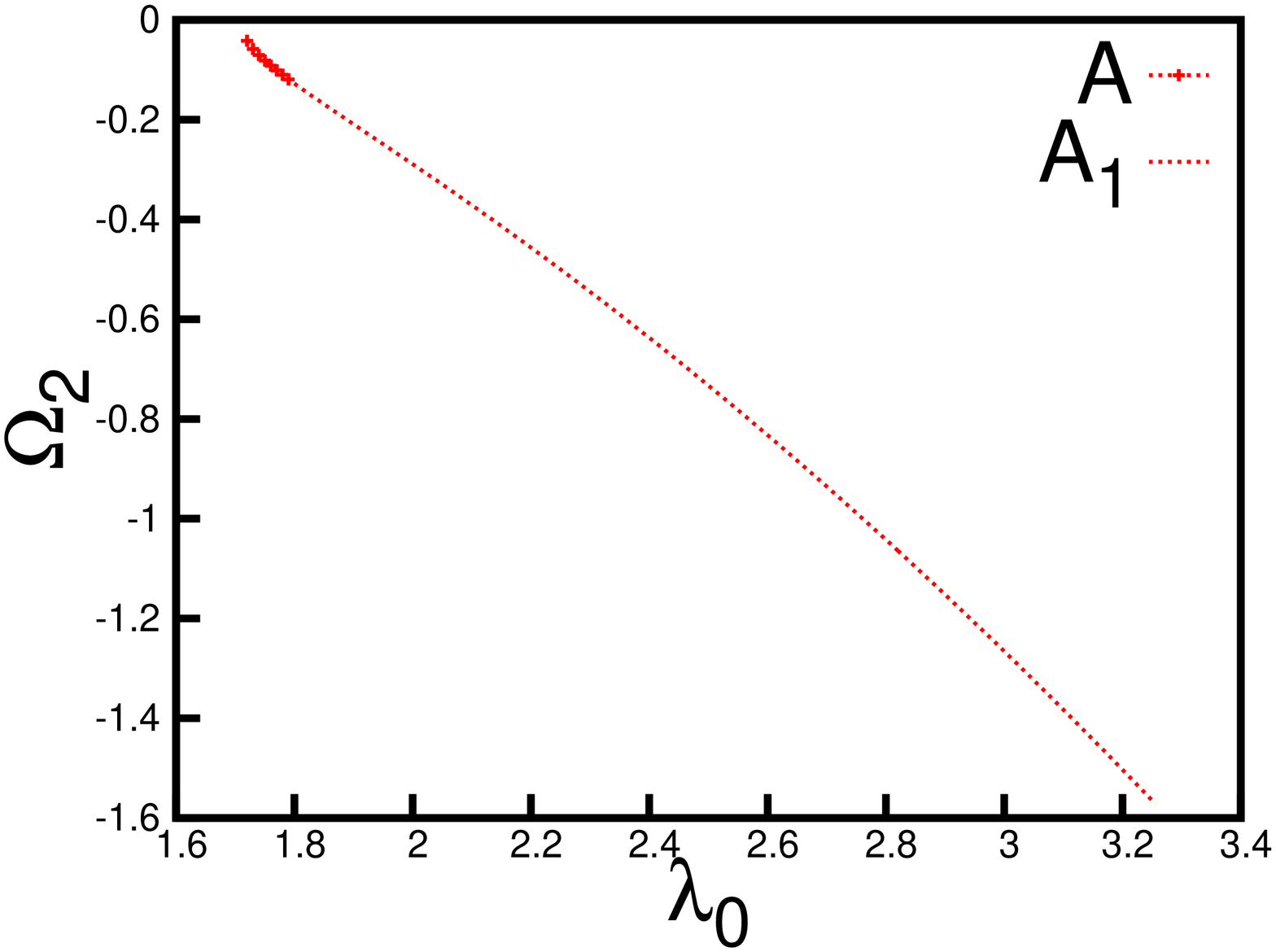,angle=270,width=1.5in}&
\epsfig{file=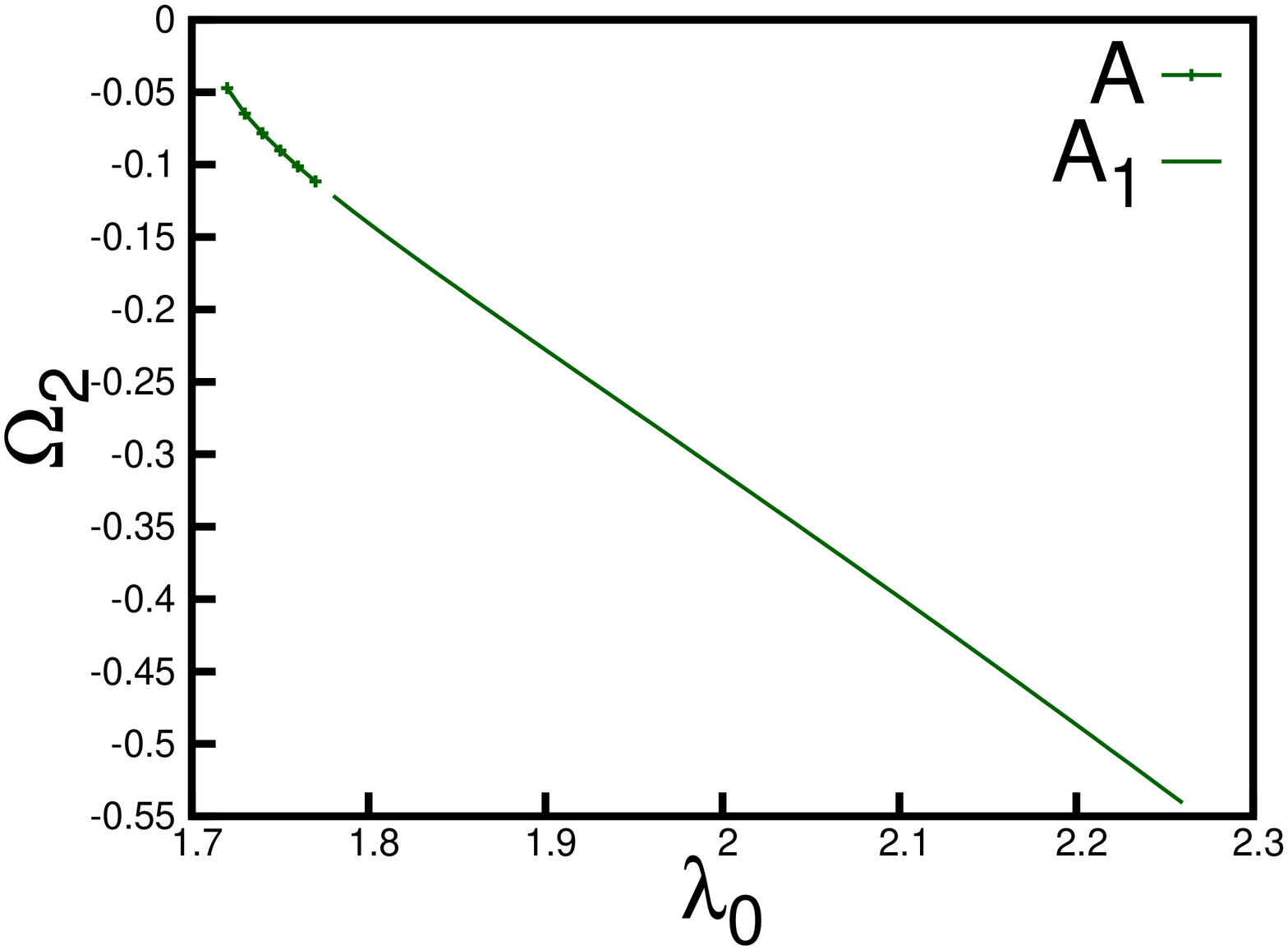,angle=270,width=1.5in}&
\epsfig{file=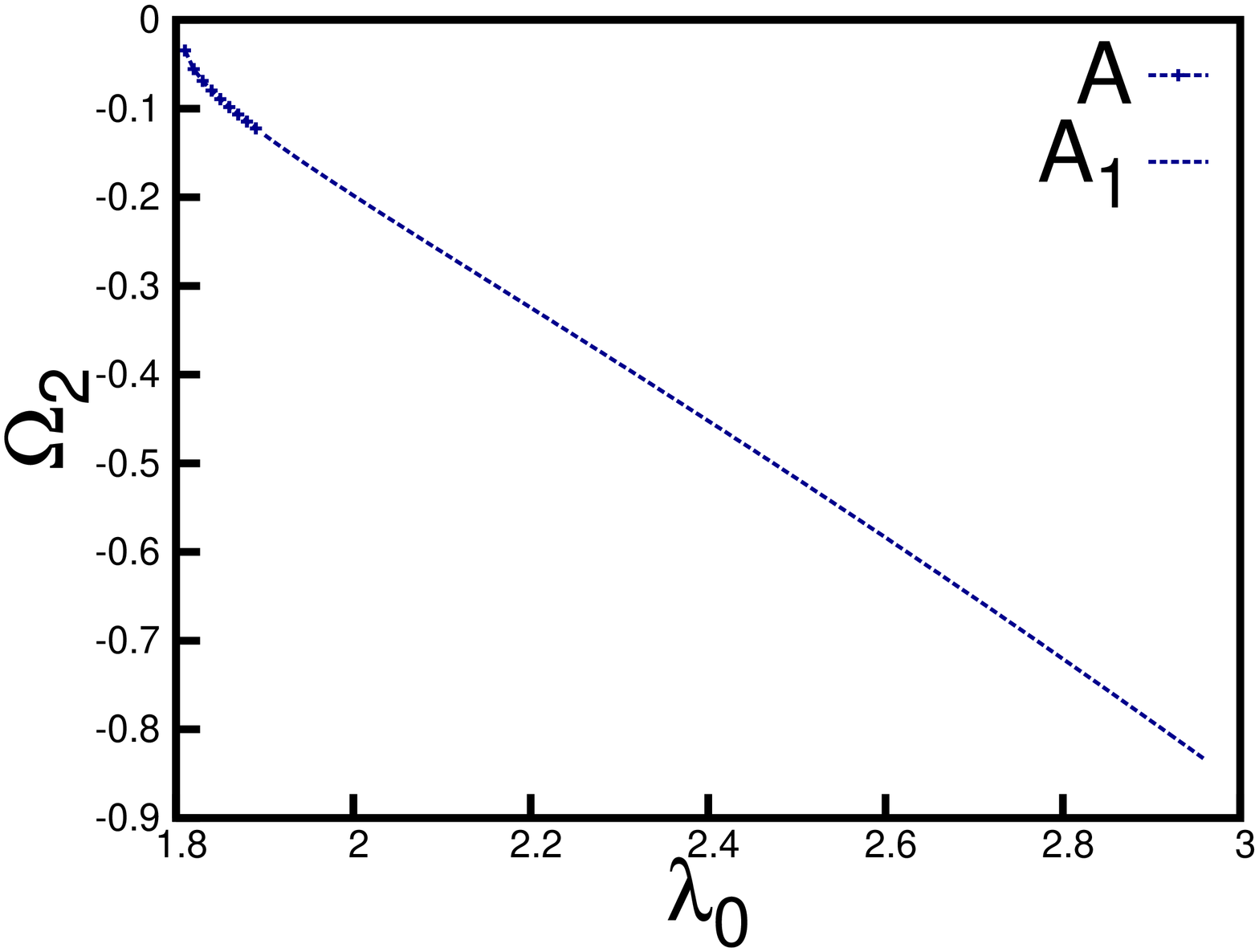,angle=270,width=1.5in}\\
\hline
&
\epsfig{file=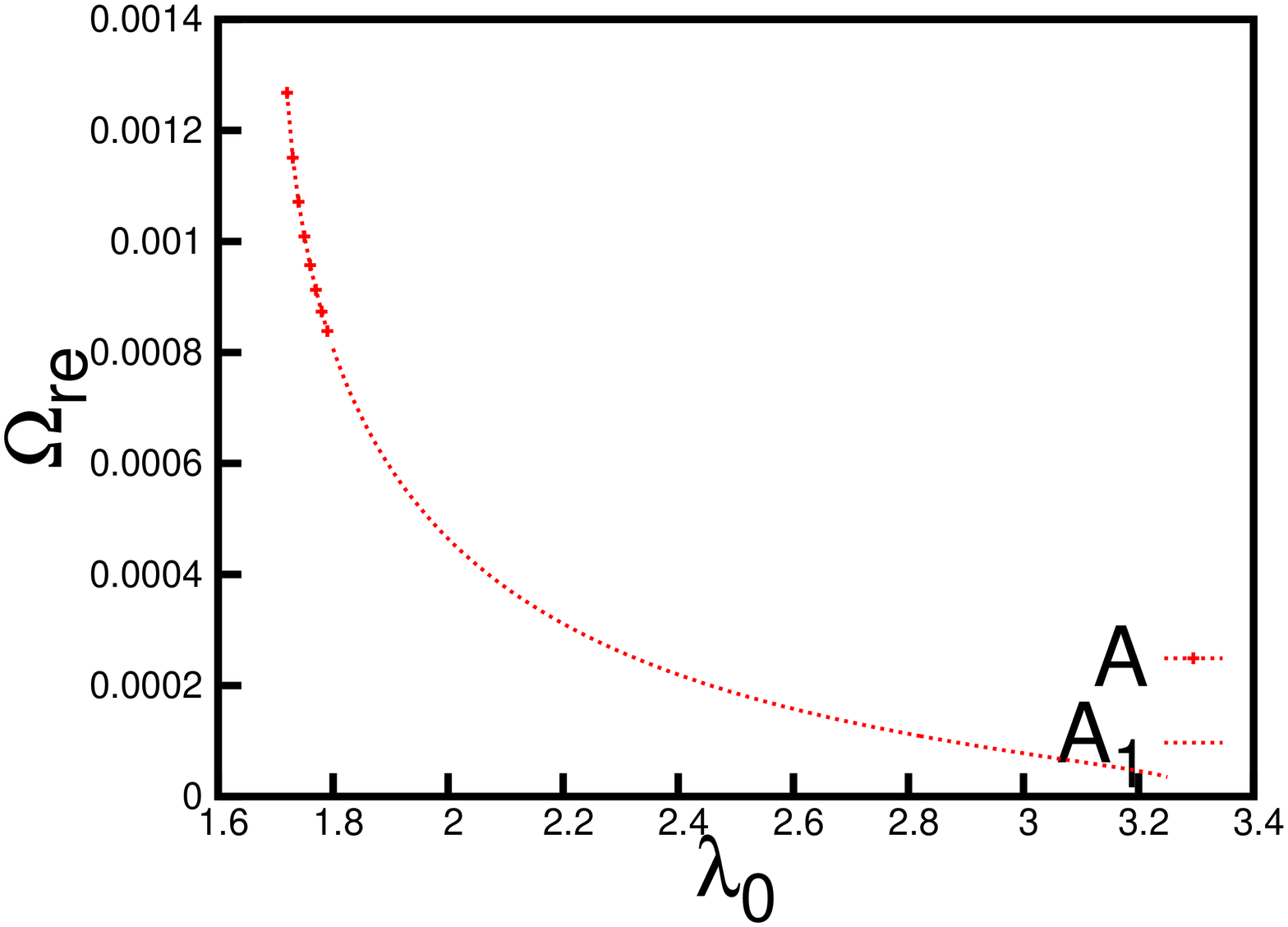,angle=270,width=1.5in}&
\epsfig{file=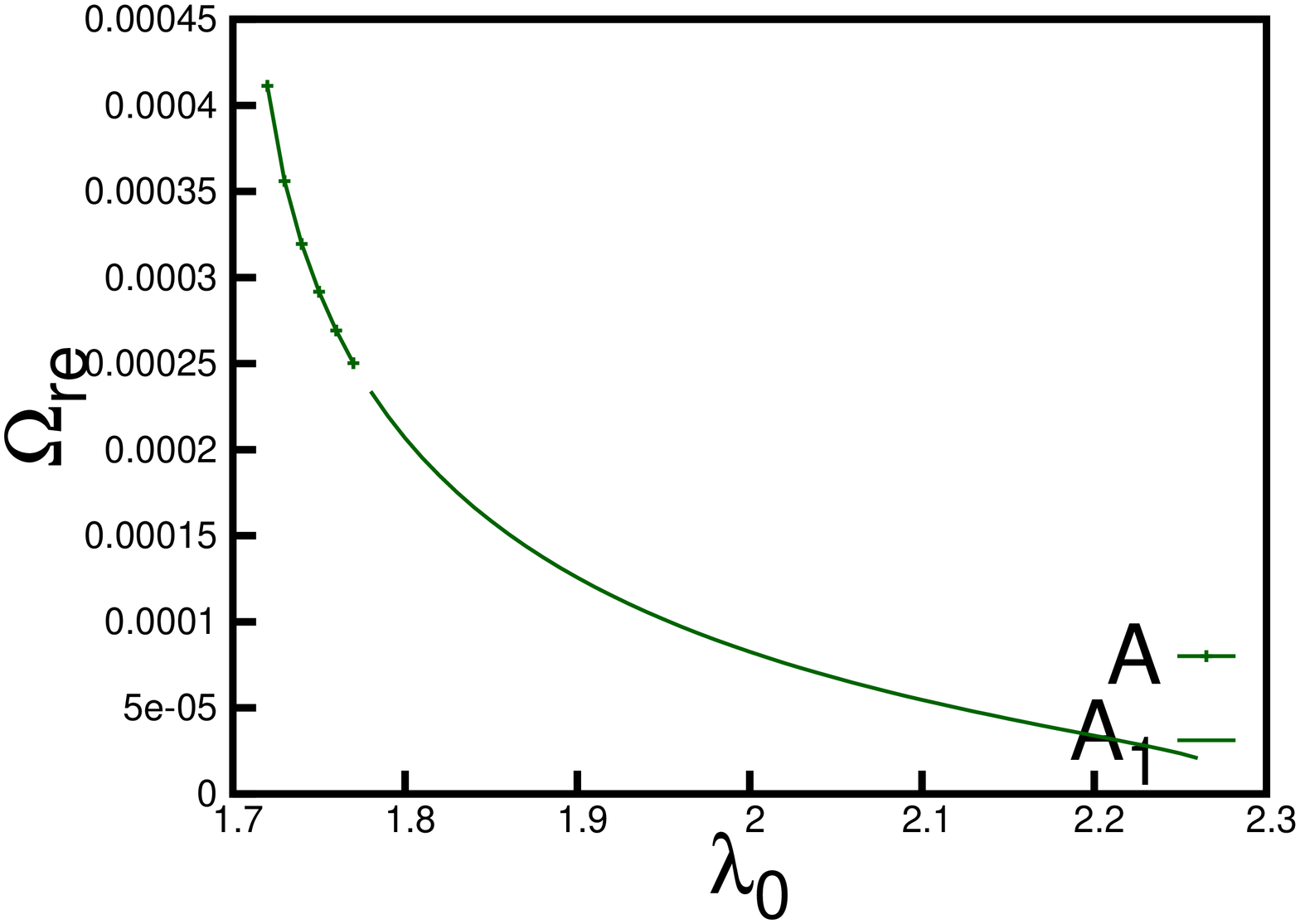,angle=270,width=1.5in}&
\epsfig{file=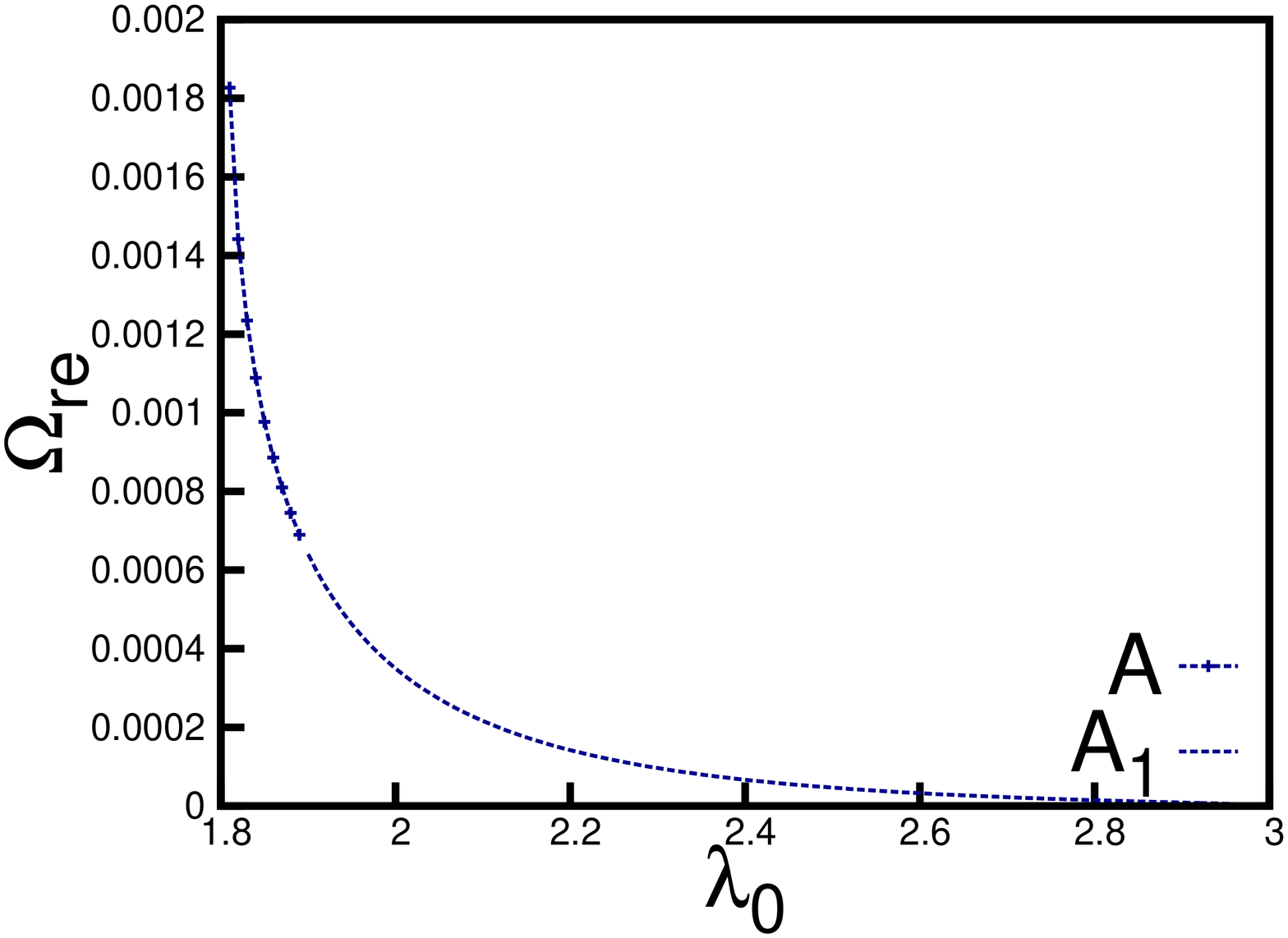,angle=270,width=1.5in}\\
Middle &
\epsfig{file=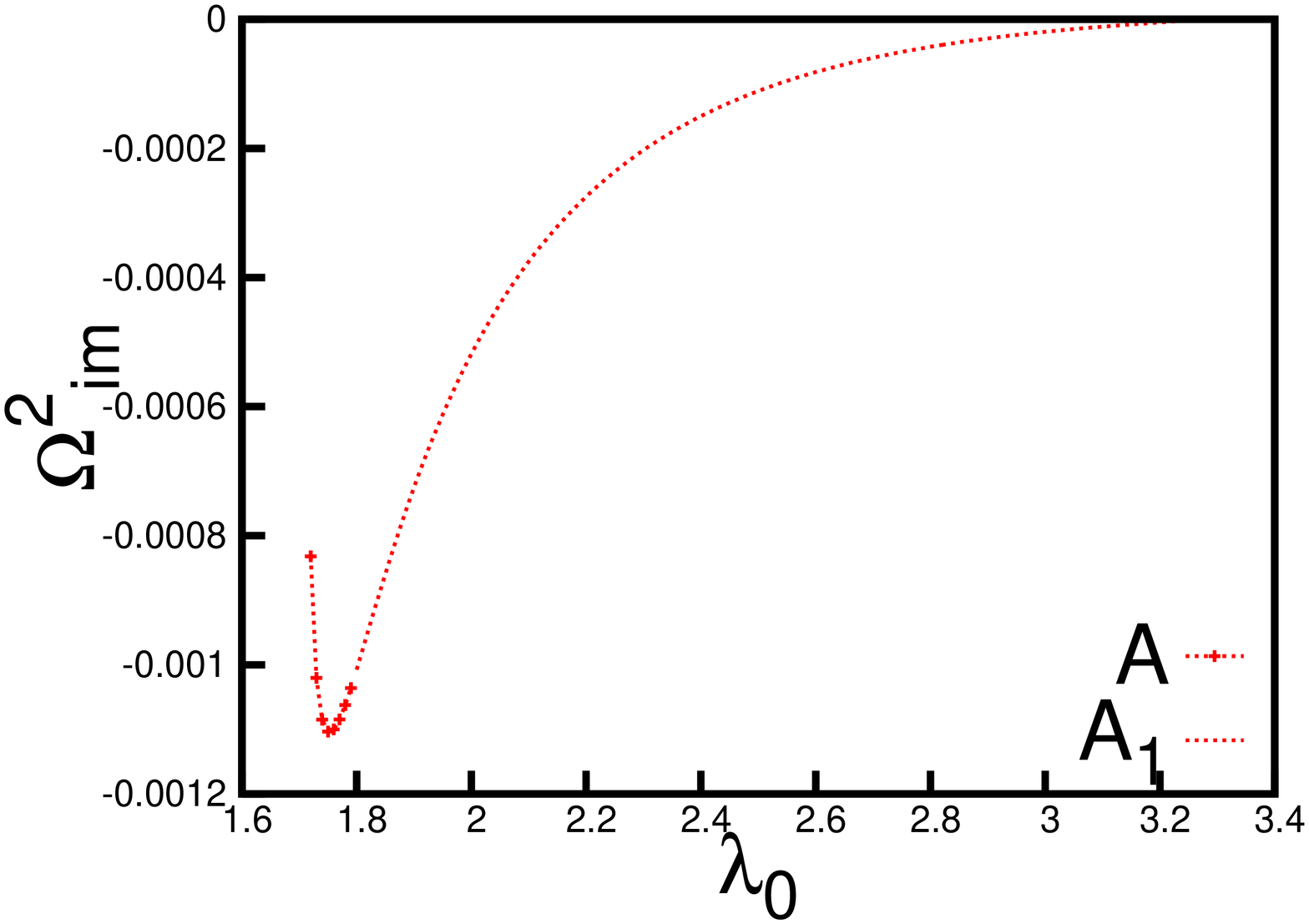,angle=270,width=1.5in}&
\epsfig{file=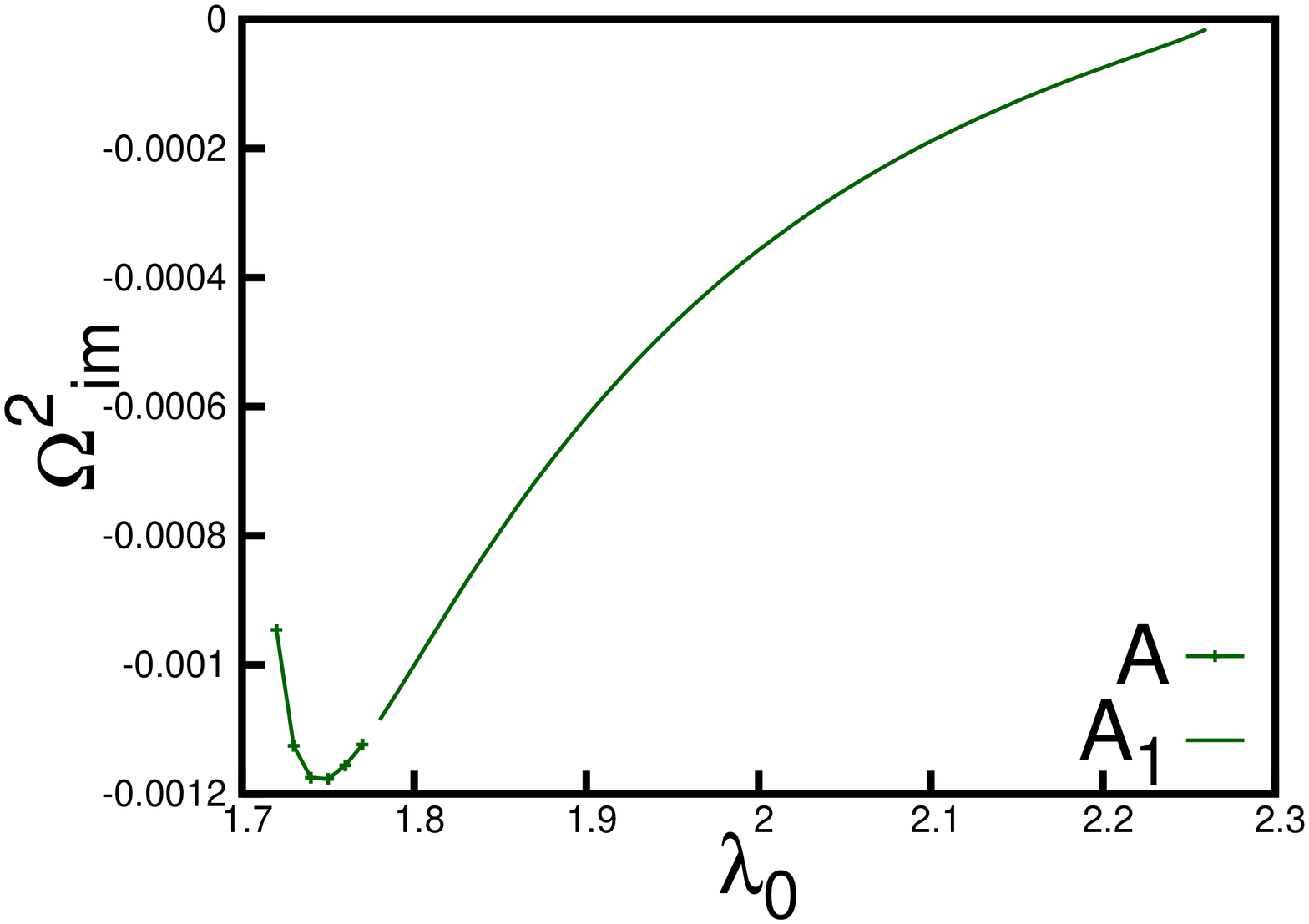,angle=270,width=1.5in}&
\epsfig{file=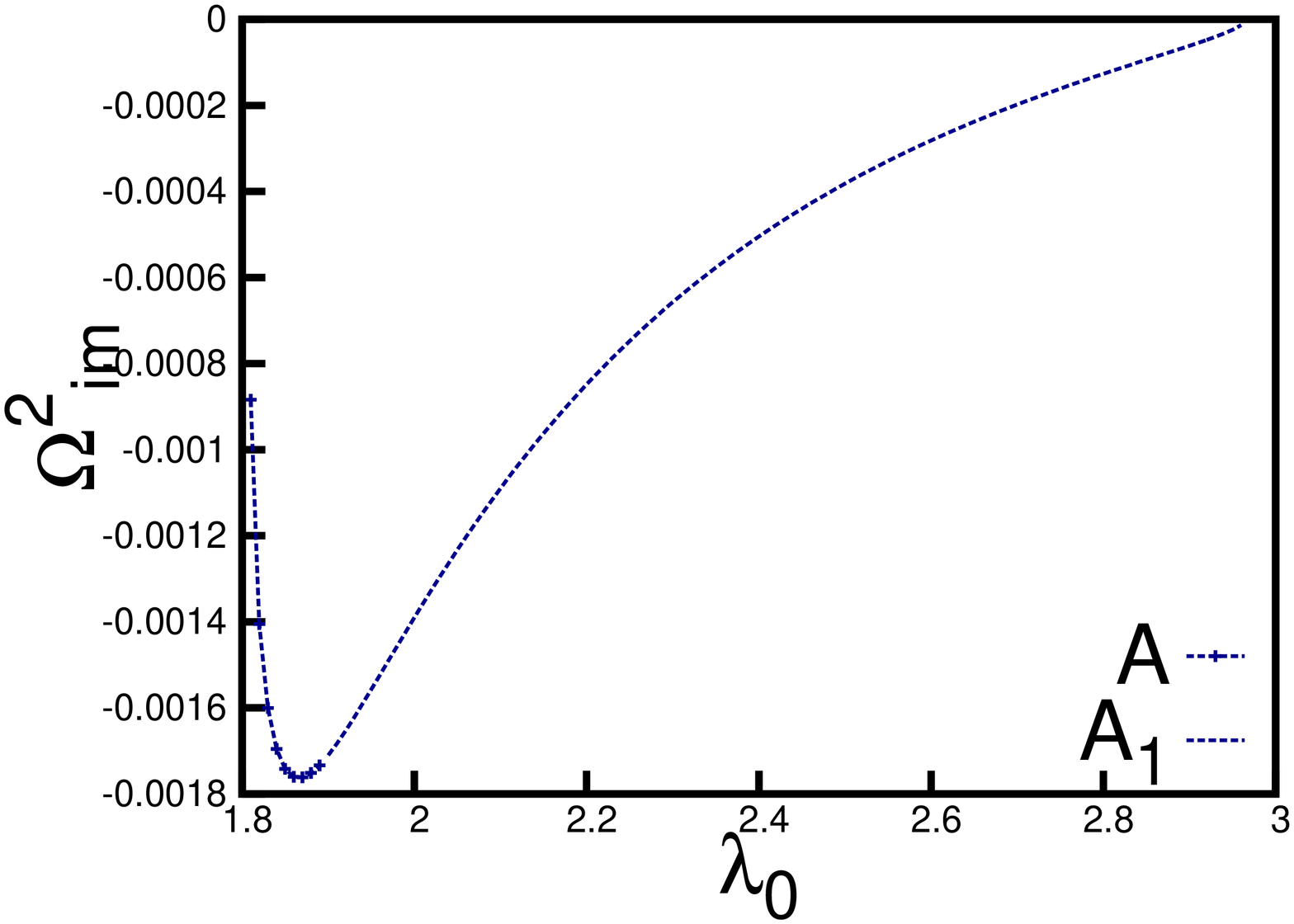,angle=270,width=1.5in}\\
%
\hline
&
\epsfig{file=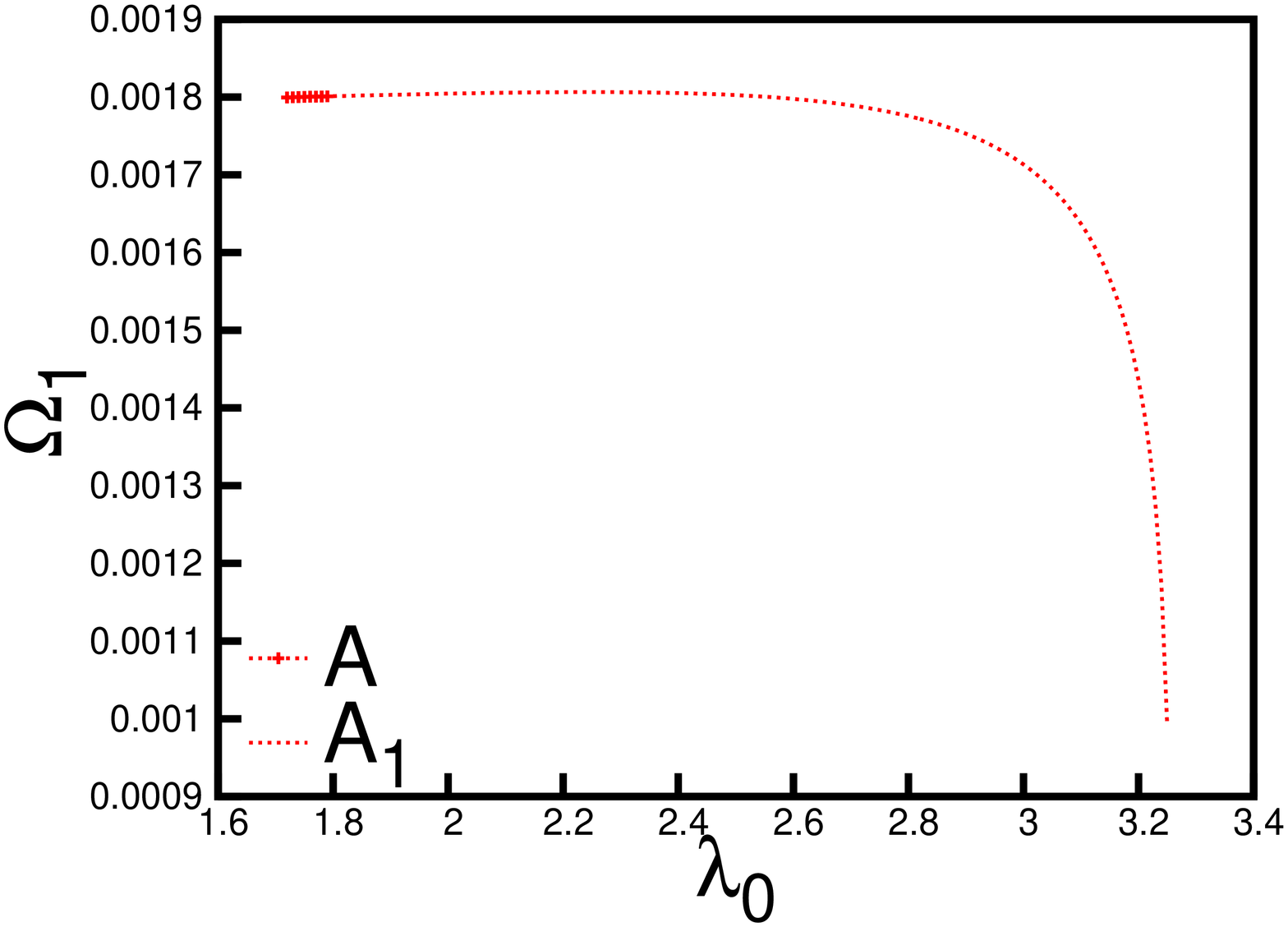,angle=270,width=1.5in}&
\epsfig{file=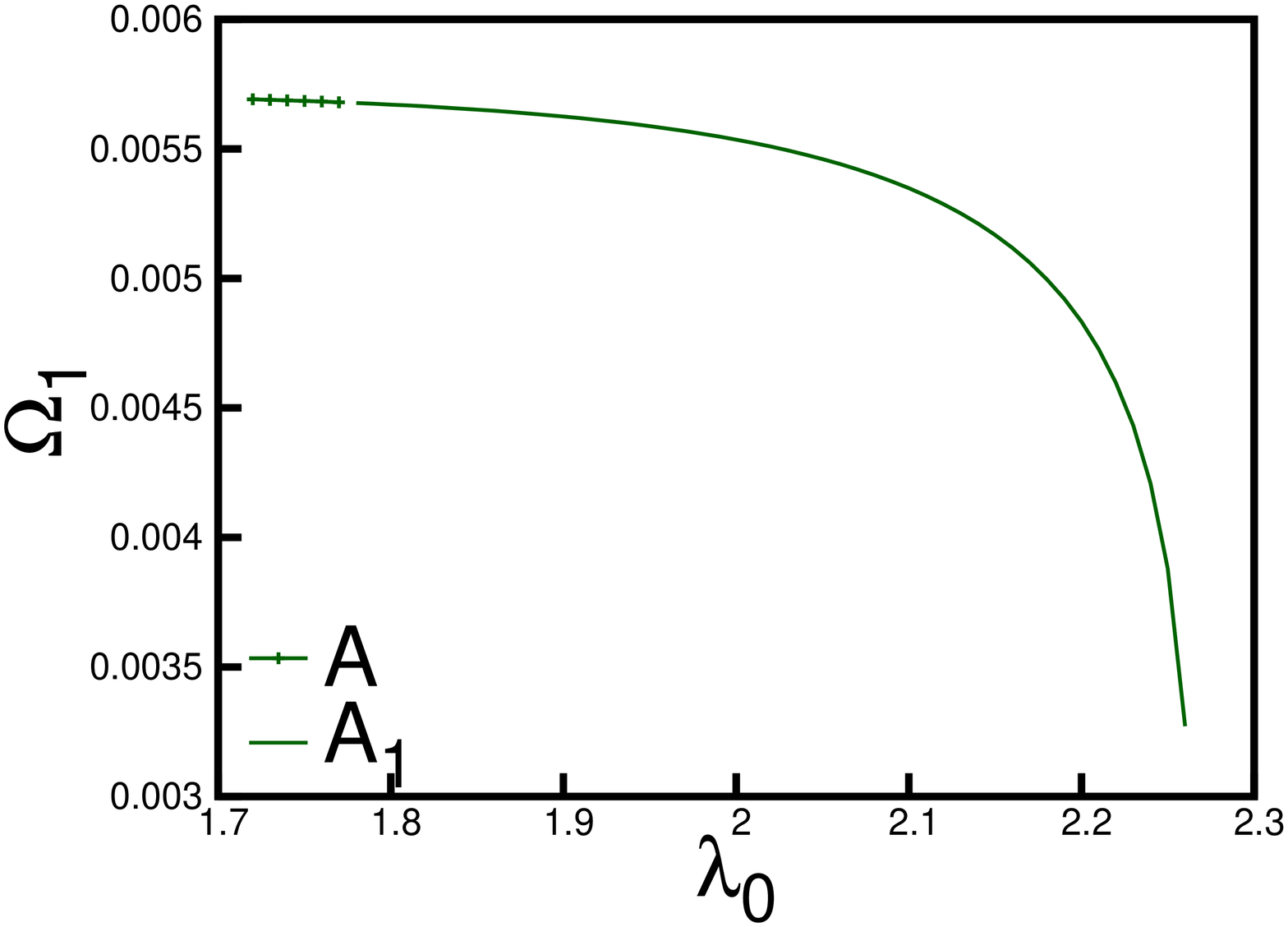,angle=270,width=1.5in}&
\epsfig{file=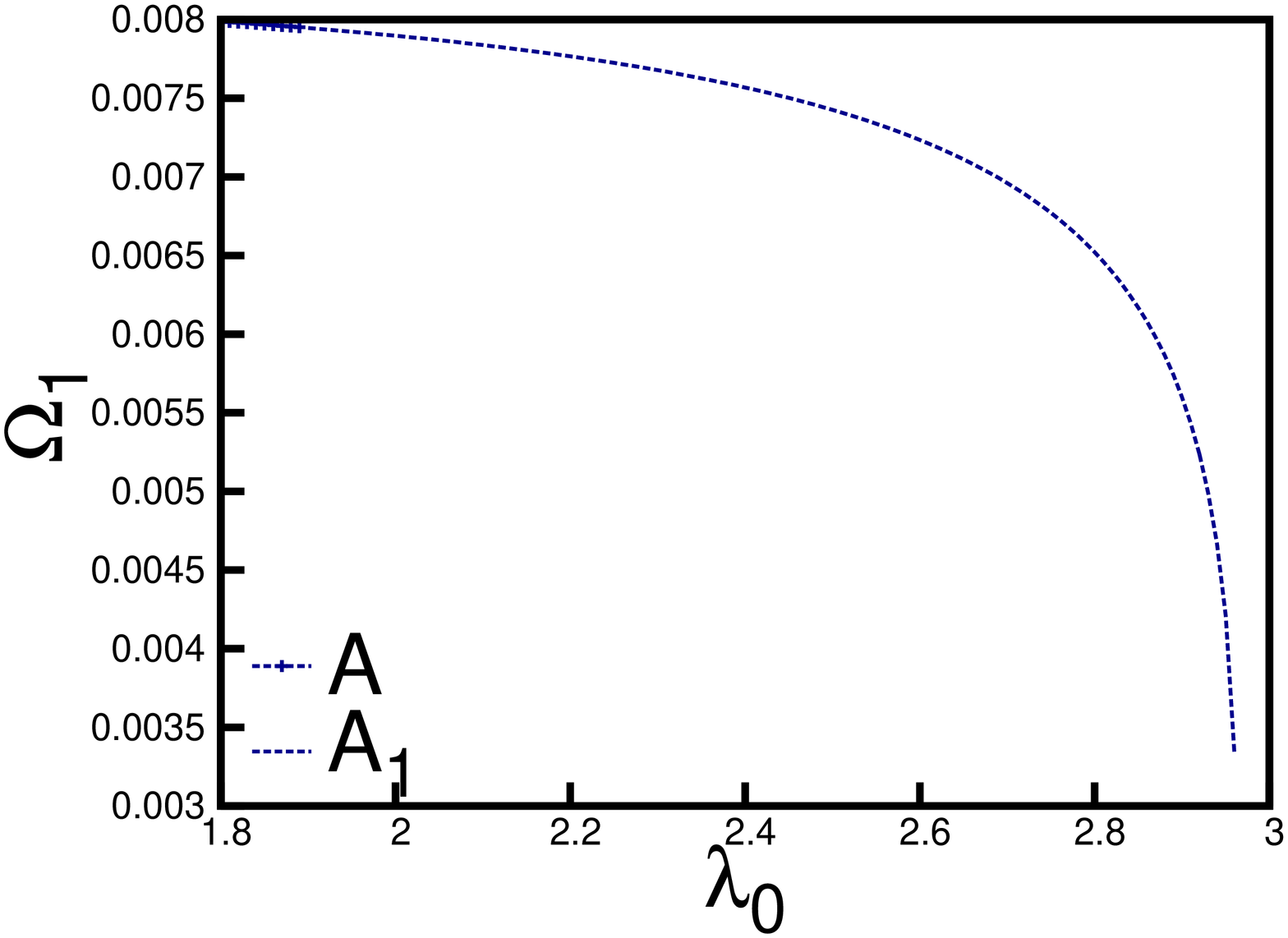,angle=270,width=1.5in}\\
Outer &
\epsfig{file=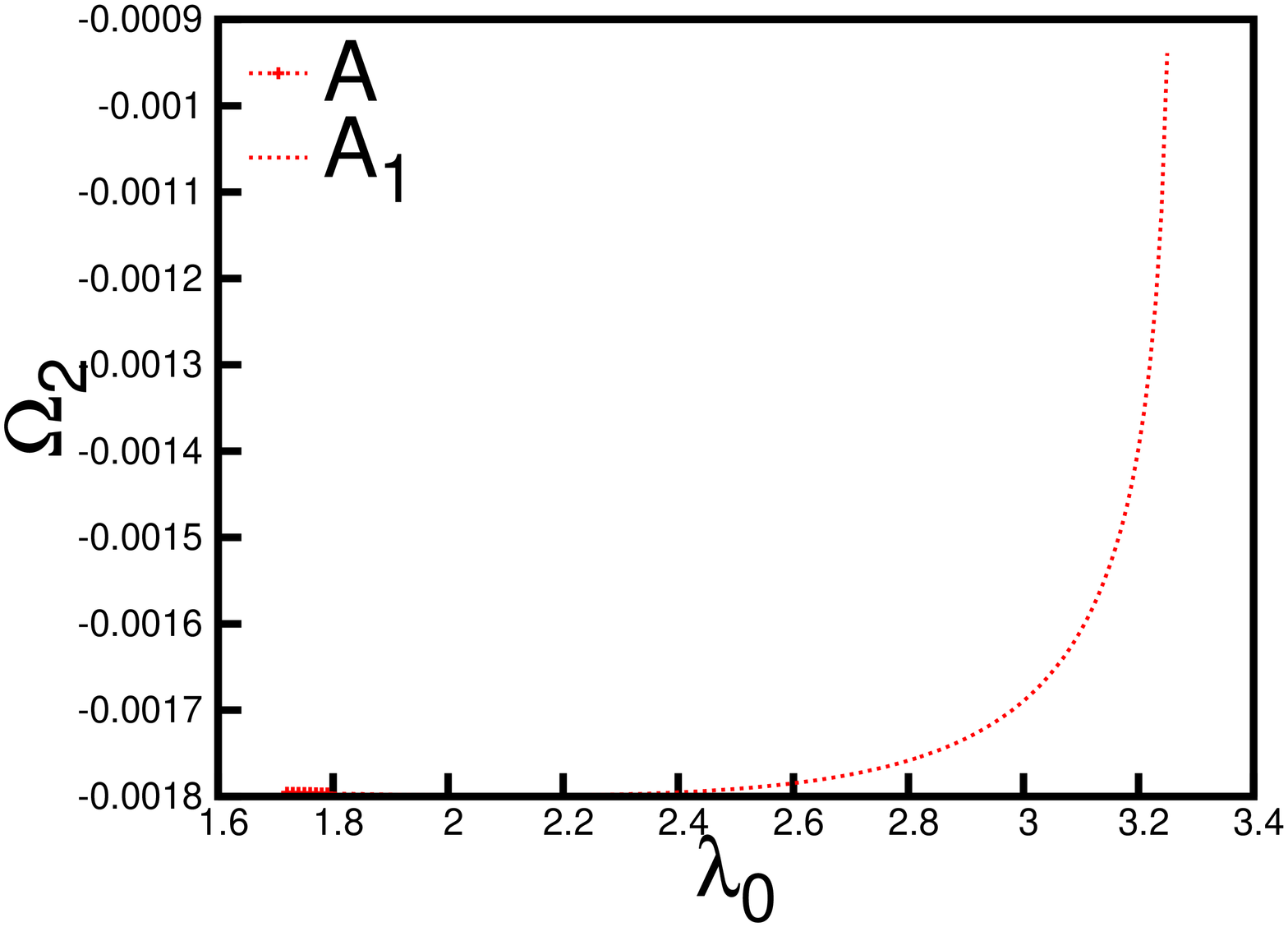,angle=270,width=1.5in}&
\epsfig{file=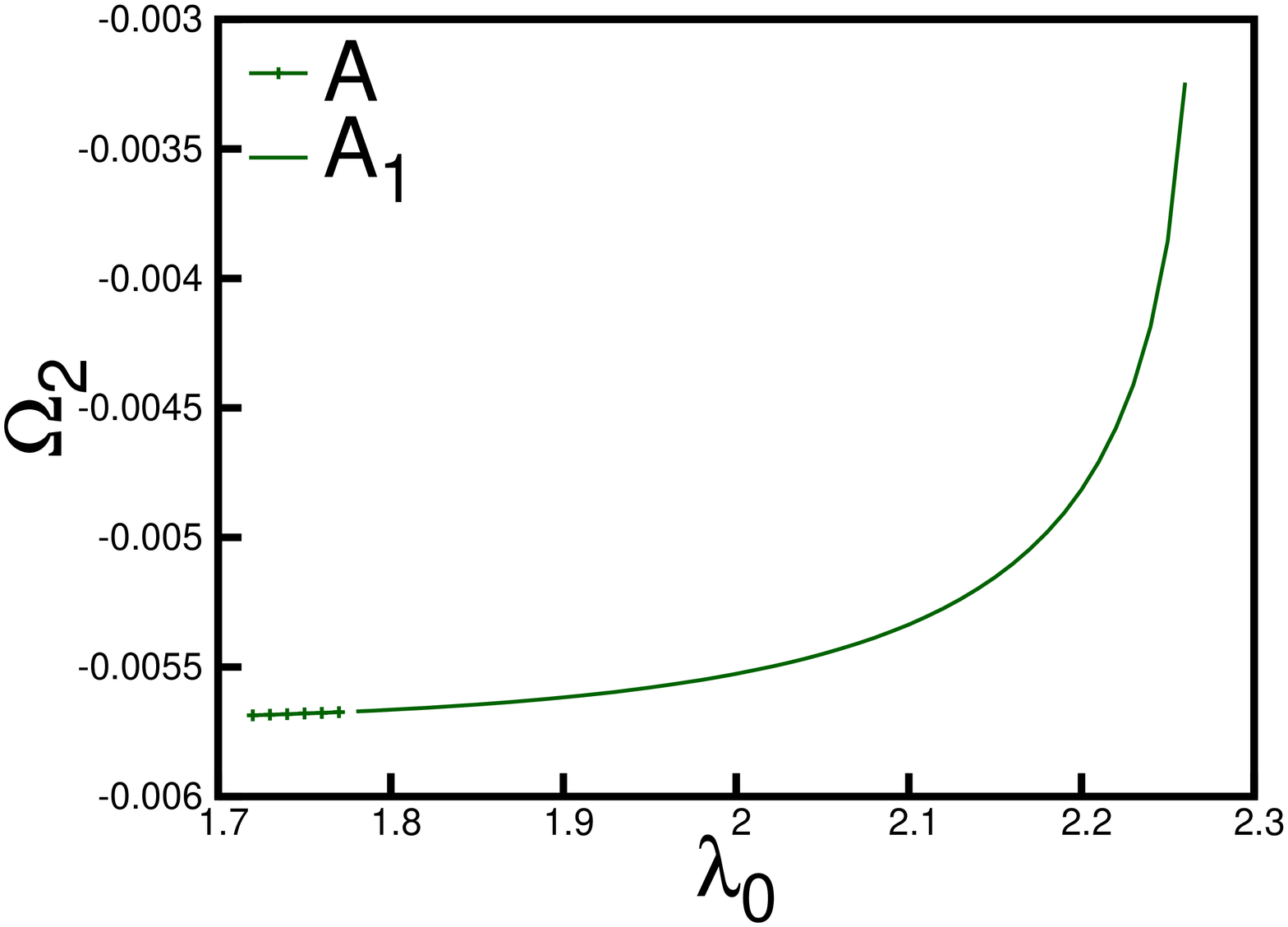,angle=270,width=1.5in}&
\epsfig{file=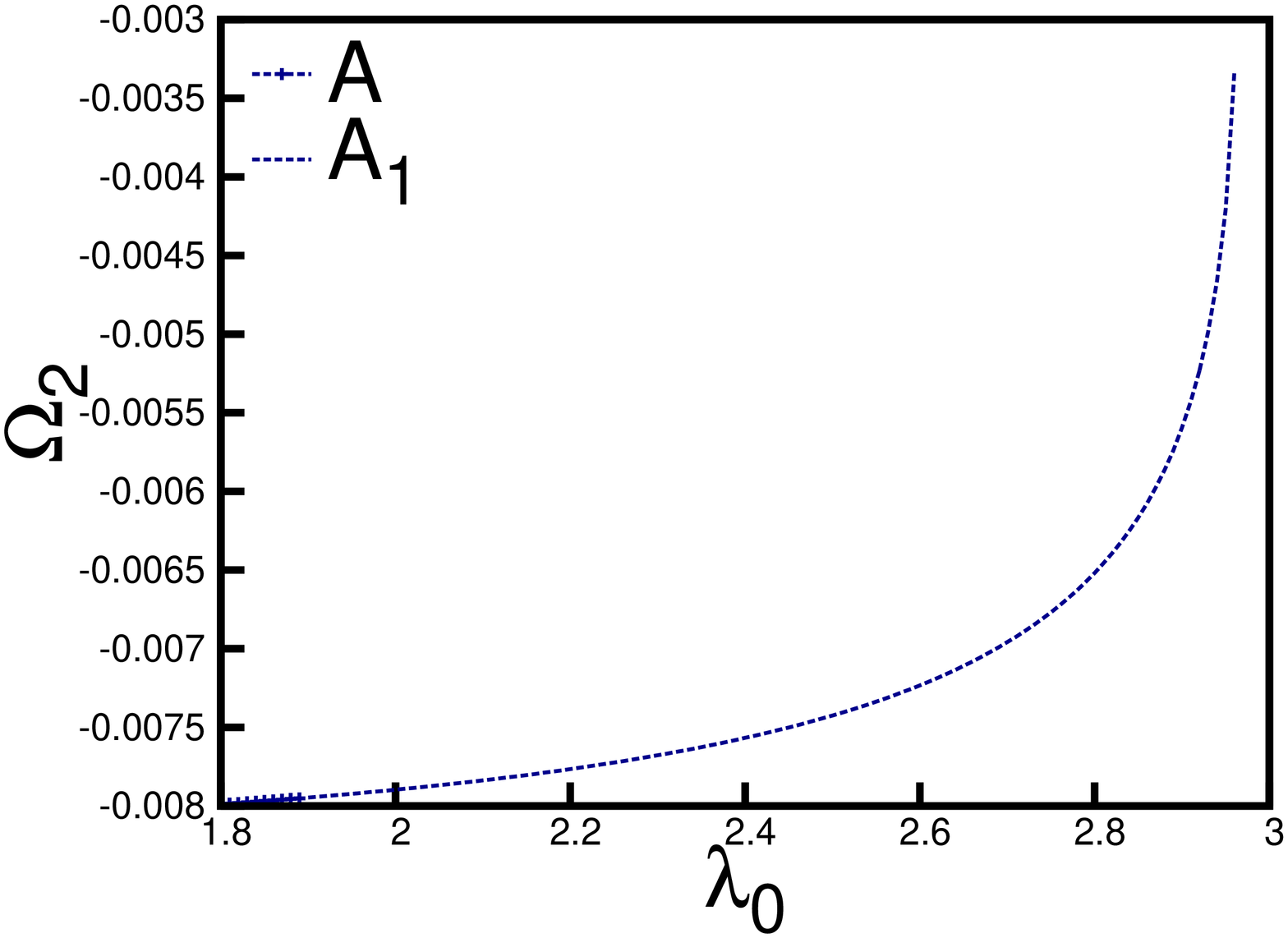,angle=270,width=1.5in}\\
%
\hline
\end{tabular*}
\caption{\small Variation of eigenvalues for the inner, middle and outer critical points (CP) for three different disk models (V, C \& H) under $\alpha$ = 0.01 for $\gamma=\frac{4}{3}$ and $\cal \dot{M}$ = $2 \times 10^{-5}$. $\rm A$ and $\rm A_1$ indicate multicritical (single critical) accretion solutions respectively. The dotted-red lines (dotted-red lines with points) represent multicritical (single critical) accretion solutions for vertical equilibrium geometry (V), the solid-green lines (solid-green lines with points)  represent multicritical (single critical) accretion solutions for conical geometry (C), and the
dashed-blue lines (dashed-blue lines with points) represent multicritical (single critical) accretion solutions for the constant-height disk geometry (H).}
\label{Fig:o2_a_0.01}
\end{figure*}

\begin{figure*}[h!]
\centering
\begin{tabular*}{1.0\linewidth}{@{\extracolsep{\fill}}|cccc|}
\hline
{\bf CP} for $\alpha=0.1$ & {\bf V} & {\bf C} & {\bf H} \\
%
\hline
&
\epsfig{file=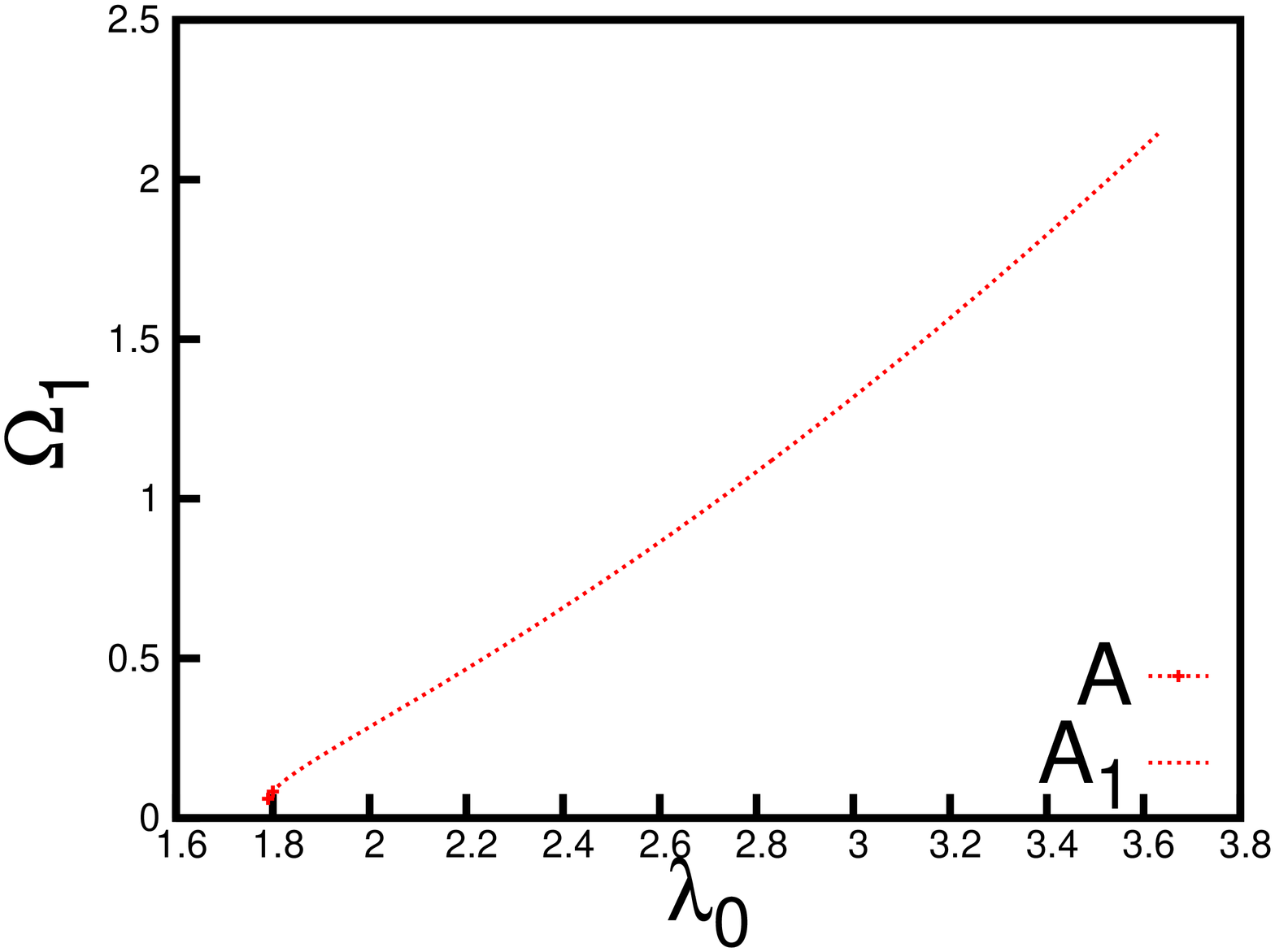,angle=270,width=1.5in}&
\epsfig{file=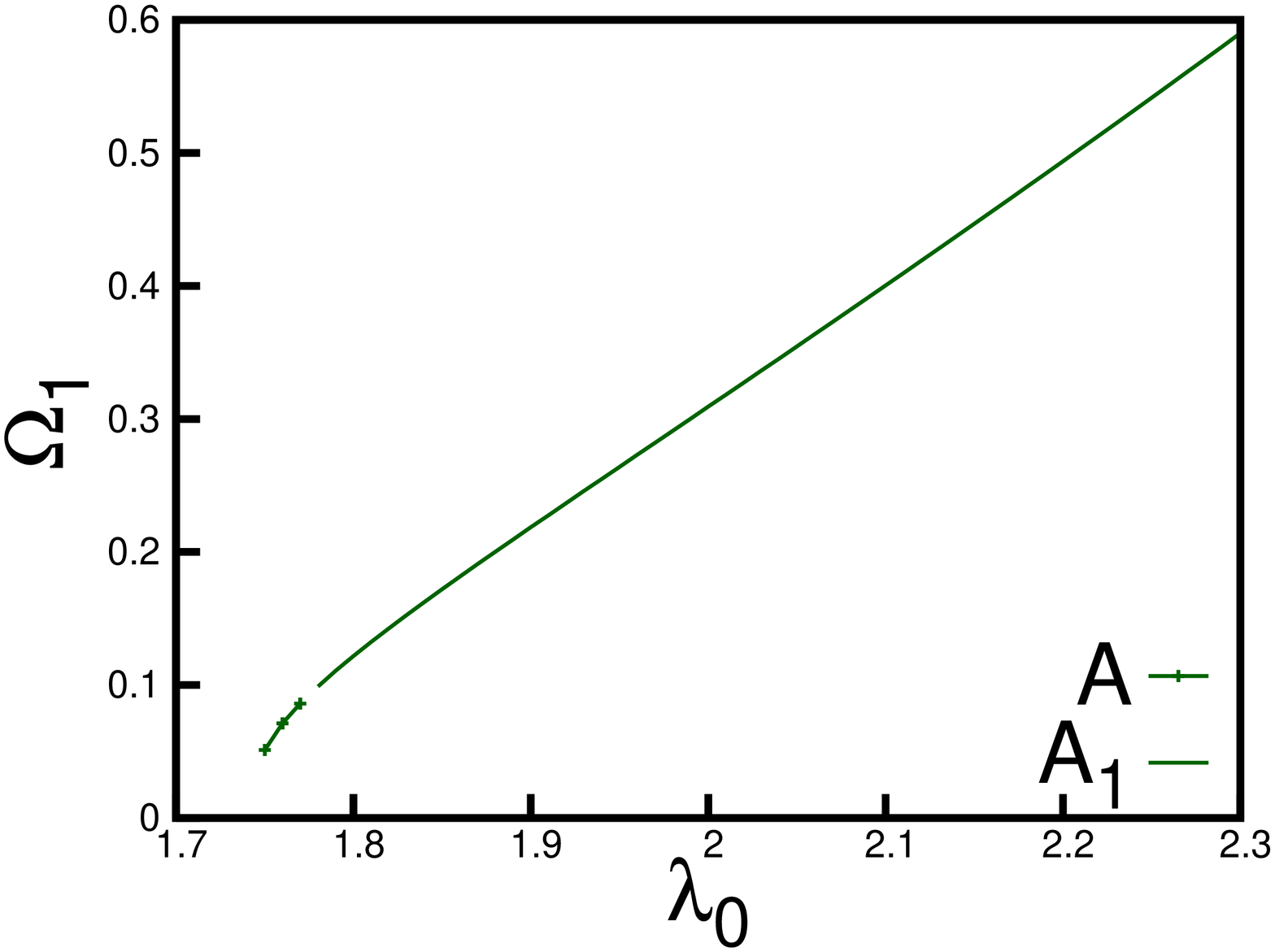,angle=270,width=1.5in}&
\epsfig{file=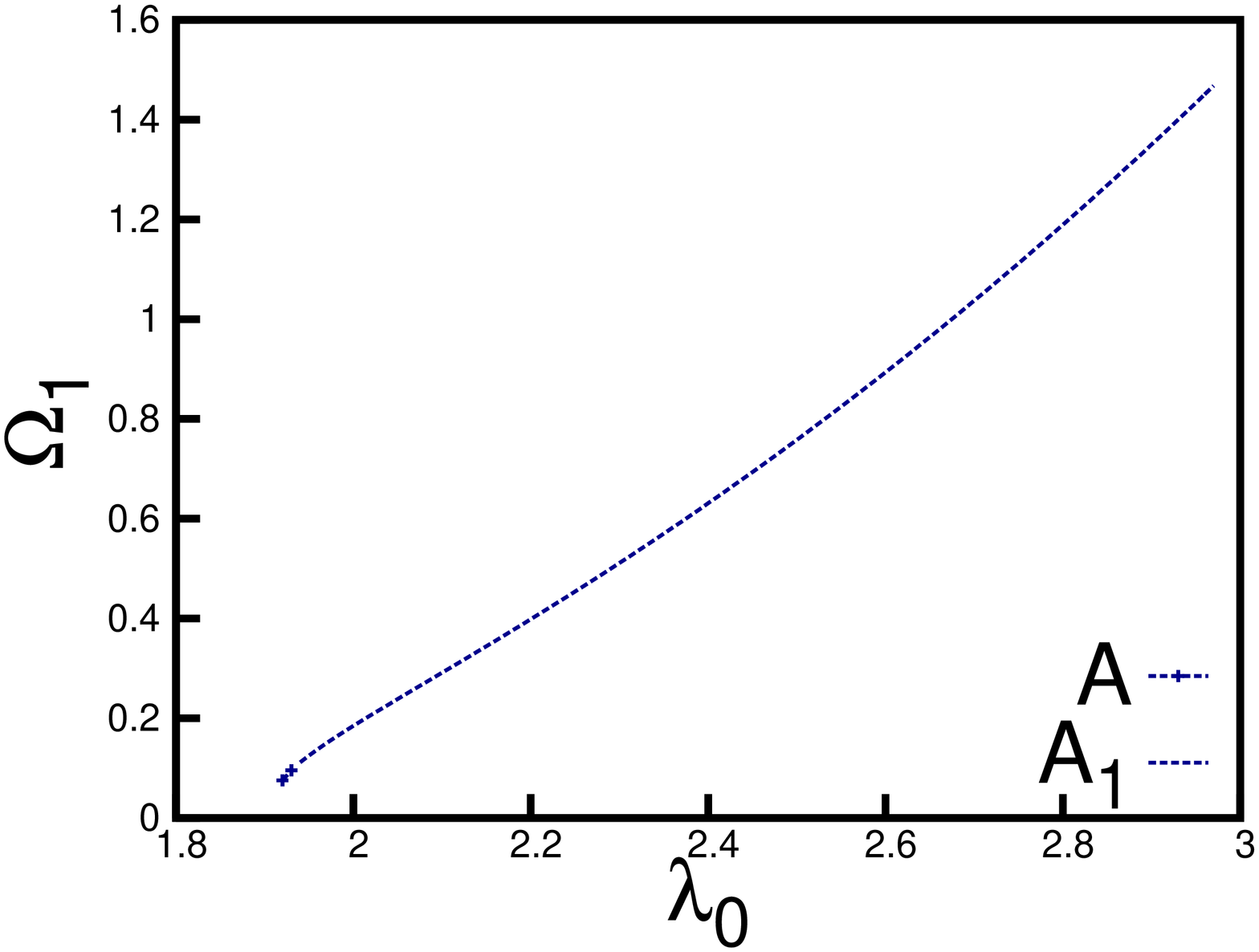,angle=270,width=1.5in}\\
%
 Inner&
\epsfig{file=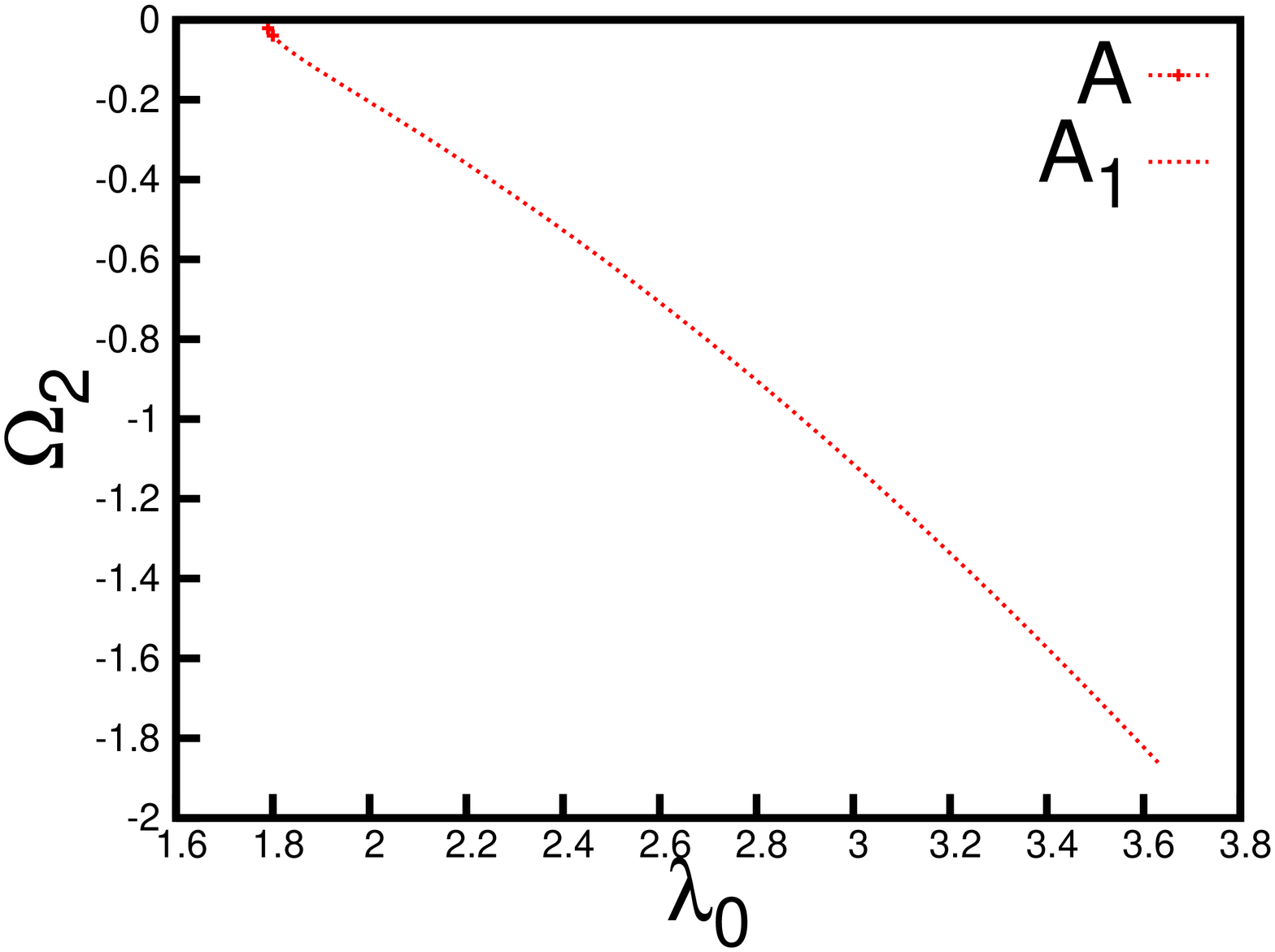,angle=270,width=1.5in}&
\epsfig{file=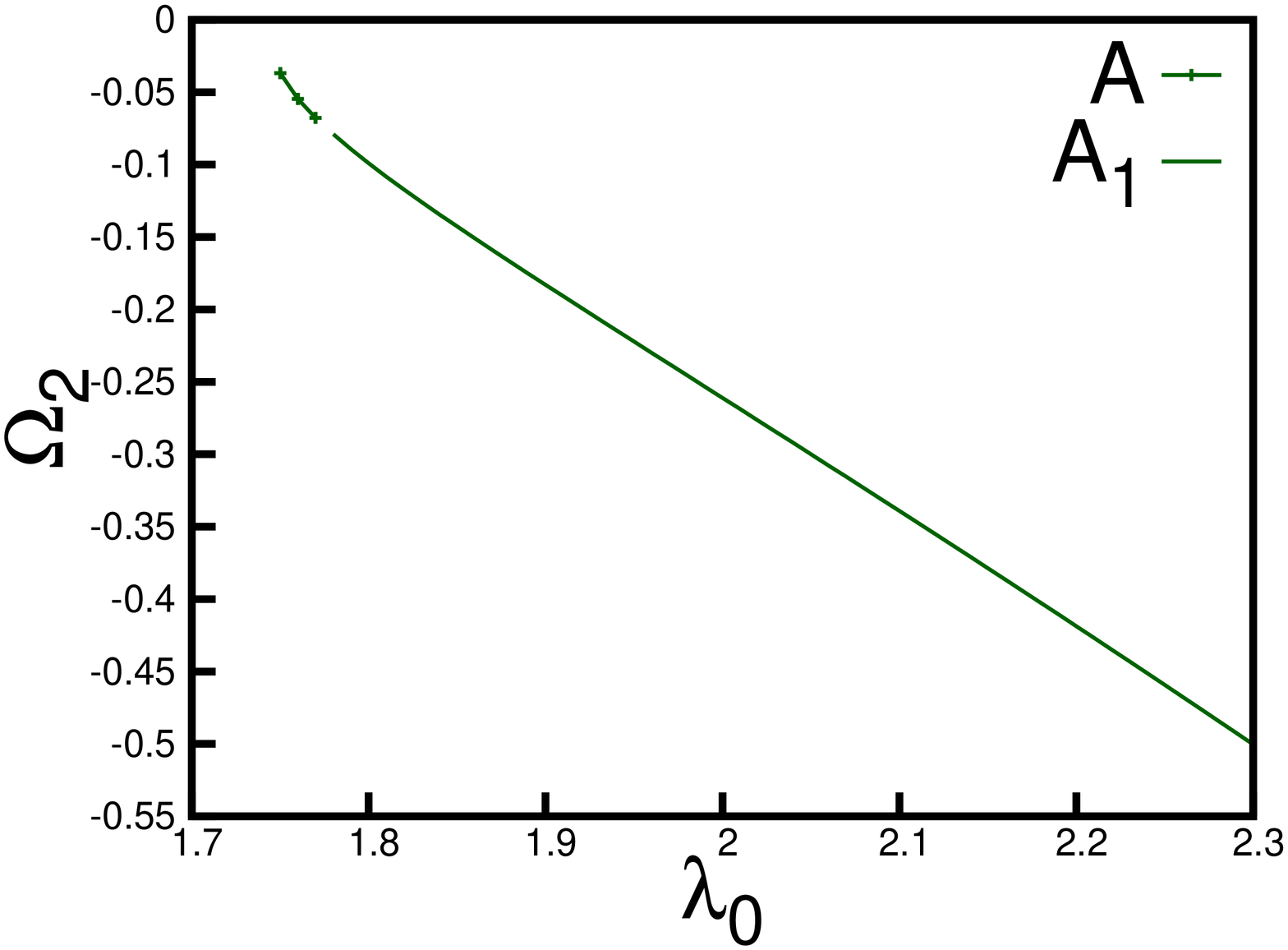,angle=270,width=1.5in}&
\epsfig{file=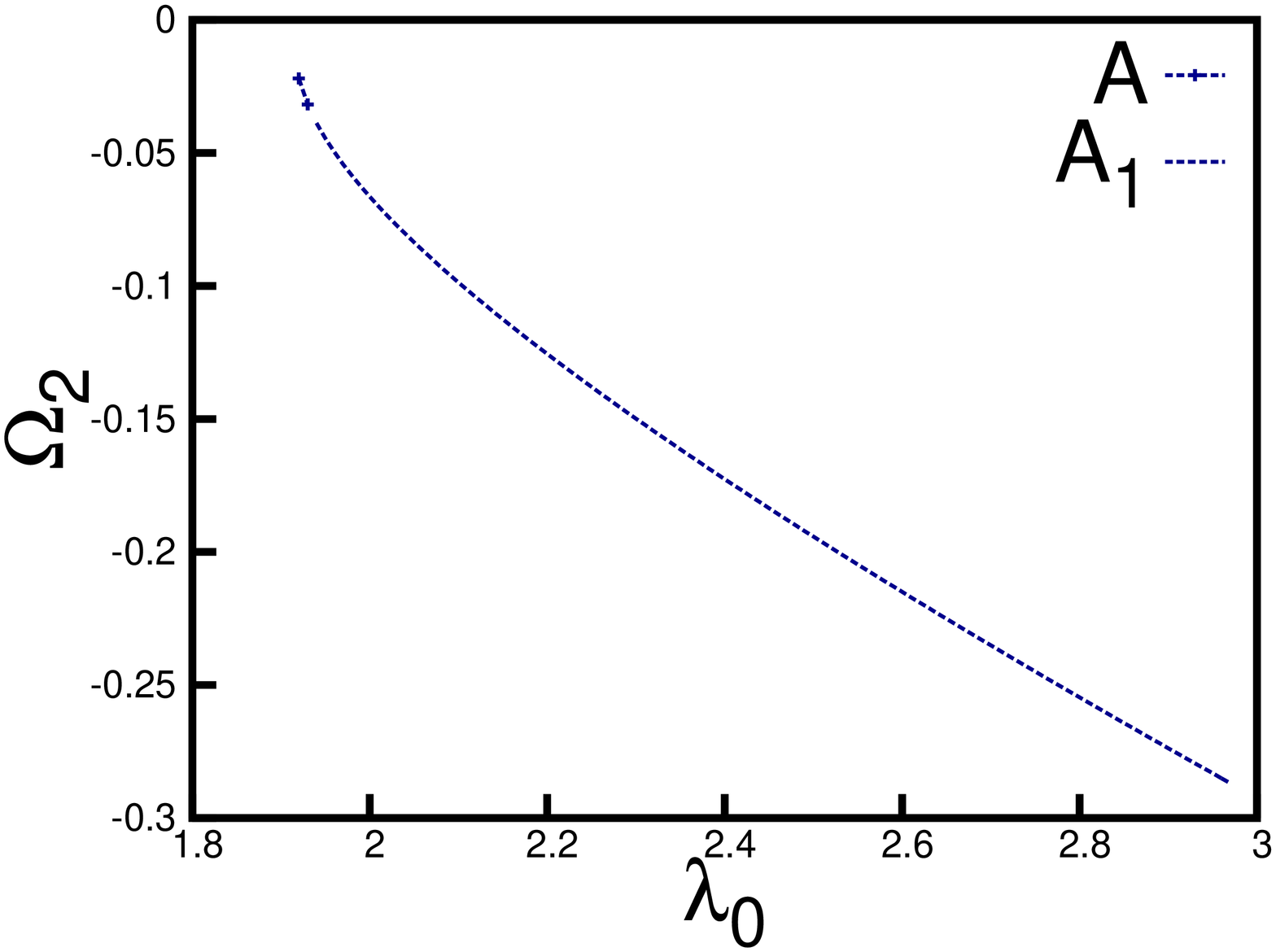,angle=270,width=1.5in}\\
\hline
&
\epsfig{file=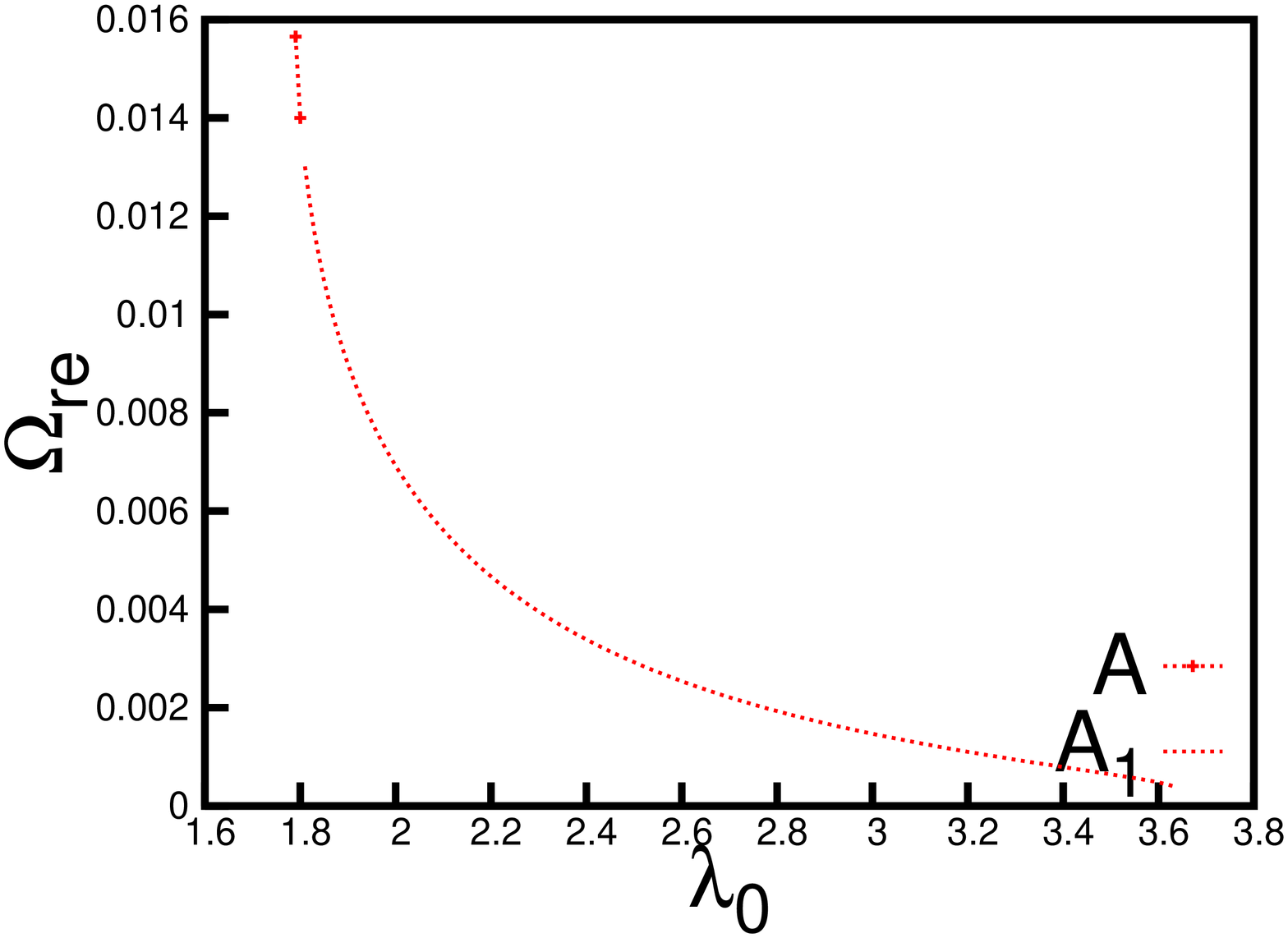,angle=270,width=1.5in}&
\epsfig{file=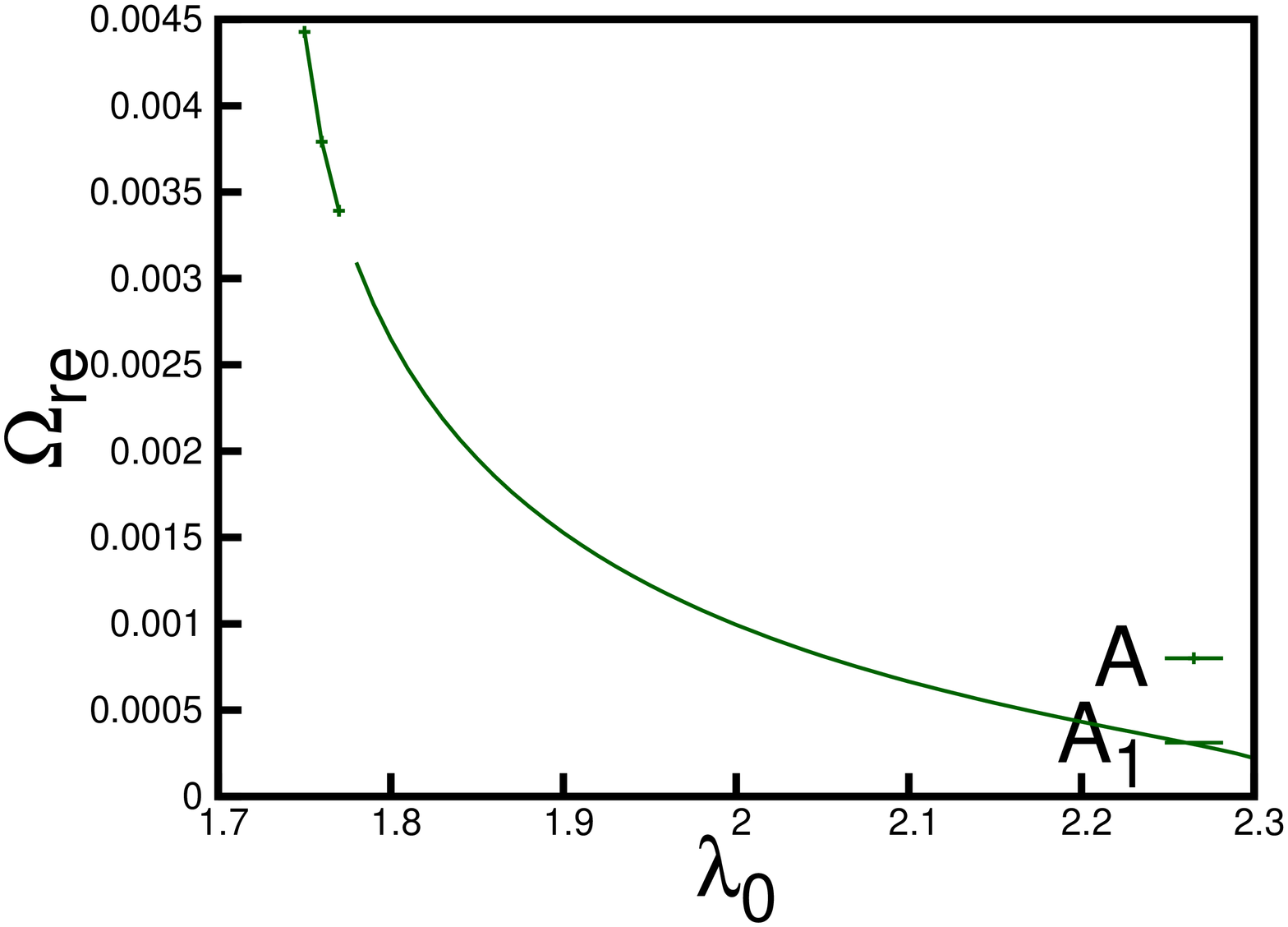,angle=270,width=1.5in}&
\epsfig{file=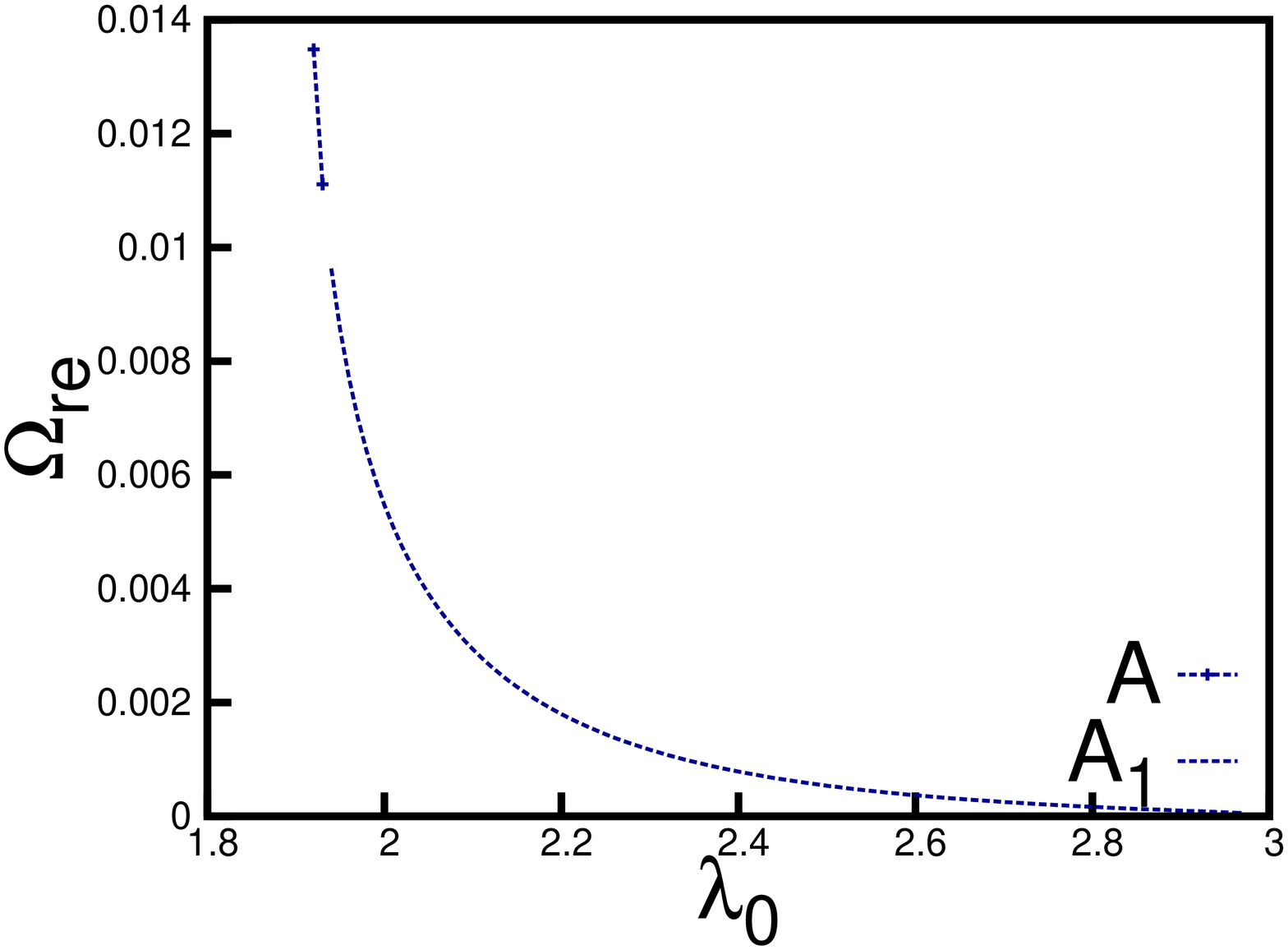,angle=270,width=1.5in}\\
Middle &
\epsfig{file=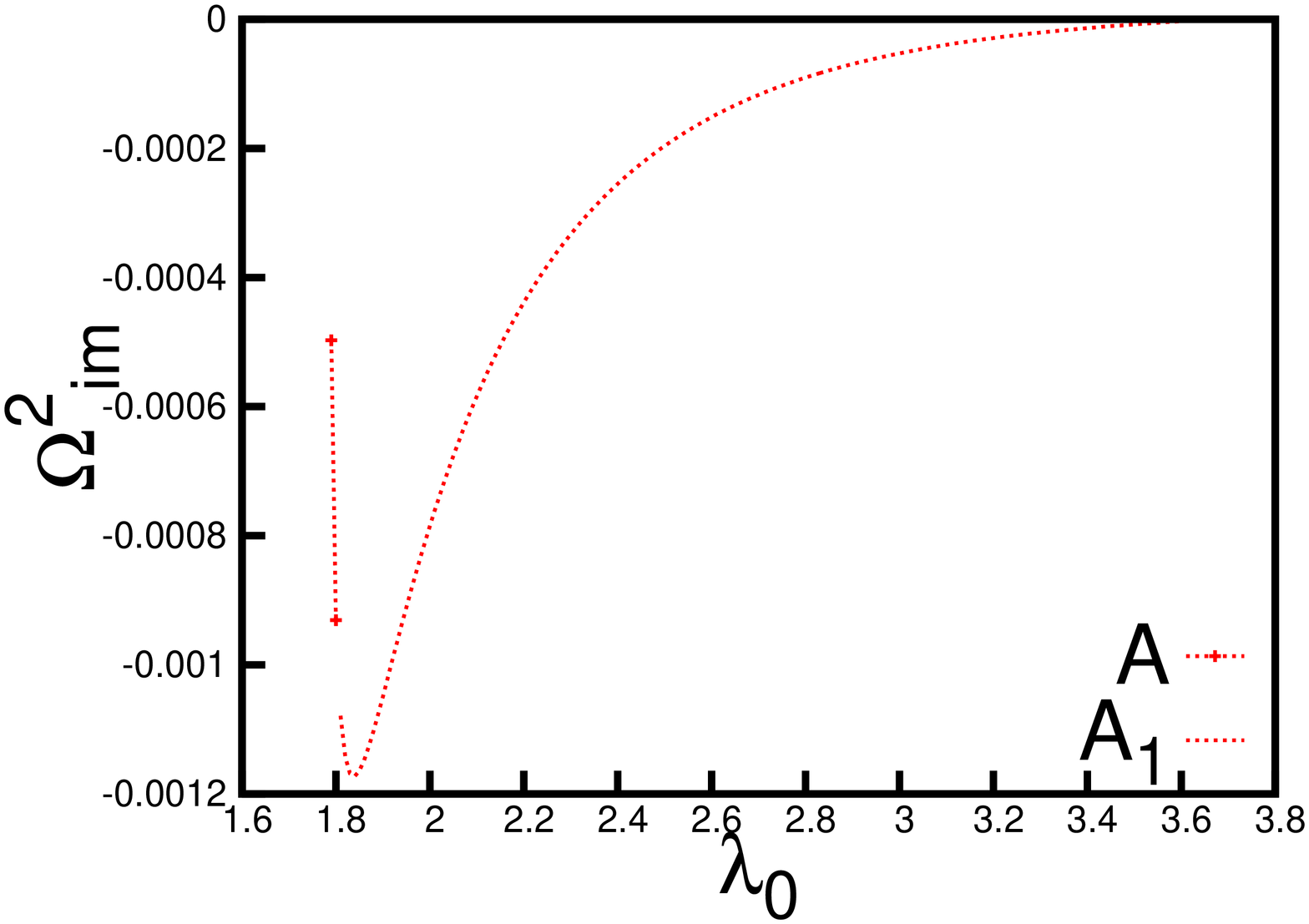,angle=270,width=1.5in}&
\epsfig{file=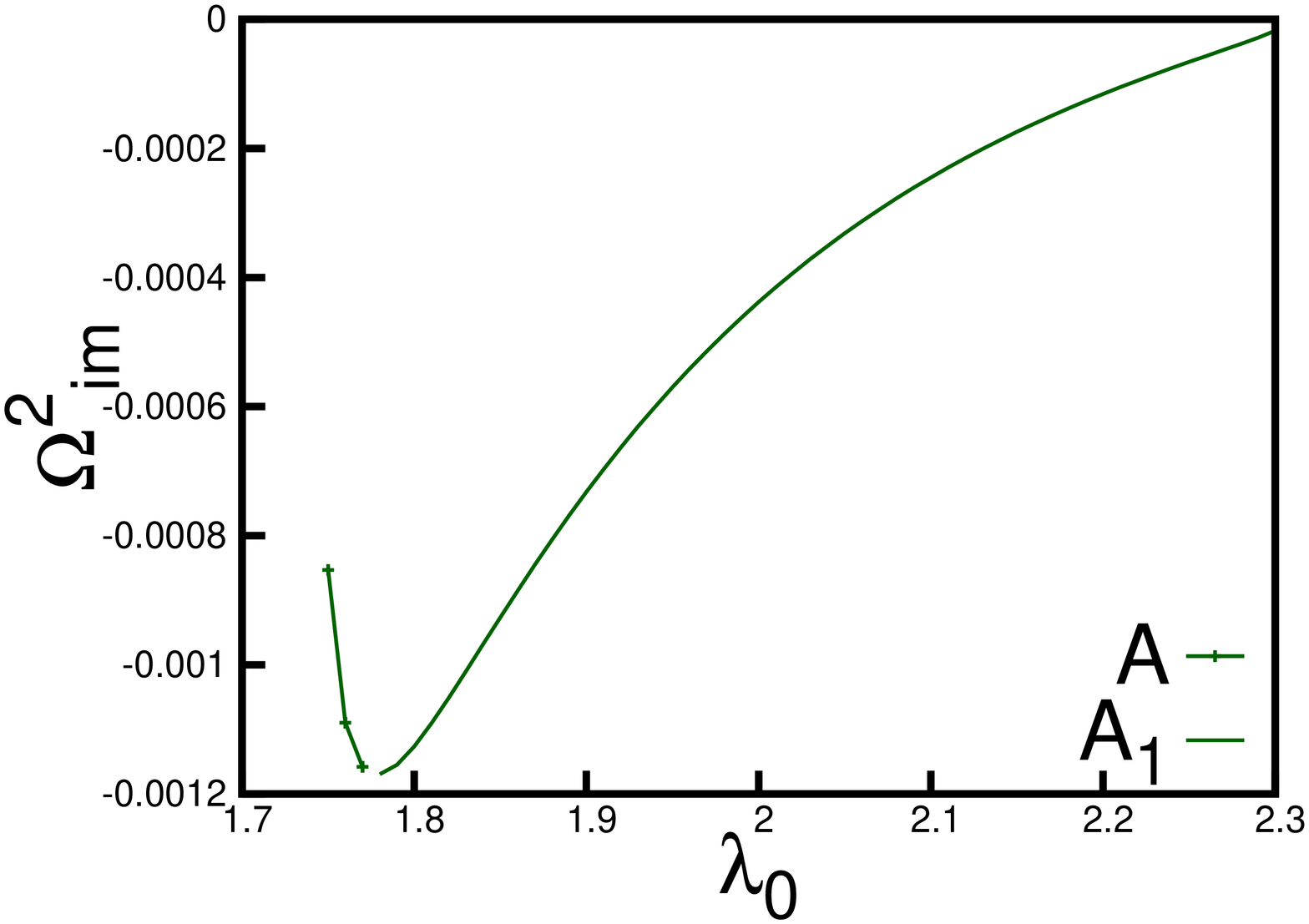,angle=270,width=1.5in}&
\epsfig{file=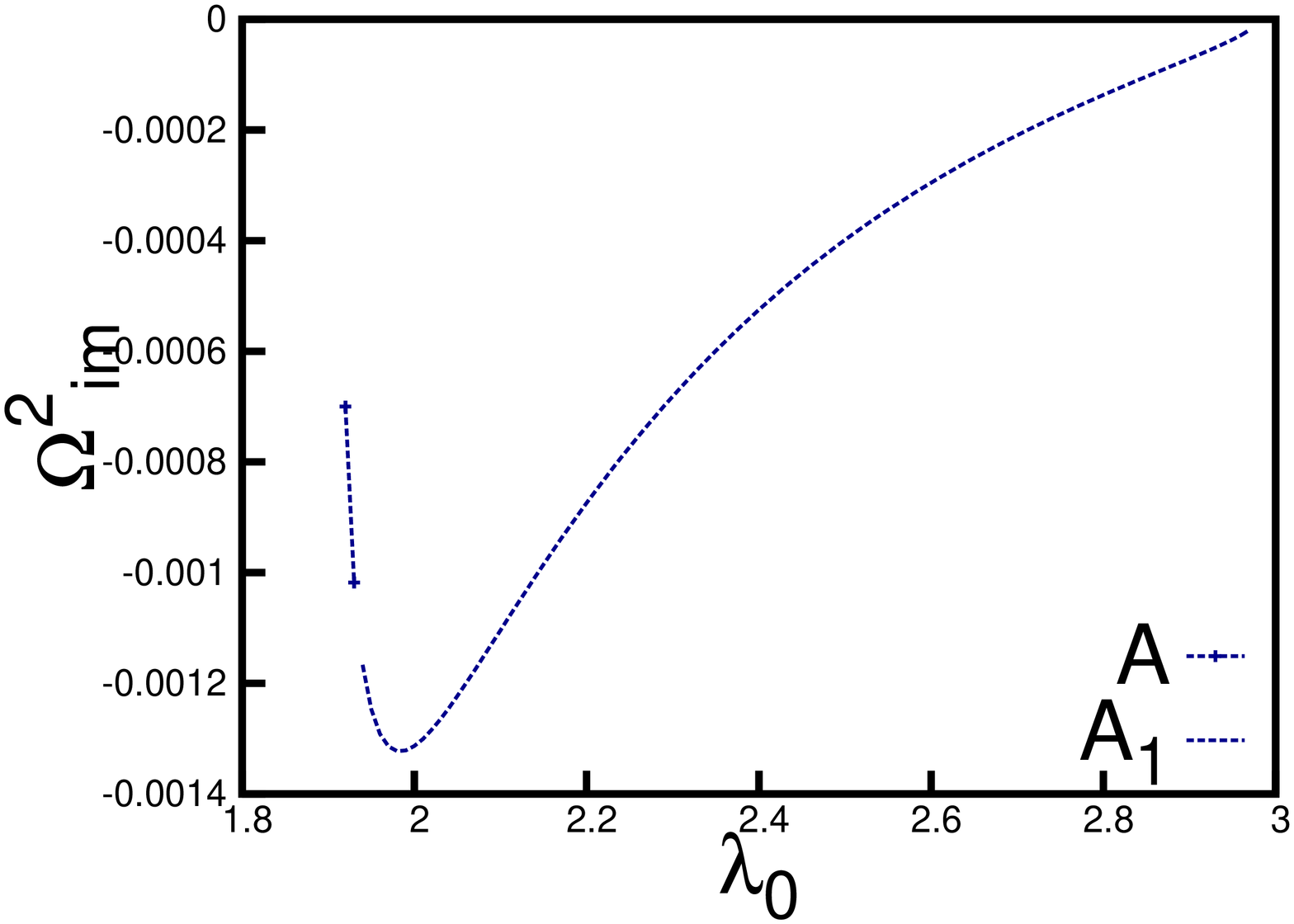,angle=270,width=1.5in}\\
%
\hline
&
\epsfig{file=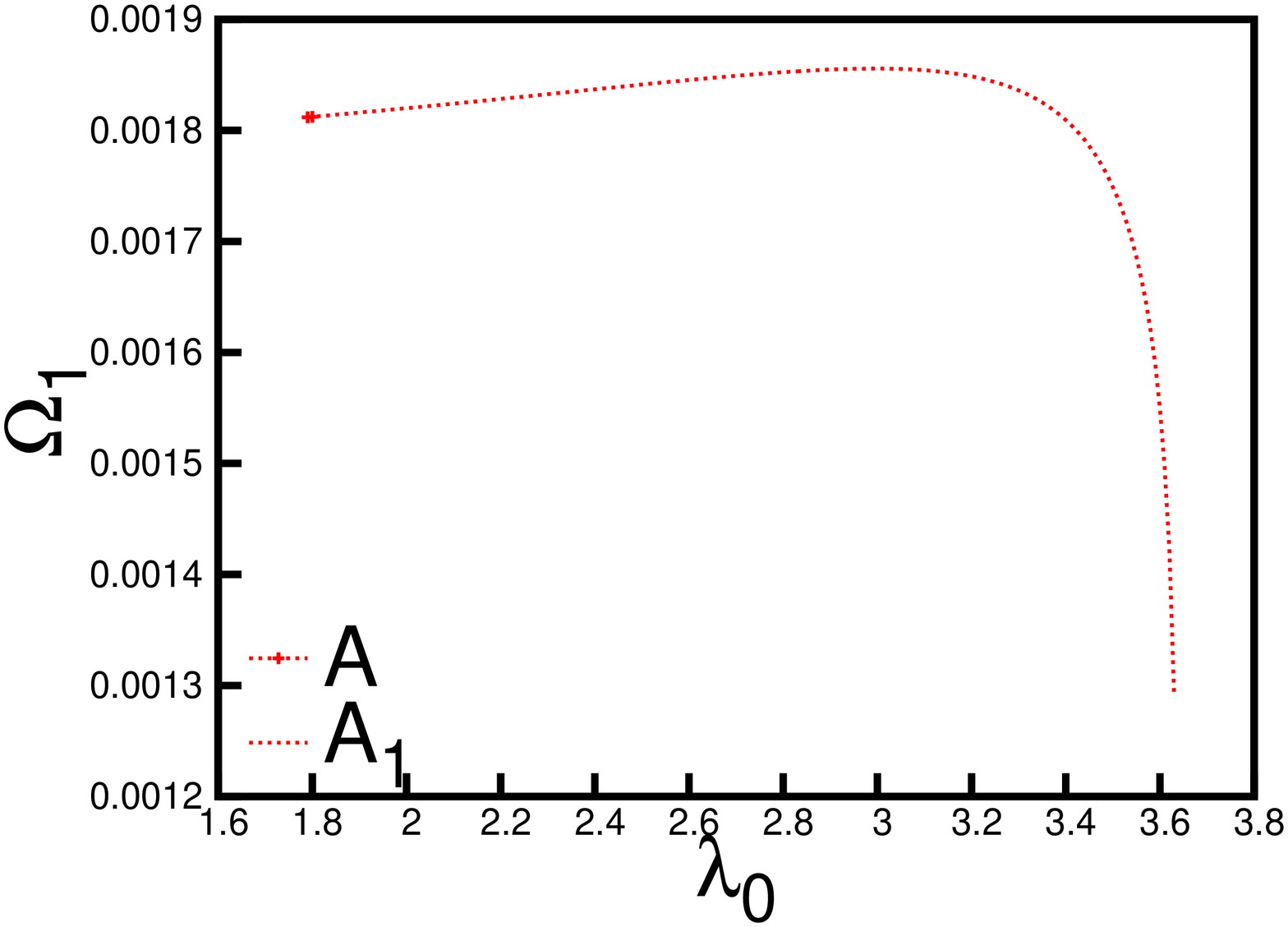,angle=270,width=1.5in}&
\epsfig{file=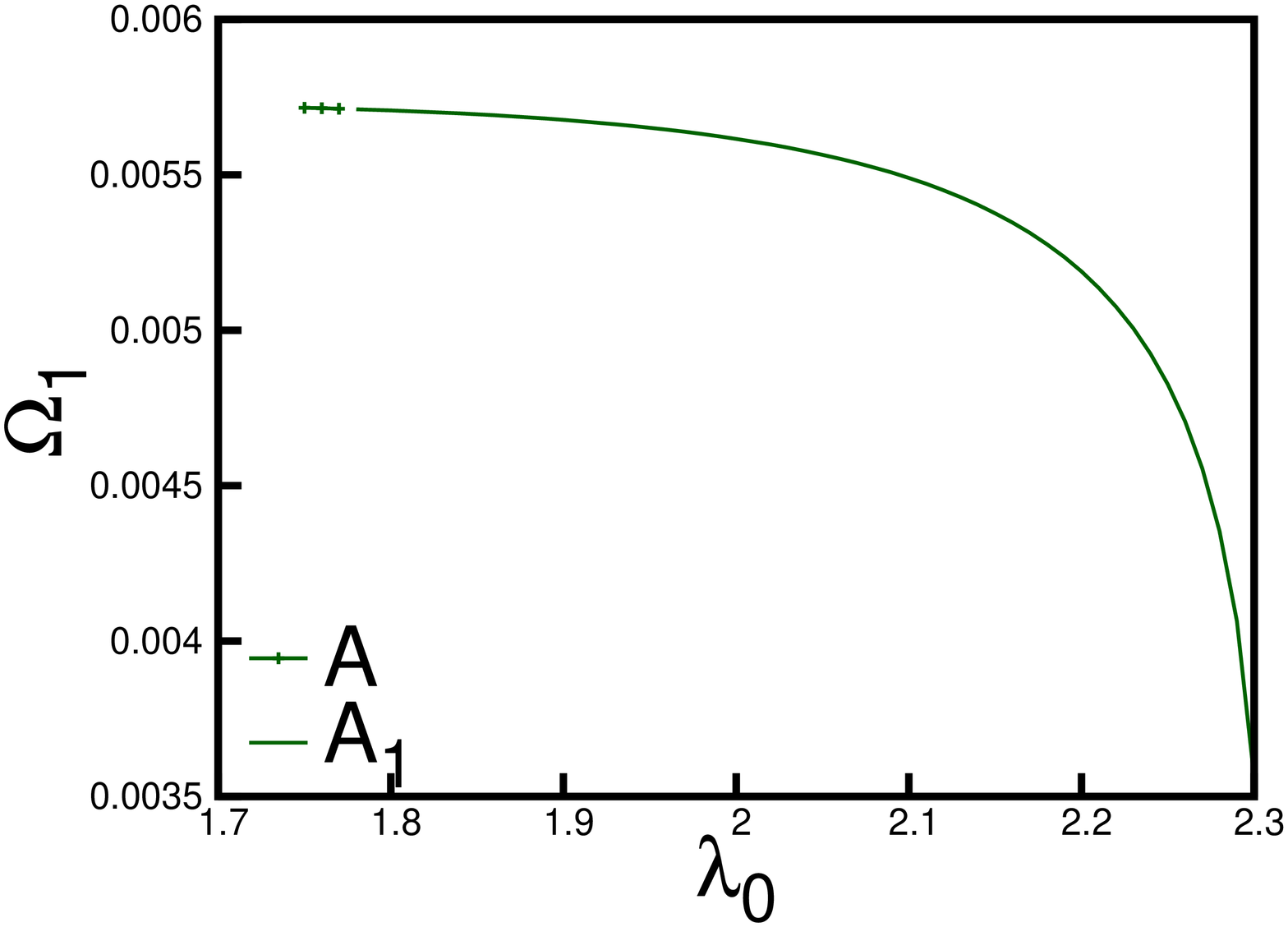,angle=270,width=1.5in}&
\epsfig{file=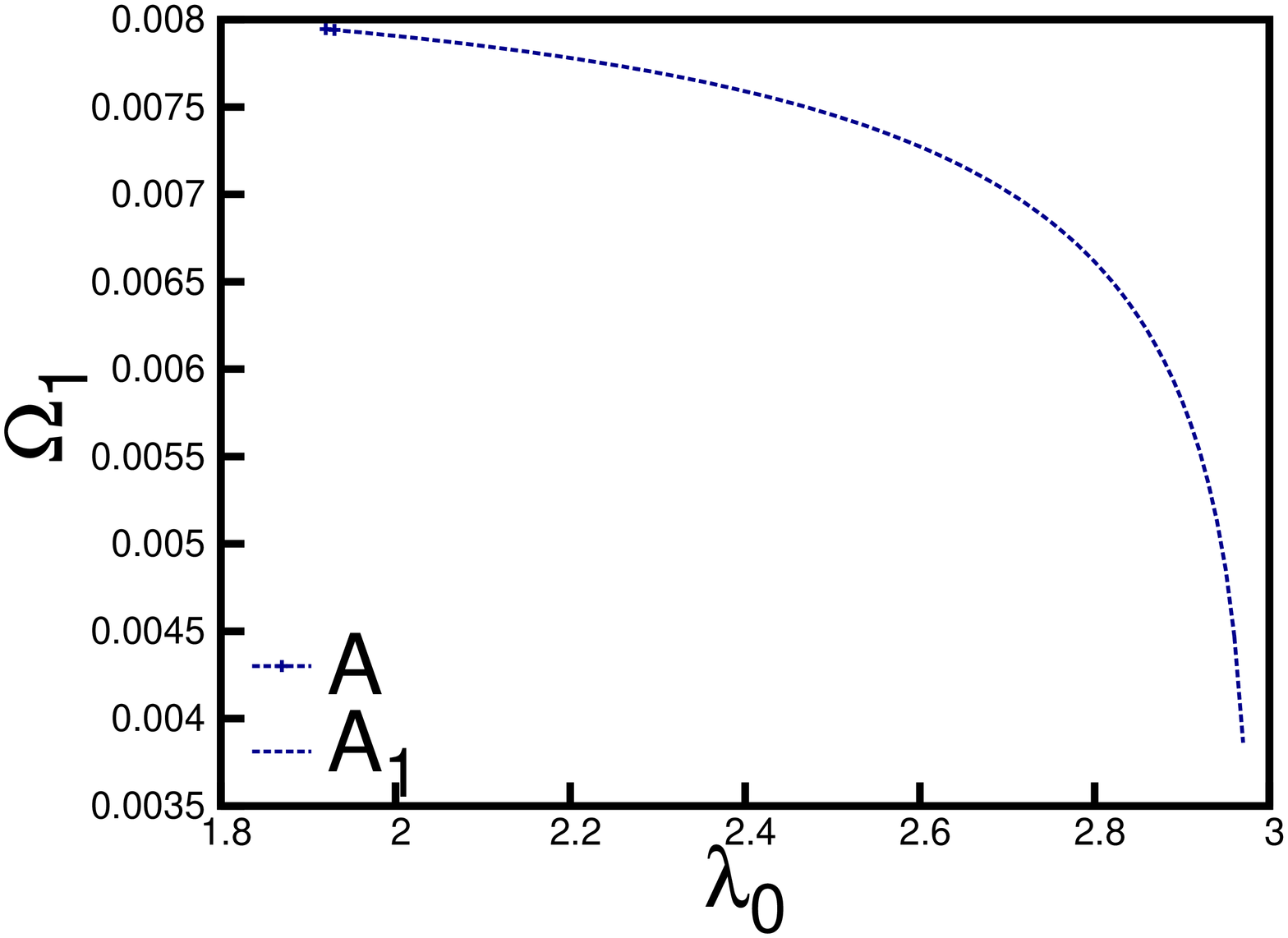,angle=270,width=1.5in}\\
Outer &
\epsfig{file=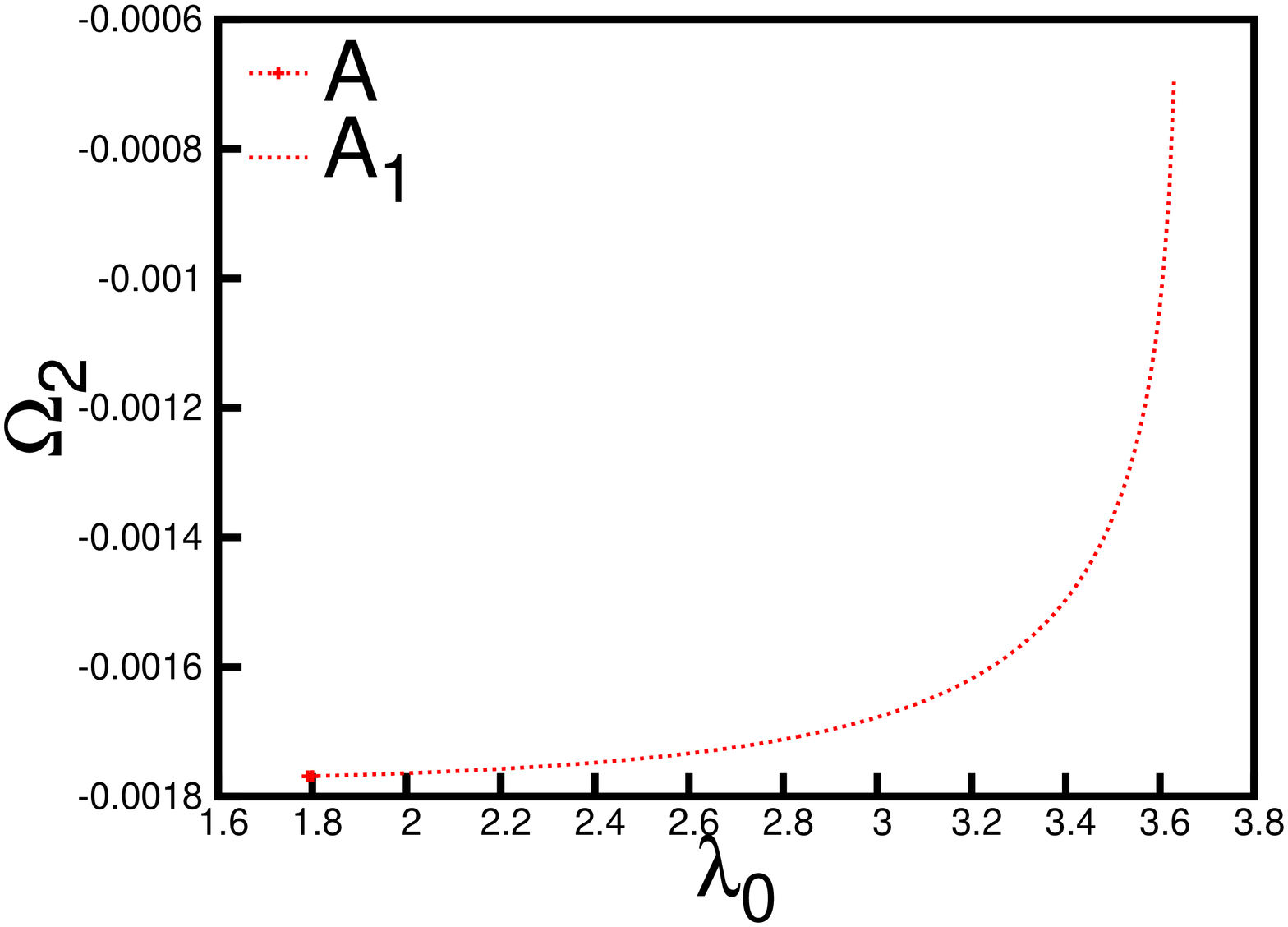,angle=270,width=1.5in}&
\epsfig{file=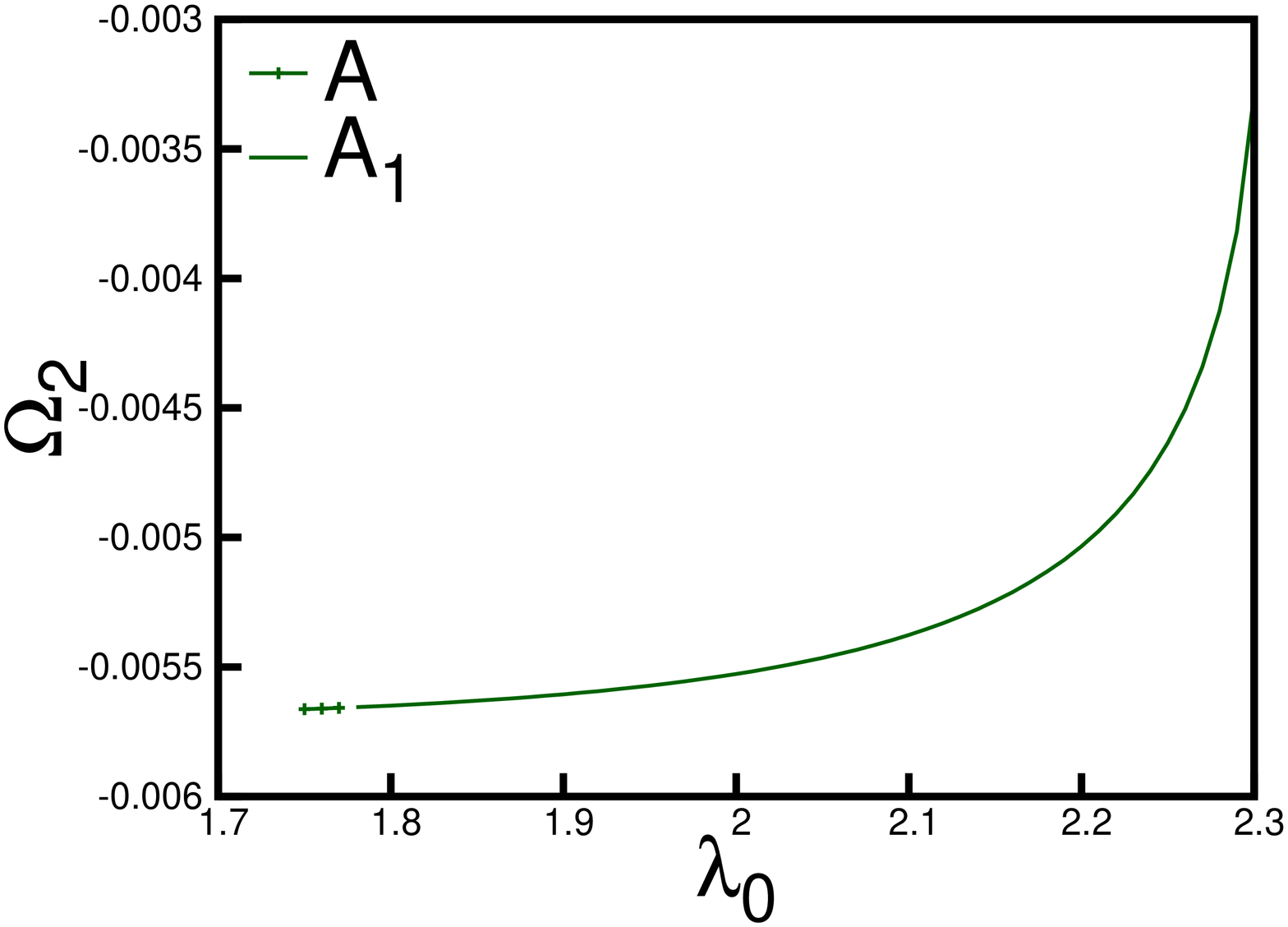,angle=270,width=1.5in}&
\epsfig{file=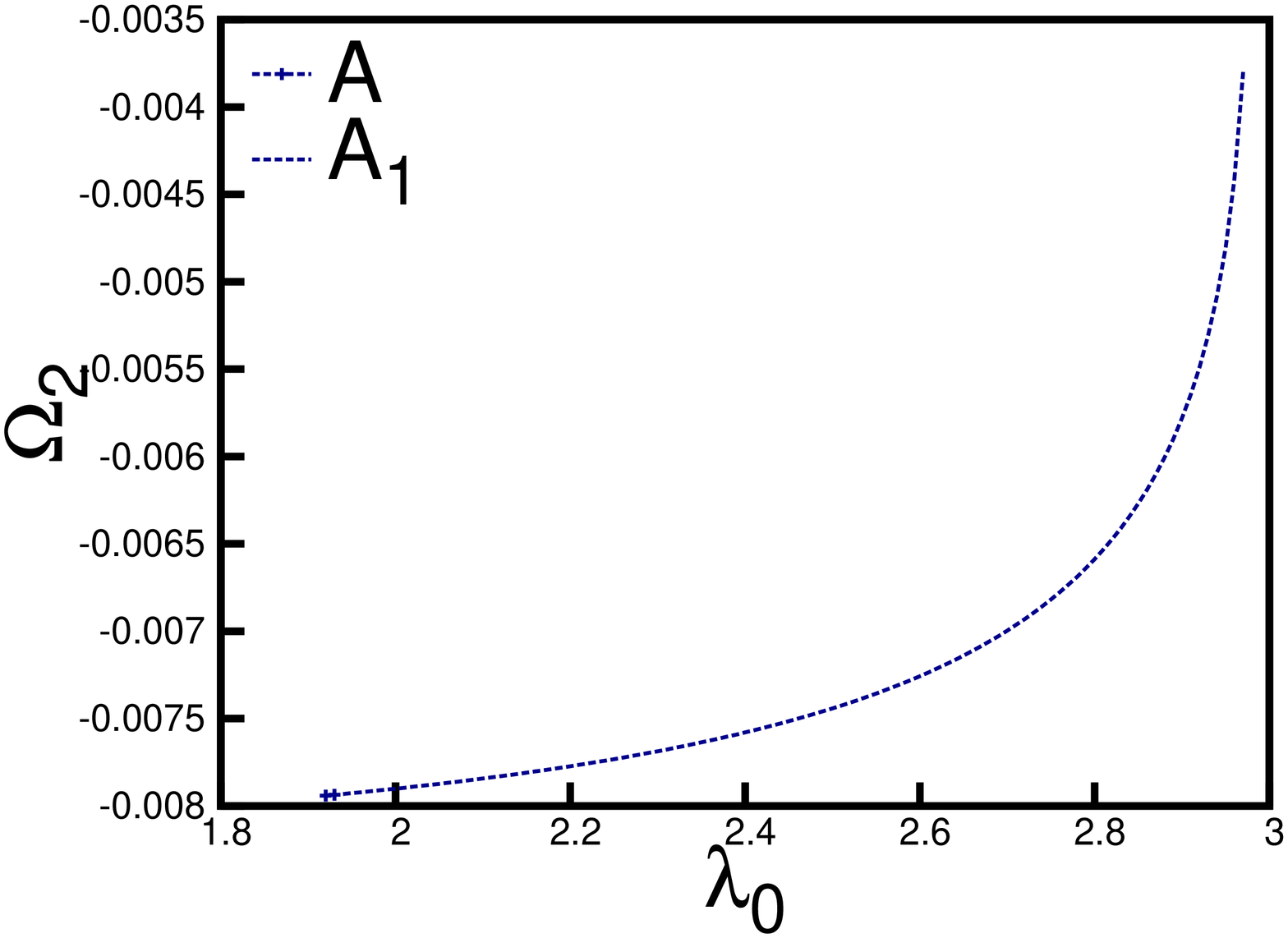,angle=270,width=1.5in}\\
%
\hline
\end{tabular*}
\caption{\small Variation of eigenvalues for the inner, middle and outer critical points (CP) for three different disk models (V, C \& H) under $\alpha$ = 0.1 for $\gamma=\frac{4}{3}$ and $\cal \dot{M}$ = $2 \times 10^{-5}$. $\rm A$ and $\rm A_1$ indicate multicritical (single critical) accretion solutions respectively. The dotted-red lines (dotted-red lines with points) represent multicritical (single critical) accretion solutions for vertical equilibrium geometry (V), the solid-green lines (solid-green lines with points)  represent multicritical (single critical) accretion solutions for conical geometry (C), and the
dashed-blue lines (dashed-blue lines with points) represent multicritical (single critical) accretion solutions for the constant-height disk geometry (H).}
\label{Fig:o2_a_0.1}
\end{figure*}

Next, we have followed the procedure described in section \ref{CP_nature} to assess the nature of the critical points (to visualise the flow topology) by calculating the $\Omega$ values for each critical point obtained by using equations \eqref{coeff_A} - \eqref{delta_defi}. In figures (\ref{Fig:o2_a_0.0} - \ref{Fig:o2_a_0.1}), we have presented the variation in eigenvalues against $\lambda_0$ for three critical points appearing for the range of $\lambda_0$ shown there
(in both the regions $\rm A$ and $\rm A_1$) for all the disk models (V, C \& H) for 
$0\leq\alpha\leq0.1$ with $\gamma=\frac{4}{3}$ and $\cal \dot{M}$=$2 \times 10^{-5}$ units. As expected, we obtain two equal and opposite real roots for $\Omega$ (i.e., $\rm \Omega_{1}$ and $\rm \Omega_{2}$) signifying saddles as inner and outer critical points and purely imaginary roots (i.e., $ \Omega_{\rm re}$ = 0 and $ \Omega^2_{\rm im}<$ 0) indicating centres as middle critical point with $\alpha$ = 0 (inviscid flow) for all the disk models. However, non-zero values of
$\alpha$ produce unequal real values, with opposite signs, of 
$\Omega_{1}$ and $\Omega_{2}$ for the saddles and $ \Omega_{\rm re} \ne 0 $ 
alongwith $\rm \Omega^2_{\rm im}<$ 0 indicating the switching over from the centre type middle critical points to spirals, according to the prescription already mentioned in the section~\ref{CP_nature}.

\newpage
\newpage

In the usual inviscid formalism, the stationary solutions are flows passing through the saddles or closed paths about a center \citep{Sonali_14,Chaudhury_2006,Mondal_2007,Goswami_2007}. However, the slightest presence of viscous perturbation alters the centers to spirals . Mathematically, $\alpha \ne 0$ implies 
$P\ne 0$  and vice versa. For any value of $P$, we obtain $Q<0$ indicating a saddle and $Q>0$ with $\Delta<0$ implying a centre (for $P = 0$) or a spiral (for $P\ne 0$) . 
Therefore, all the outer or inner critical points have been found to be saddle and no simple node (with $\Delta>0$ and $Q>0$) have been obtained for our accreting system. Nodal points, however, cannot be ruled out completely as \citep{Afshordi_Paczynski_2003} have shown. Therefore, the resultant picture that emerges as far as the phase portrait of the flow is concerned is that there exists adjacent saddle points and  spiral points, one type located after another.

Once the nature of all the physically relevant critical points have been understood, a complete qualitative picture of the flow solutions passing through these points (if they are saddle points), or in the neighbourhood of these points (if they are centre/ spiral points), can be constructed, along with an impression of the direction that these solutions can have in the phase portrait of the flow. A physical multitransonic flow can result connecting the inner and outer saddle points via formation of standing shocks \citep{Sonali_13,Chakra_1989,Chakra_Das_2004,Das_2007,
Fuku_Kaza_2007,Lanza_2008}. Under shock formation, the multicritical accretion solution implies that the flow connects infinity to event horizon passing through both the outer and inner critical points via shock formation which allows a discontinuous jump from the supersonic branch of the seperatrix for outer critical point to the subsonic part of the seperatrix of the inner saddle point for a physically allowed accretion process.

\section{Introduction of Time Dependent Perturbation}
So far we have analysed the characteristics of the stationary flows in this quasi-viscous scheme without bothering about the sustainability of such a flow. But for such a flow to exist, one has to investigate the stability of such a flow from the time dependent flow equations. In this section thus we attempt to study the linear stability analysis of the quasi-viscous accretion disk about its stationary solutions. Let us introduce small time dependent perturbations (marked by a \texttildelow$\:$ sign at the top of the respective variables) on the stationary solution of the flow variables (denoted by suffix $0$) at each radial distance $r$ in order to track the time evolution of the perturbation and make predictions about the stability of the background stationary configuration, determined by it. The perturbed quantities appear as,

\begin{subequations}
\begin{eqnarray}
v(r,t) = v_{0}(r)+ \tilde{v}(r,t), \\
\rho(r,t) = \rho_{0}(r)+ \tilde{\rho}(r,t), \\
f(r,t) = f_{0}(r)+ \tilde{f}(r,t), \\
g_{1}(r,t) = g_{10}(r)+ \tilde{g_{1}}(r,t), \\
g_{2}(r,t) = g_{20}(r)+ \tilde{g_{2}}(r,t), \\
F(r,t) = F_{0}(r)+ \tilde{F}(r,t).
\end{eqnarray}
\label{pert}
\end{subequations} 

To investigate the stability of the stationary solution, one has to identify whether these perturbing terms grow with time or not. Substituting these expressions given in equation \eqref{pert} into equation~\eqref{euler_eq1} one gets, keeping up to the linear perturbing terms,
\begin{equation}
\dfrac{\partial \tilde{v}}{\partial t} +  \dfrac{\partial}{\partial r}\left( v_{0} \tilde{v} + \dfrac{c_{s0}^{2} \tilde{\rho}}{\rho} \right) -\dfrac{2\alpha \lambda_{0}^2F}{r^{3}}=0.
\end{equation}
and the equation \eqref{modcon} gives,
\begin{equation}
\dfrac{\partial \tilde{g}_{1}}{\partial t} = -g_{20}\dfrac{\partial \tilde{f}}{\partial r} - \tilde{g_{2}} \dfrac{\partial f_{0}}{\partial r}.
\label{g1t}
\end{equation}

Again, 
\begin{equation*}
\dfrac{\partial \tilde{g}_{1}}{\partial t} = \dfrac{\partial g_{1}}{\partial t} = \dfrac{\mathrm d g_{1}}{\mathrm d \rho} \dfrac{ \partial \tilde{\rho}}{\partial t}
\end{equation*}
Now from \eqref{conf},\[\frac{\partial f_0}{\partial r}=\left.\frac{\dd f}{\dd r}\right|_0=0.\]
Hence from equation~\eqref{g1t},
\begin{equation}
\dfrac{\partial \tilde{\rho}}{\partial t} = -\dfrac{g_{20}}{\left.\dfrac{\mathrm d g_{1}}{\mathrm d \rho}\right|_{0}}\dfrac{\partial \tilde{f}}{\partial r}\label{rhot}
\end{equation}
From the relation $f( \rho ,v,r) = \rho vrH$ we get,\[\dfrac{\partial\tilde{f}}{\partial t}=\dfrac{\partial f}{\partial\rho}\dfrac{\tilde{\rho}}{\partial t}+\dfrac{\partial f}{\partial v}\dfrac{\tilde{v}}{\partial t};\] and hence,
\begin{equation}
\dfrac{\partial \tilde{\rho}}{\partial t} = \dfrac{1}{\left.\dfrac{\partial f}{ \partial \rho} \right|_{0}}\dfrac{\partial \tilde{f}}{ \partial t} - \dfrac{\left.\dfrac{\partial f}{\partial v_{0}} \right|_{0}}{\left.\dfrac{\partial f}{\partial v_{0}} \right|_{0}}\dfrac{\partial \tilde{\rho}}{\partial t}
\label{rhotilt}
\end{equation}
--- where $\partial f/\partial v|_0=\rho_0 rH_0=f_0/v_0$ and $\partial f/\partial\rho|_0=v_0rH_0=f_0/\rho_0$.
From the equations~\eqref{rhot} and~\eqref{rhotilt} we get
\begin{equation}
\dfrac{\partial \tilde{v}}{\partial t} = \dfrac{v_{0}}{f_{0}} \dfrac{ \partial \tilde{f}}{\partial t} + \dfrac{v_{0}^{2}}{f_{0	}}\dfrac{\partial \tilde{f}}{\partial r}.\label{vtilt}
\end{equation}
Finally after taking another $\partial/\partial t$ on the last equation and using equations~\eqref{rhot},\eqref{rhotilt} and \eqref{vtilt}, one gets an inhomogeneous wave equation,
\begin{equation}
\dfrac{\partial } {\partial t} \left[ \dfrac{v_{0}}{f_{0}} \dfrac{ \partial \tilde{f}}{\partial t}\right] + \dfrac{\partial}{\partial t} \left[\dfrac{v_{0}^{2}}{f_{0}} \dfrac{\partial \tilde{f}}{\partial r} \right] + \dfrac{\partial}{\partial r}\left[\dfrac{v_{0}^{2}}{f_{0}}\dfrac{\partial \tilde{f}}{\partial t} \right] + \dfrac{\partial}{\partial r} \left[\dfrac{v_{0}}{f_{0}} \left( v_{0}^{2}-\dfrac{c_{s0}^{2}}{1+\varepsilon} \right)\dfrac{\partial \tilde{f}}{\partial r} \right]-\dfrac{2 \alpha \lambda_{0}^{2}}{r^{3}}\dfrac{\partial \tilde{F}}{\partial t}=0\label{wave}
\end{equation}
\textbf{Calculation of $\tilde{F}$ and $\dfrac{\partial \tilde{F}}{\partial t}$}

$F$ can be decomposed into two parts such as $F = F_{s}+F_{N}$,
where $F_{s}$ is already defined in equation~\eqref{Fs} and
\begin{equation}
F_{N}= -\int \dfrac{2f_{1}}{f r^{3}}\left( \dfrac{1}{f}\dfrac{\partial f}{\partial r}\right) dr \label{Fn}
\end{equation} being the additional contribution under no-stationary condition.
Now after perturbation, the perturbed quantity $F_{s}=F_{s0}+\tilde{F_{s0}}$ and $F_{N}=\tilde{F_{N}}$; because at steady state $F_{N0}=0$\\
From equation~\eqref{Fs} we get
\begin{equation}
\dfrac{\tilde{F_{s}}}{\tilde{F_{s0}}} = \dfrac{\tilde{f}}{f_{0}}+\dfrac{\tilde{c}_{s}}{c_{s0}}-\dfrac{2\tilde{v}}{v_{0}}-\dfrac{\tilde{\rho}}{\rho_{0}}
\end{equation}
Since $\dfrac{\tilde{c_{s}}}{c_{s0}}=\dfrac{\gamma -1}{2} \dfrac{\tilde{\rho}}{\rho_{0}}$
\begin{equation}
\dfrac{\tilde{F_{s}}}{\tilde{F_{s0}}} = \dfrac{\tilde{f}}{f_{0}}-\dfrac{2\tilde{v}}{v_{0}}-\dfrac{\gamma -3}{2}\dfrac{\tilde{\rho}}{\rho_{0}}\label{Fstil}
\end{equation}

From equation~\eqref{rhotilt} it follows that 
\begin{equation}
\dfrac{\partial \tilde{\rho}}{\partial t} = -\dfrac{\rho_{0} v_{0}}{(1+ \varepsilon) f_{0}} \dfrac{\partial \tilde{f}}{\partial r}\label{rhotilN}
\end{equation}
Substituting equations~\eqref{rhotilN} and~\eqref{vtilt} in equation ~\eqref{Fstil}
\begin{equation}
\dfrac{\partial \tilde{F_{s}}}{\partial t} = -\dfrac{F_{s0}}{f_{0}} \left[ \dfrac{\partial \tilde{f}}{\partial t} + v_{0} \left( \dfrac{1+ \gamma +4 \varepsilon}{2( 1 + \varepsilon )} \right) \right]\label{FstilN}
\end{equation}

From equation~\eqref{Fn} we can write
\begin{equation}
\dfrac{\partial \tilde{F_{N}}}{\partial t}= -\left[ \dfrac{2f_{10}}{f_{0}^{2} r^{3}} \dfrac{\partial \tilde{f}}{\partial t} -2 \int \dfrac{\partial}{\partial r}\left( \dfrac{f_{10}}{f_{0}^{2} r^{3}} \right)\dfrac{\partial \tilde{f}}{\partial t} dr \right] \label{FntilN}
\end{equation}

Now combining equations~\eqref{Fs}, \eqref{FstilN} and~\eqref{FntilN},
\begin{equation}
\dfrac{\partial \tilde{F}}{\partial t} = \dfrac{c_{s0}}{\rho_{0} v_{0} r^{2}} \dfrac{(1+ \gamma + 4 \varepsilon)}{(1+ \varepsilon)} \dfrac{\partial \tilde{f}}{\partial r} + 2 \int \dfrac{\partial}{\partial r} \left( \dfrac{c_{s0}}{\rho_{0} v_{0}^{2}r^{2}}\right) \dfrac{\partial \tilde{f}}{\partial t} dr.
\end{equation}
\subsection{Solution of the Wave Equation}
Now choosing a trial wave solution,
\begin{equation}
\tilde{f}(r,t) = g_{\omega}(r)e^{-i \omega t},
\end{equation} one gets may explore the behavior of the spatial part.  Hence 
substituting the above in equation \eqref{wave}, the equation becomes,
\begin{equation}
\begin{split}
\omega^{2}v_{0}g_{\omega}^{2}+v_{0}^{2}i \omega \dfrac{\mathrm d g_{\omega}}{\mathrm dr} g_{\omega} + i\omega \dfrac{\mathrm d}{\mathrm dr}(v_{0}^{2}g_{\omega})g_{\omega} - g_{\omega} \dfrac{\mathrm d}{\mathrm dr}\left( v_{0} \left( v_{0}^{2} - \dfrac{c_{s0}^{2}}{1+ \varepsilon}\right) \dfrac{\mathrm d g_{\omega}}{\mathrm dr} \right) + &\\
 \dfrac{2 \alpha \lambda_{0}^{2} c_{s0}f_{0}g_{w}}{\rho_{0} v_{0} r^{5}}\left(\dfrac{1+ \gamma + 4 \varepsilon}{1+ \varepsilon} \right) \dfrac{\mathrm dg_{\omega}}{\mathrm dr} - \dfrac{4 \alpha \lambda_{0}^{2}i \omega f_{0}g_{\omega}}{r^{3}} \int \dfrac{\mathrm d}{\mathrm dr}\left( \dfrac{c_{s0}}{\rho_{0} v_{0}^{2}r^{2}}\right) g_{\omega} \mathrm dr &= 0.\label{wv}
\end{split}
\end{equation}
The solution of this equation may be of two types. One is of standing wave type, where $\tilde{f}$ takes the form $A_{\omega}(r,t)\exp(-i\omega_r t)$ where $A_{\omega}(r,t)$ is real and $\omega$ may be complex with a real part $\omega_r$. The other is of travelling wave type where $\tilde{f}$ takes the form $A_{\omega}(r,t)\exp\left[i(s_r(r)-\omega t)\right]$, both $A_{\omega}(r,t)$ and $\omega$ being real though $s_r(r)$ may be the imaginary part of a spatially dependent complex exponent $i s(r)$ in the solution.


\subsection{Standing Wave Analysis}
The stationary transonic solution connects the outer boundary at infinity to inner boundary at event horizon pass through the critical point (saddle type). Complying the respective boundary conditions, there may always be a rapid time variation or fluctuations at different points, which may be captured by a standing wave solution. 

Upon integrating the above equation~\eqref{wv} so that the surface integral terms, evaluated at the boundaries,  vanish we get finally something like
\begin{equation}
A \omega^{2}+B \omega+C = 0
\end{equation}
where
\begin{subequations}
\begin{eqnarray}
A &=& \int v_{0} g_{\omega}^{2} \mathrm dr, \\
B &=& - 4i \alpha \lambda_{0}^{2} \int \dfrac{g_{\omega} f_{0}}{r^{3}}\left[ \int g_{\omega} \dfrac{\mathrm d}{\mathrm dr} \left( \dfrac{c_{s0}}{\rho_{0} v_{0}^{2}r^{2}}\right) \mathrm dr\right] \mathrm dr, \\
C &=& \int v_{0} \left(v_{0}^{2} - \dfrac{c_{s0}^{2}}{1+ \varepsilon} \right) \left(\dfrac{\mathrm d g_{\omega}}{\mathrm dr} \right)^{2} + 2 \alpha \lambda_{0}^{2} \int \dfrac{c_{s0}f_{0}}{\rho_{0} v_{0} r^{5}}\left( \dfrac{1+ \gamma + 4 \varepsilon}{1+ \varepsilon} \right)g_{\omega} \dfrac{\mathrm d g_{\omega}}{\mathrm dr} \mathrm dr,
\end{eqnarray}
\end{subequations}
such that
\begin{equation}
Re(-i \omega) = \alpha \left [ \int v_{0} g_{\omega}^{2} \xi (r) \mathrm dr \right ] \left [ v_{0} g_{\omega}^{2} \mathrm dr \right ]^{-1} \sim \alpha \xi (r);
\end{equation}
where 
\begin{equation}
\xi (r) = \dfrac{2 \lambda_{0}^{2}}{r^{3} g_{w}} \dfrac{g_{10}}{g_{20}} \int g_{w} \dfrac{\mathrm d}{\mathrm dr} \left( \dfrac{c_{s0} g_{10}^{\left( \dfrac{1+2 \varepsilon}{1+ \varepsilon} \right) }}{g_{20}^{2}f_{0}^{2}r^{2}} \right) \mathrm dr.
\end{equation}
Evidently the $\exp(\alpha\xi(r))$ enters into $A_{\omega}(r,t)$ and the solution diverges making the situation unstable if $\xi(r)$ diverges to positive infinity at any particular length scale.

\subsection{Travelling Wave Analysis}
Disturbance may propagate too, which may be of travelling wave type.
Rearrangement of equation~\eqref{wv} gives
\begin{equation}
P \dfrac{\mathrm d^{2}g_{\omega}}{\mathrm d r^{2}} + Q \dfrac{\mathrm d g_{\omega}}{\mathrm d r} - R g_{\omega} + T \int g_{\omega} \dfrac{\mathrm d \sigma}{\mathrm dr} \mathrm dr = 0 \label{trwv}
\end{equation}
where 
\begin{subequations}
\begin{eqnarray}
P &=& v_{0}^{2}- \dfrac{c_{s0}^{2}}{1+\varepsilon} \\
Q &=&  \dfrac{1}{v_{0}} \dfrac{\mathrm d}{\mathrm dr} \left[ v_{0} \left( v_{0}^{2} - \dfrac{c_{s0}^{2}}{1+ \varepsilon} \right) \right] - 2i \omega v_{0} - \dfrac{2 \alpha \lambda_{0}^{2}c_{s0} f_{0}}{\rho_{0} v_{0}^{2} r^{5}} \left( \dfrac{1+ \gamma + 4 \varepsilon}{1+ \varepsilon}\right)\\
R &=& 2i \omega \dfrac{\mathrm dv_{0}}{\mathrm dr}+ \omega^{2}\\
T &=&  \dfrac{4 \alpha \lambda_{0}^{2}i \omega}{v_{0} r^{3}}
\end{eqnarray}
\end{subequations}
and
\begin{equation}
\sigma = \dfrac{c_{s0} f_{0}}{\rho_{0} v_{0}^{2}r^{2}} = -\dfrac{F_{s}}{2} = \dfrac{c_s}{vr}
\end{equation}
Taking the spatial part of the perturbation as 
\begin{equation}
g_{\omega} (r) = exp(is(r))
\end{equation}
where the function $s(r)$ itself is represented as a power series of the form 
\begin{equation}
i s(r) =\sum_{n=-1}^{\infty} \dfrac{k_{n}(r)}{\omega^{n}}
\end{equation}
The integral term in equation~\eqref{trwv} can, through some suitable algebraic substitutions, be recast as
\begin{equation}
\int g_{\omega} \dfrac{\mathrm d \sigma}{\mathrm dr} \mathrm dr = \int exp(s) \dfrac{\mathrm d \sigma}{ds} ds = g_{\omega} (r) \textit{S}
\end{equation}
where $\textit{S}$ itself is given by another power series as
\begin{equation}
\textit{S} = \sum_{n=-1}^{\infty} (-1)^{m+1} \dfrac{\mathrm d^{m} \sigma}{ds^{m}}
\end{equation}
To a leading order 
\begin{equation}
\textit{S} \sim \dfrac{\mathrm d \sigma}{\mathrm ds} \simeq \dfrac{\mathrm d \sigma}{\mathrm dr}\left( \omega \dfrac{\mathrm dk_{-1}}{\mathrm dr}\right)^{-1}
\end{equation}
Substituting for $g_{\omega}$ in equation~\eqref{trwv} and collecting the same powers of $\omega$ gives us coefficients of $\omega^{2}$,$\omega$ and $\omega^{0}$\\
Making coefficient of $\omega^{2} = 0$ gives
\begin{equation}
\left( v_{0}^{2} - \dfrac{c_{s0}^{2}}{1+ \varepsilon } \right) \left( \dfrac{\mathrm d k_{-1}}{\mathrm dr} \right)^{2}-2iv_{0} \dfrac{k_{-1}}{\mathrm dr}-1=0
\end{equation}
which leads to,
\begin{equation}
k_{-1} = \int \dfrac{i}{\dfrac{f_{0} g_{20}(r)}{g_{10} ( \rho )} \pm \dfrac{c_{s0}}{\sqrt{1+ \varepsilon}}} \mathrm dr
\end{equation}
Similarly making coefficient of $\omega = 0$ gives

\begin{equation}
\begin{split}
\left( v_{0}^{2} - \dfrac{c_{s0}^{2}}{1+ \varepsilon}\right) \left[ \dfrac{-i}{\left( v_{0} \pm \dfrac{c_{s0}}{\sqrt{1+ \varepsilon}}\right)^{2}} \left( \dfrac{\mathrm d v_{0}}{\mathrm dr} \pm \dfrac{\mathrm dc_{s0}}{\mathrm dr}\pm \dfrac{1}{\sqrt{1+ \varepsilon}}\right) + \dfrac{2i}{\left( v_{0} \pm \dfrac{c_{s0}}{\sqrt{1+ \varepsilon}}\right)} \dfrac{\mathrm dk_{0}}{\mathrm dr} \right]- 2iv_{0} \dfrac{\mathrm dk_{0}}{\mathrm dr}-\\
2i\dfrac{\mathrm dv_{0}}{\mathrm dr} - \dfrac{2 \alpha \lambda_{0}^{2} c_{s0} f_{0}}{\rho_{0} v_{0}^{2} r^{5}} \left(\dfrac{1+ \gamma + 4 \varepsilon}{1+ \varepsilon} \right) \dfrac{i}{\left( v_{0} \pm \dfrac{c_{s0}}{\sqrt{1+ \varepsilon}}\right)} +\\
 \dfrac{1}{v_{0}} \dfrac{i}{\left( v_{0} \pm \dfrac{c_{s0}}{\sqrt{1+ \varepsilon}}\right)} \left[ 2v_{0}^{2} \dfrac{\mathrm dv_{0}}{\mathrm dr}- 
\dfrac{2c_{s0}v_{0}}{1+ \varepsilon} \dfrac{\mathrm d c_{s0}}{\mathrm dr} + v_{0}^{2} \dfrac{\mathrm dv_{0}}{\mathrm dr} - \dfrac{c_{s0}^{2}}{1+ \varepsilon} \dfrac{\mathrm d v_{0}}{\mathrm dr} \right] = 0
\end{split}
\end{equation}
which gives us
\begin{equation}
k_{0} = -\dfrac{1}{2}ln\left(\dfrac{f_{0} g_{20}(r) c_{s0}}{\sqrt{1+ \varepsilon} g_{10} (\rho )} \right) \pm \int \alpha \lambda_{0}^{2} c_{s0}^{2}f_{0} \left( \dfrac{1 + \gamma + 4 \varepsilon}{\sqrt{1+ \varepsilon}} \right) \left(\dfrac{f_{0} g_{20}(r)}{g_{10}(\rho)} \mp \dfrac{c_{so}}{\sqrt{1+ \varepsilon}} \right)^{-1} \mathrm dr
\end{equation}
With large enough value of $\omega$, characteristic to any noise as perturbation, these leading order terms determine the nature of the solution.

\subsection{Stability under perturbations}
For standing wave analysis the behavior of $\xi (r)$ determines how rapidly the perturbation grows. Now after putting the specific  forms of $g_1$, $g_2$ and $\epsilon$, for the standing wave solutions one gets for H model, $\xi (r) \sim r^{-2}$ while for C, $\xi (r) \sim r^{1}$ and for V, $\xi (r) \sim r^{5/2}$; which readily implies  that at large radial distance for C \& V the perturbation rapidly grows with time making the disk unstable but for H it diminishes with radial distance making the disk practically stable for large amount of time. 
Thus in the case of standing wave solution the asymptotic behaviour of $\xi (r)$ shows the disk may become highly unstable if it spans large enough radial distances; though there is a question of time scale that ought to be within astrophysically realisable timescale. If $\xi(r)$ diverges at some $r$, the disk eventually gets destabilized at that distance for whatever small value of $\alpha$ under the present quasi-viscous scheme. 

But actually this disk analysis is valid only under the condition $\alpha F<<1$. As,\[F\approx F_s=\dfrac{-2c_sH}{vr}= \dfrac{-2 \rho^{2\epsilon+1}h^2c_s\bar g^2(r)}{f},\] it is evident that $\xi (r) \sim {\bar g^{3}(r)}/{r^2}$ cannot diverge within this limit, irrespective of the flow geometry (the minus sign comes because $v$ and $f$, both are negative for accretion).

In the case of traveling wave too, it is easy to see that $k_{-1} \sim r$ and for $\alpha = 0$, $k_{0} \sim \ln(r)$. But if $\alpha \neq 0$, $k_{0} \sim \alpha \bar g(r)$ which may make $i s(r)$ diverge for $r \rightarrow \infty$. But again $\alpha F<<1$ puts restriction on that. The rest of the terms, with higher powers of $1/\omega$-- where $\omega$ is generally large enough ($\gg 1$), are not expected to make the whole thing diverge. So here too within the region where the present scheme of viscosity remains valid, there is no imminent threat to the stability of the disk.

Here this analysis is valid only under the condition when the disk is formed within $\alpha F_s<<1$ and only under this condition the disk is stable. As, \[F_s=\dfrac{-2 \rho^{2\epsilon+1}h^2c_s\bar g^2(r)}{f},\] the validity of the model is restricted within the region where, 
\[\bar g^2(r)<<\dfrac{f}{-2 \rho^{2\epsilon+1}h^2c_s}\] which may be estimated by plugging in corresponding values of the quantities in the right hand side. 

\section{Concluding Remarks}
In the Sakura-Sunayev model, viscosity is introduced in the entire set of equations governing the disk dynamics to understand the flow profile. But, it takes a lot of mathematical difficulties to solve the equations explicitly. On the other hand in this present quasi-viscous approach, viscosity is introduced as a perturbative correction to the constant specific angular momentum in the radial Euler equation for inviscid flow, which saves a lot of mathematical complexities. Our quasi-viscous approach, however, as shown is valid within a limited region of the disk depending upon the value of viscosity parameter, $\alpha$. But, within this region we compared the qualitative nature of the transonic accretion flow topologies over three different disk geometries. The formulation of the problem and the stability analysis was done in a single parametrized form for all the three geometries. This analysis too has been done here without explicit numerical integration by using the methodology of dynamical system analysis. The stationary solutions conform with the earlier studies using Sakura-Sunayev scheme \citep{Chakra_Das_2004} and quasi viscous scheme \citep{JKB_2009}-- both of which dealt with vertical hydrostatic equilibrium disk model only. 

In some earlier works \citep{JKB_2007,JKB_2009} the respective authors identified the existence of a secular instability under linear perturbation about the stationary solutions using the time dependent equations raising doubt about the sustainability of such a stationary flow. Here too, we identified the same sort of secular instability -- at least in another disk geometry, along with the previously identified one -- asymptotically at a large length scale. But, we have shown that such divergence in the time evolution of the perturbation cannot take place within the length scale in which the model itself is applicable. That conclusion is valid both in stationary and travelling wave type perturbations on the stationary flow background. Hence we conclude that the stationary flow sustains at least for a considerable period of time for sufficiently low value of viscosity parameter ($0<\alpha\leq0.1$) for which the present approach stands. 

\vskip 10mm

\noindent
{\bf Acknowledgements}

SN and DA would like to acknowledge the kind hospitality provided by HRI, Allahabad,
India, for several visits. The work of SN has been partially
supported by the UGC MRP grant (sanction no: F. PSW -- 163/13-14). The work of TD and SC has been partially supported by the
astrophysics project fund under the XII th plan at HRI, Allahabad.



\end{document}